\documentclass{amsart}

\usepackage{style}
\usepackage[font=footnotesize]{caption, subcaption}
\usepackage{xcolor}
\usepackage{soul}
\usepackage{todonotes}
\DeclarePairedDelimiter{\norm}{\lVert}{\rVert}

\title[FV-based ROMs for Incompressible Flows in Parametrized Domains]{A Comparative Study of Finite-Volume-based Coupled and Segregated Reduced-Order Models for Incompressible Flows in Parametrized Domains}
\author[A. Buffolini, D. Oberto, G. Rozza]{Andrea Buffolini$^{\dagger}$, Davide Oberto$^{\dagger}$, Gianluigi Rozza$^{\dagger}$}
\date{\today}

\address{
$^{\dagger}$ Mathlab, Mathematics area, SISSA, Via Bonomea 265, 34136, Trieste, Italy.}
\email{\{abuffoli, doberto, grozza\}@sissa.it}

\begin{document}

\begin{abstract}
This work presents a comparative analysis of Reduced-Order Models (ROMs) applied to incompressible fluid dynamics within geometrically parametrized domains. Two distinct reduced-order solution strategies are investigated and compared: a monolithic coupled solver and a segregated SIMPLE-based algorithm. Their performance is assessed on two steady, two-dimensional benchmark cases: a lid-driven cavity flow and a flow past a cylindrical obstacle. The two algorithms are compared in terms of fields evaluation and aerodynamic coefficients prediction. The computational results highlight a fundamental trade-off between accuracy and numerical efficiency. On the one hand, after an opportune supremizers enrichment, the coupled approach guarantees a faster convergence, despite the need of a larger number of degrees of freedom. On the other hand, the segregated SIMPLE algorithm yields superior reconstruction accuracy, particularly for lower-dimensional reduced spaces, at the cost of a slower convergence rate.
\end{abstract}
\maketitle
\section{Introduction}

A large variety of mathematical models is based on a Partial Differential Equation (PDE) formulation. Nonetheless, for most of these problems, looking for analytical solutions may be a really challenging, or even impossible, task. Therefore, numerical methods for PDEs have widely spread in the last decades, proposing strategies and algorithms capable of providing discretized solutions to the continuous models.

Despite the massive computational resources currently available, numerical problems requiring the repeated evaluation of a model under varying configurations (the so-called \textit{many-query} context) or demanding real-time solutions, still pose a significant challenge for classical numerical techniques. In these scenarios, Reduced-Order Models (ROMs) \cite{quarteroni2015reduced, quarteroni2014reduced, hesthaven2016certified} play a crucial role by drastically lowering the computational burden while preserving a high degree of accuracy.

A major area of application for ROMs is the solution of parametric PDEs, where parameters may govern physical properties (e.g., fluid viscosity \cite{stabile2018finite}) or geometrical variations (e.g., the shape of a domain \cite{stabile2020efficient}). Specifically, ROMs are extensively adopted in Computational Fluid Dynamics (CFD) \cite{HIJAZI2020109513, Zancanaro_2022_laminar, Zancanaro_2022_turbulent,Ivagnes2025,Oberto2026,Ivagnes_Khamlich_Siena}, where \textit{full-order} numerical simulations of the Navier-Stokes equations typically yield prohibitive computational costs and CPU time.

Following classical CFD strategies \cite{ferziger2019computational, patankar2018numerical}, different numerical techniques can be implemented at the reduced level. These fundamentally divide into \textit{monolithic} (or \textit{coupled}) \cite{stabile2018finite} and \textit{segregated} \cite{Zancanaro_2022_laminar} approaches. The former typically offer robust pressure-velocity coupling at the cost of higher dimensionality and the need for stabilization techniques (e.g., supremizer enrichment \cite{ballarin2015supremizer}). Nonetheless, coupled approaches ensure faster convergence, in terms of non-linear iterations. Conversely, segregated ones decouple the fields, potentially reducing the computational burden but introducing numerical sensitivities \cite{stabile2020efficient}. Particularly, the under-relaxation coefficients adopted in the iterative numerical scheme have to be opportunely tuned and this may imply a slower convergence rate.\\
To the best of the authors' knowledge, the existing literature lacks a direct and explicit comparison between these two methodological families within a reduced-order framework: filling this gap constitutes the primary novelty of the present work.

Geometric deformations are investigated in two distinct benchmark cases: a lid-driven cavity flow \cite{Ghia1982-dn}, where the cavity domain is randomly perturbed, and a flow past a cylindrical obstacle \cite{fornberg1980numerical}, whose cross-section is geometrically parametrized. To the best of the authors' knowledge, the application of reduced-order techniques to the geometric parametrization of the latter benchmark has not yet been explored in the literature.

At the full-order level, the Finite Volume (FV) method \cite{ferziger2019computational} is employed, coupled with Radial Basis Function (RBF) interpolation to handle the internal mesh motion \cite{DEBOER2007784}. Subsequently, POD-Galerkin projections are adopted to define the reduced-order formulations of the governing equations.

The main contributions of the present work consist of:
\begin{itemize}
    \item a structured comparison between two distinct reduced-order architectures, namely a coupled and a segregated approach, evaluated through two benchmark test cases subject to geometric deformations: a lid-driven cavity flow and a flow past a cylindrical obstacle;
    \item a comprehensive analysis of the supremizer enrichment within the coupled framework, in the FV context, to achieve the highest possible stability for the numerical scheme;
    \item an investigation into the under-relaxation coefficients of the reduced SIMPLE algorithm to maximize the convergence rate of the segregated scheme;
\end{itemize}

The manuscript is structured as follows: Sections \ref{mathematical background} and \ref{ROMs theory} introduce the theoretical and numerical foundations of the paper. Specifically, the former details the full-order formulation of the problem, focusing on the FV discretization and the SIMPLE algorithm. Conversely, the latter addresses the reduced counterpart, providing a brief overview of ROMs and presenting the reduced algorithms adopted for the numerical experiments. In Section \ref{lid driven cavity}, the lid-driven cavity flow is investigated, whereas Section \ref{cylinder} explores the cylindrical obstacle benchmark. Finally, Section \ref{conclusions} draws the main conclusions and outlines potential future developments.
\section{Mathematical and numerical background}\label{mathematical background}
\subsection{The geometrically-parametrized steady Navier-Stokes equations}
Steady incompressible Navier-Stokes equations are considered as mathematical model, as they represent a valid approximation for the study of laminar flows \cite{chorin1990mathematical}. They can be expressed as
\begin{equation}\label{incompressible steady N-S}
    \begin{dcases}
            (\boldsymbol{u} \cdot \nabla) \boldsymbol{u} = -\dfrac{\nabla p}{\rho} + \nu \Delta \boldsymbol{u} \: , \qquad \mathrm{in} \: \Omega(\mub)\\
            \nabla \cdot \boldsymbol{u} = 0 \: , \qquad \mathrm{in} \: \Omega(\mub) \\
            \boldsymbol{u} (\boldsymbol{x}) = f(x) \:, \qquad \mathrm{on} \: \Gamma_D(\mub) \\
            (\nu \nabla \boldsymbol{u} - p \boldsymbol{I}) \boldsymbol{n} = 0 \:, \qquad \mathrm{on} \: \Gamma_{N}(\mub)
    \end{dcases}
\end{equation}
where:
\begin{itemize}
    \item $\mub \in \mathcal{P} \subseteq \mathbb{R}^P$ are (geometric) parameters, expressing the variation of the considered domain $\Omega(\mub) \subseteq \mathbb{R}^d$ with respect to an undeformed reference one $\Omega\subseteq \mathbb{R}^d$;
    \item $\boldsymbol{u} : \Omega(\mub) \longrightarrow \mathbb{R}^d$ ($d = 2,3$) is the fluid velocity field;
    \item $p : \Omega(\mub)\longrightarrow \mathbb{R}$ stands for the pressure field;
    \item $\Gamma_D(\mub)$, $\Gamma_N(\mub) \subseteq \partial\Omega(\mub)$ are the Dirichlet and Neumann boundaries, such that $\Gamma_D(\mub)\cup \Gamma_N(\mub) = \partial\Omega(\mub)$ and $\Gamma_D(\mub)\cap \Gamma_N(\mub) = \emptyset$;
    \item $\rho$, $\nu \in\mathbb{R}$ are the fluid density and the kinematic viscosity respectively, both assumed to be constant hereinafter.
\end{itemize}
\subsection{The Finite Volume discretization}\label{fvm discretization FOM algorithm} 
\begin{figure}[!htb]
        \centering
        \includegraphics[width=0.7\linewidth]{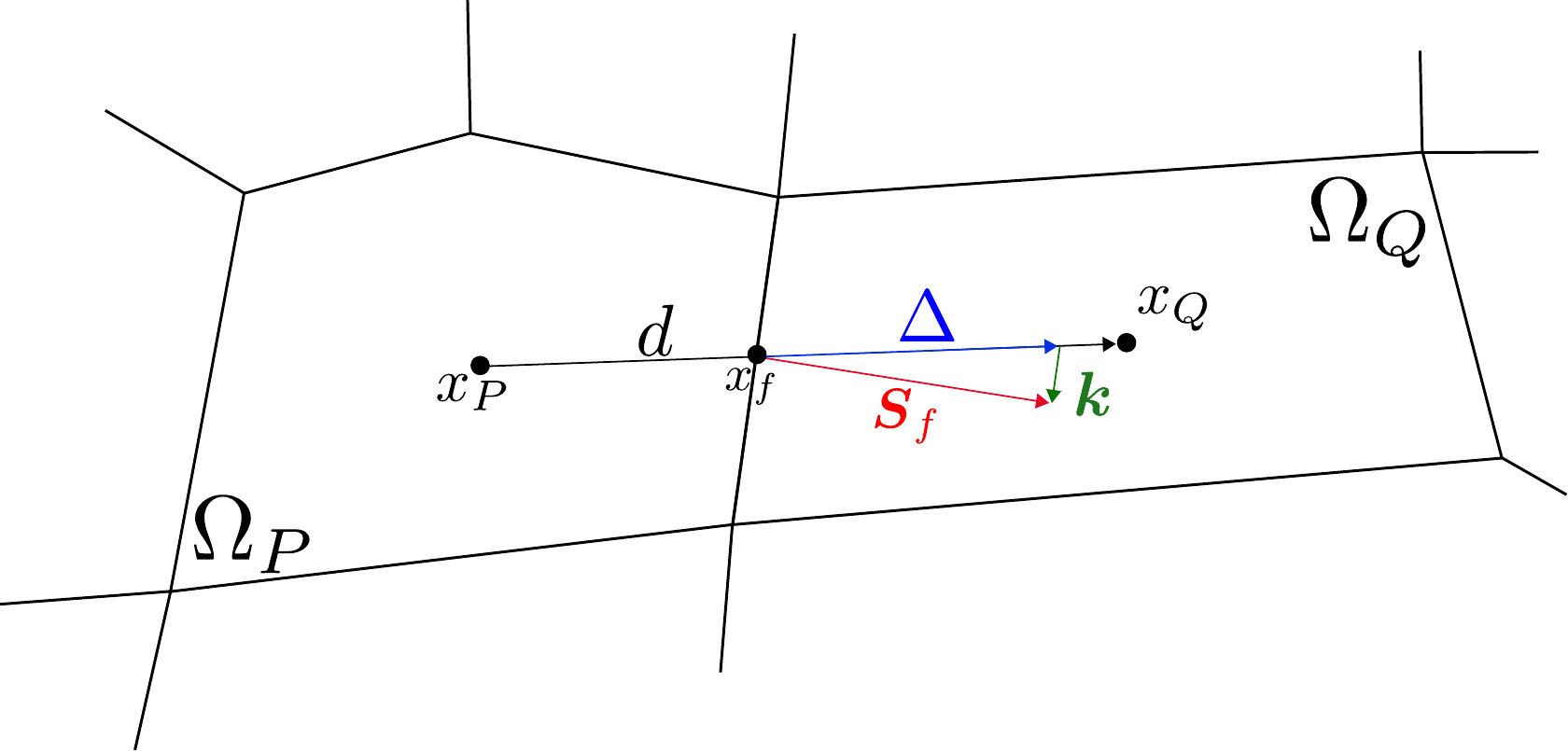}
    \caption{Two computational cells $\Omega_P$ and $\Omega_Q$ sharing the face $f$. The vector $\boldsymbol{d}$ links the two cell centres. The surface vector $\boldsymbol{S}_f$ is split as $\boldsymbol{S}_f = \boldsymbol{\Delta} + \boldsymbol{k}$, being $\boldsymbol{\Delta} \parallel \boldsymbol{d}$. }
    \label{FV_schema}
\end{figure}

The continuous problem of Equation \eqref{incompressible steady N-S} is discretized through the \textit{Finite Volumes} (FV) method \cite{ferziger2019computational}. Unlike projection-based methods, FV does not refer to a weak formulation of the continuous problem, but takes into account an integral form of the latter.\\
Consider $\{\Omega_i\}_{i=1}^{N_h}$, covering of the generic domain $\Omega$ (we omit the parameter dependence in this section for notational simplicity), such that $\underset{i }{\bigcup}\overline{\Omega}_i = \overline{\Omega}$ and $\Omega_i \cap \Omega_j = \emptyset$, for $i \neq j$. For each of the \textit{computational cells} $\Omega_i$, consider the integral form of \eqref{incompressible steady N-S}:
\begin{equation}\label{integral N-S}
\begin{dcases}
    \displaystyle\int_{\Omega_i}(\boldsymbol{u} \cdot \nabla) \boldsymbol{u} \: \mathrm{d}\Omega = \displaystyle\int_{\Omega_i}-\dfrac{\nabla p}{\rho} + \nu \Delta \boldsymbol{u} \: \mathrm{d} \Omega\\
    \displaystyle\int_{\Omega_i}\nabla \cdot \boldsymbol{u} \: \mathrm{d}\Omega = 0
\end{dcases}
\end{equation}
Using the incompressibility condition $\nabla \cdot \boldsymbol{u} = 0$, the transport term $(\boldsymbol{u} \cdot \nabla) \boldsymbol{u}$ can be written as $\nabla \cdot (\boldsymbol{u} \otimes \boldsymbol{u})$. Moreover, considering $\Delta \boldsymbol{u} = \nabla \cdot \nabla\boldsymbol{u}$ and using the divergence theorem, Equation \eqref{integral N-S} can be rewritten as:
\begin{equation}
\begin{dcases}
    \displaystyle \int_{\partial\Omega_i}(\boldsymbol{u} \otimes \boldsymbol{u}) \boldsymbol{n} \, \mathrm{d}\Gamma = - \displaystyle \dfrac{1}{\rho}\int_{\partial \Omega_i} p \boldsymbol{n}\, \mathrm{d} \Gamma + \nu \int_{\partial \Omega_i}(\nabla \boldsymbol{u}) \boldsymbol{n} \, \mathrm{d}\Gamma\\
    \displaystyle\int_{\partial\Omega_i}\boldsymbol{u} \cdot \boldsymbol{n} \:\mathrm{d}\Gamma = 0
\end{dcases}
\end{equation}
where $\boldsymbol{n} \in \mathbb{R}^d$ is the outer normal direction with respect to $\partial\Omega_i$.\\
Assume $\Omega_i$ to be polygons (in 2D) or polyhedra (in 3D). The surface integrals can be split into the contributions given by each face $\partial \Omega_{if}$:
\begin{equation}\label{N-S flux form}
\begin{dcases}
    \sum_f \displaystyle \int_{\partial\Omega_{if}}(\boldsymbol{u} \otimes \boldsymbol{u}) \boldsymbol{n}_f \, \mathrm{d}\Gamma = - \displaystyle \dfrac{1}{\rho}\sum_f\int_{\partial \Omega_{if}} p \boldsymbol{n}_f\, \mathrm{d} \Gamma + \nu \sum_f \int_{\partial \Omega_{if}}(\nabla \boldsymbol{u}) \boldsymbol{n}_f \, \mathrm{d}\Gamma\\
    \sum_f \displaystyle \int_{\partial \Omega_{if}}\boldsymbol{u} \cdot \boldsymbol{n}_f \: \mathrm{d}\Gamma = 0
\end{dcases}
\end{equation}
A common choice \cite{PhD_thesis} is to assume the generic unknown field $\phi$ (namely the velocity $\ub$ and the pressure $p$) to have linear variation around the centroid of each control volume and surface:
\begin{align*}
    \phi(\boldsymbol{x}) = \phi\left(\boldsymbol{x}_P\right) + \nabla \phi(\boldsymbol{x}_P) \cdot (\boldsymbol{x} - \boldsymbol{x}_P) \\
    \phi(\boldsymbol{x}) = \phi\left(\boldsymbol{x}_f\right) + \nabla \phi(\boldsymbol{x}_f) \cdot (\boldsymbol{x} - \boldsymbol{x}_f),
\end{align*}
where 
\begin{equation*}
    \boldsymbol{x}_P = \dfrac{\int_{\Omega_{i}} \boldsymbol{x} \, \mathrm{d}\Omega}{\int_{\Omega_i} \mathrm{d}\Omega}, \qquad
    \boldsymbol{x}_f = \dfrac{\int_{\partial\Omega_{if}} \boldsymbol{x} \, \mathrm{d}\Gamma}{\int_{\partial\Omega_{if}} \mathrm{d}\Gamma}.
\end{equation*}
Using these linearity assumptions and the definition of $\boldsymbol{x}_f$, Equation \eqref{N-S flux form} yields the following form:
\begin{equation}\label{N-S pseudo-discretized}
\begin{dcases}
    \sum_f \boldsymbol{S}_f \cdot \boldsymbol{u}_f \otimes \boldsymbol{u}_f = - \dfrac{1}{\rho}\sum_f p_f \boldsymbol{S}_f + \nu \sum_f (\nabla \boldsymbol{u})_f \boldsymbol{S}_f\\
    \sum_f \boldsymbol{u}_f\cdot \boldsymbol{S}_f = 0
\end{dcases}
\end{equation}
where the subscript $\square_f$ denotes the evaluation of the fields at the point $\boldsymbol{x}_f$ (e.g. $\boldsymbol{u}_f = \boldsymbol{u}(\boldsymbol{x}_f)$). Conversely, $\boldsymbol{S}_f$ stands for the surface vector of the face $f$, i.e. the vector proportional to $\boldsymbol{n}_f$, with magnitude equal to the area of the face $f$.\\
Referring to \cite{PhD_thesis} for more details, Equation \eqref{N-S pseudo-discretized} can be rewritten as
\begin{equation}\label{FULLY discretized N-S}
    \begin{dcases}
        \sum_f F_f \boldsymbol{u}_f - \nu \sum_f \left(\norm{\boldsymbol{\Delta}} \dfrac{\boldsymbol{u}_{Q_f} - \boldsymbol{u}_P}{\norm{\boldsymbol{d}}}  +   \left(\nabla \boldsymbol{u}\right)_f \boldsymbol{k}\right) + \dfrac{1}{\rho}\sum_f p_f \boldsymbol{S}_f = 0 \\ 
        \sum_f F_f = 0
    \end{dcases}
\end{equation}
where $F_f \coloneqq \boldsymbol{S}_f \cdot \boldsymbol{u}_f$ denotes the volumetric face flux through the face $f$. Moreover, $(\nabla\boldsymbol{u})_f$ is split into orthogonal and parallel contributions with respect to the face $f$, as depicted in Figure \ref{FV_schema}. Specifically, $\boldsymbol{d}$ represents the distance vector connecting the cell centre $\boldsymbol{x}_P$ to the neighbouring cell centre $\boldsymbol{x}_{Q_f}$, while $\boldsymbol{\Delta}$ and $\boldsymbol{k}$ denote the orthogonal and non-orthogonal decomposition vectors, respectively. This framework is essential to handle mesh non-orthogonality and skewness effects, as discussed in \cite{PhD_thesis, Bruno_2022}.
\subsection{The SIMPLE algorithm}\label{FOM SIMPLE}
The numerical solution of Equation \eqref{FULLY discretized N-S} requires solving a large non-linear system.

On the one hand, \textit{monolithic approaches} solve the complete system of equations simultaneously. Equation \eqref{incompressible steady N-S} can be rewritten in matrix form as
\begin{equation}\label{N-S continuos matrix form}
    \begin{bmatrix}
        A_u & \nabla(\cdot) \\
        \nabla \cdot (\cdot) & 0
    \end{bmatrix}
    \begin{bmatrix}
        \boldsymbol{u} \\
        p
    \end{bmatrix}
    =
    \begin{bmatrix}
        0 \\
        0
    \end{bmatrix},
\end{equation}
where
\begin{equation*}
    A_u \boldsymbol{u} = \nabla \cdot (\boldsymbol{u} \otimes \boldsymbol{u}) - \nu\Delta \boldsymbol{u}.
\end{equation*}
This assumes the form of a saddle-point problem \cite{ballarin2015supremizer, stabile2018finite}, which is well-known hard to be solved via a \textit{coupled approach}.

On the other hand, in the \textit{segregated approach}, equations are solved iteratively. In the present work, numerical simulations are run through this method and specifically the SIMPLE (Semi-Implicit Method for Pressure Linked Equations) algorithm \cite{patankar2018numerical, PhD_thesis, ferziger2019computational, caretto2007two} is employed.\\
Equations \eqref{FULLY discretized N-S} can be rewritten, introducing a so-called \textit{pressure equation}. For each cell centre $P$, the momentum equation is split into the contributions coming from $P$ and its neighbours $Q$:
\begin{equation*}\label{SIMPLE N-S only velocity discretized}
    a_P \boldsymbol{u}_P = \sum_Q a_Q \boldsymbol{u}_Q - \nabla p.
\end{equation*}
With a slight abuse of notation, $\nabla p$ stands for the pressure gradient contribution on the computational cell $P$, i.e. it represents the integral of the same quantity on the computational cell $\Omega_P$. Introducing the off-diagonal term $H(\boldsymbol{u}) \coloneqq \underset{Q}{\sum}a_q\boldsymbol{u}_Q$,
\begin{equation}\label{velocity in P}
    \boldsymbol{u}_P = \dfrac{1}{a_P}\Big[H(\boldsymbol{u}) - \nabla p \Big].
\end{equation}
By the incompressibility condition in Equation \eqref{N-S pseudo-discretized}, Equation \eqref{velocity in P} reads as
\begin{equation*}
    \sum_f \boldsymbol{n}_f \cdot \left(\dfrac{\nabla p}{a_P}\right)_f =  \sum_f \boldsymbol{n}_f \cdot \left(\dfrac{H(\boldsymbol{u})}{a_P} \right)_f
\end{equation*}
Using the divergence theorem on the contribution of $\nabla p$, the final form of the discretized incompressible Navier-Stokes system becomes
\begin{equation}\label{SIMPLE discretized N-S}
    \begin{dcases}
        a_P \boldsymbol{u}_P = \sum_Q a_Q \boldsymbol{u}_Q - \sum_f \boldsymbol{n}_f \,p_f \\
        \sum_f \boldsymbol{n}_f \cdot \left(\dfrac{\nabla p}{a_P}\right)_f =  \sum_f \boldsymbol{n}_f \cdot \left( \dfrac{H(\boldsymbol{u})}{a_P} \right)_f
    \end{dcases}
\end{equation}
Referring e.g. to \cite{patankar2018numerical, ferziger2019computational, PhD_thesis} for a deeper discussion, a numerical algorithm can be extrapolated from the above procedure.\\
The first equation in \eqref{SIMPLE discretized N-S} is under-relaxed by a coefficient $\alpha_{\boldsymbol{u}} \in (0,1]$ and is solved for a velocity field $\boldsymbol{u}^*$, with respect to a guess pressure $p^{(k-1)}$. The latter may come from the previous iteration or from an initial guess $p^{(0)}$. In general, the intermediate velocity $\boldsymbol{u}^*$ is not necessarily divergence-free. Therefore, a correction $\boldsymbol{u}^{(k)}_{cor}$ is found by imposing the incompressibility condition
\begin{equation}\label{correction divergence}
    \nabla \cdot (\boldsymbol{u}^{*} + \boldsymbol{u}_{cor}^{(k)}) = 0.
\end{equation}
As shown above, this can be interpreted as a pressure equation for a correction pressure term $p^{(k)}_{cor}$, which is solved under the so-called \textit{semi-implicit} condition
\begin{equation*}
    H(\boldsymbol{u}_{cor}^{(k)}) \simeq 0.
\end{equation*}
Once the pressure correction is computed, an under-relaxation step via a coefficient $\alpha_p \in (0,1]$ is performed to update the pressure field:
\begin{equation*}
    p^{(k)} = p^{(k-1)} + \alpha_p p^{(k)}_{cor}.
\end{equation*}
$p^{(k)}$ is then inserted in the momentum equation to compute the divergence-free velocity field $\boldsymbol{u}^{(k)}$. Additionally, the corresponding flux term $F_f^{(k)}$ is updated at each iteration as well.\\
The iterative scheme continues until reaching a desired tolerance or a maximum number of iterations.
\section{The reduced order models}\label{ROMs theory}
Numerical solution of continuous problems usually consists of formulating and solving very large (non-)linear systems. For instance, in our framework, Equation \eqref{FULLY discretized N-S} can be rewritten as
\begin{equation}\label{FOM compact form}
    M_h(\mub)\phi_h(\mub) = f_h(\mub),
\end{equation}
where $\phi_h(\mub) = (\boldsymbol{u}_h(\mub), p_h(\mub))^T \in \mathbb{R}^{4N_h}$ is the concatenation of the discretized velocity and pressure fields, while $N_h$ denotes the number of control volumes considered in the discretization. Solving multiple geometric configurations of the same reference problem (namely a \textit{many-query} problem) implies therefore large computational costs. \textit{Reduced-Order Models} (ROMs) \cite{quarteroni2014reduced,quarteroni2015reduced,hesthaven2016certified} aim to lower the complexity of the \textit{Full-Order Model} (FOM), by exploiting the redundancies of the investigated phenomena.

Full-order solutions define a \textit{parametric manifold}
\begin{equation}\label{parametric manifold}
    \mathcal{M}_h = \{\phi_h(\mub), \mub \in \mathcal{P}\},
\end{equation}
immersed in a $4N_h$ dimensional space. The core concept behind ROMs is to project $\mathcal{M}_h$ on a different space of dimension $N \ll N_h$, spanned by a parametric-independent basis $\{\varphi_k\}_{k=1}^N$. This can be performed in different ways: in the present work, \textit{Proper Orthogonal Decomposition} (POD) \cite{berkooz1993proper} is employed. Given a set of \textit{training parameters} $\{\mub_i, i = 1, \dots, n_s\}$, the corresponding full-order  solutions, called \textit{snapshots}, $\phi(\boldsymbol{\mub_i}) = (\boldsymbol{u}_h(\mub_i), p_h(\mub_i))^T$ are computed and collected in two different matrices
\begin{equation}
    S_{\boldsymbol{u}} = \Bigg[
        \boldsymbol{u}_h(\mub_1) \: \Big| \: \cdots \: \Big| \: \boldsymbol{u}_h(\mub_{n_s})\Bigg] \in \mathbb{R}^{3 N_h \times n_s} \qquad S_p = \Bigg[p_h(\mub_1) \: \Big| \: \cdots \: \Big| \: p_h(\mub_{n_s})\Bigg] \in \mathbb{R}^{N_h \times n_s}.
\end{equation}
For the sake of notational simplicity, we focus on the velocity case in the following. The pressure one is analogous.

\begin{figure}[!htb]
        \centering
        \includegraphics[width=\linewidth]{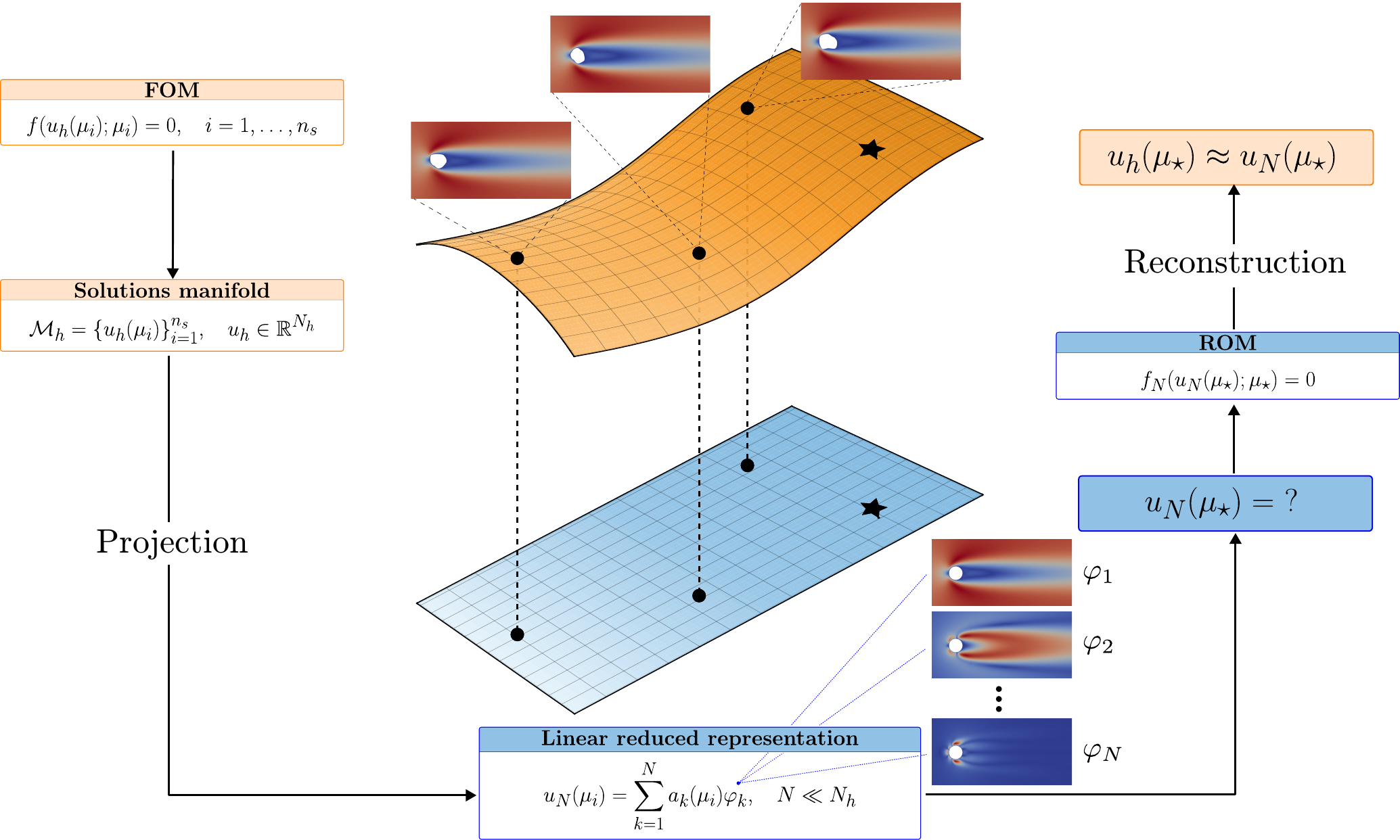}
    \caption{Model reduction scheme: given a set of parameters $\{\mub_i\}_{i=1}^{n_s}$, a POD-Galerkin projection is performed, obtaining a reduced version of the problem for the unknown parameter $\mub_{\star}$. The reduced basis $\{\varphib_i\}_{i=1}^N$ is used to reconstruct the full field approximation via linear combination.}
    \label{ROM_schema}
\end{figure}

The POD problem consists of finding, for each training parameter $\mub_i$, the best choice among the possible basis functions $\boldsymbol{\varphi}_1$, $\dots$, $\boldsymbol{\varphi}_N$ and reduced coefficients $a_1(\mub_i)$, $\dots$, $a_N(\mub_i)$ minimizing the total squared reconstruction error:
\begin{equation}\label{Reduction problem}
    E_{N_{POD}} \coloneqq \sum_{i=1}^{n_s}\Big\lVert\boldsymbol{u}_h(\mub_i) - \sum_{k=1}^{N}a_k(\mub_i)\boldsymbol{\varphi}_k\Big\rVert^2_{L^2(\Omega)},
\end{equation}
with
\begin{equation*}
    \langle\boldsymbol{\varphi}_i, \boldsymbol{\varphi}_j\rangle_{L^2(\Omega)} = \delta_{ij}\: , \quad \forall i, j \in {1, \dots, N}.
\end{equation*}
It is well known \cite{quarteroni2014reduced,quarteroni2015reduced,hesthaven2016certified} that minimizing \eqref{Reduction problem} is equivalent to solving the eigenvalue problem
\begin{equation}\label{velocioty POD eigenvalues problem}
    C_{\boldsymbol{u}}\boldsymbol{\psi}_i = \sigma_i^2 \boldsymbol{\psi}_i \:, \quad i = 1, \dots, n_s,
\end{equation}
where $C_{\boldsymbol{u}_{ij}} = \langle\boldsymbol{u}_h(\mub_i), \boldsymbol{u}_h(\mub_j)\rangle_{L^2(\Omega)}$ is the snapshots \textit{correlation matrix}, while $\sigma_i$ and $\boldsymbol{\psi}_i$ are the singular values and right singular vectors of $S_{\ub}$, respectively. The reduced \textit{modes} $\boldsymbol{\varphi}_i$ are the left singular vectors of $S_{\boldsymbol{u}}$, computed as
\begin{equation*}
    \boldsymbol{\varphi_{i}} = \dfrac{1}{\sigma_i} S_{\boldsymbol{u}}\boldsymbol{\psi}_i \:, \quad i = 1, \dots, n_s.
\end{equation*}
It is worth emphasizing that, employing FV discretization, this work deals with geometrically parametrized domains without mapping the configurations back to a common reference domain. Consequently, an intrinsic challenge arises: snapshots corresponding to different parameters belong to different physical domains. To address this issue, the correlation matrix $C_{\boldsymbol{u}}$ is evaluated by computing the $L^2$ inner products over the undeformed reference geometry $\Omega$.\\
The optimal dimension $N$ of the reduced subspace is selected based on the \textit{information content}
\begin{equation}\label{information content}
    I(N) \coloneqq \dfrac{\sum_{i=1}^N \sigma_i^2}{\sum_{i=1}^{n_s}\sigma_i^2},
\end{equation}
such that for a predefined tolerance $\varepsilon_{POD}$, the criterion $I(N) \geq 1 - \varepsilon_{POD}^2$ is satisfied.
The modes are stored column-wise in the projection matrix
\begin{equation*}
    V = \Bigg[
        \boldsymbol{\varphi}_1 \: \Big| \: \dots \: \Big| \: \boldsymbol{\varphi}_N
        \Bigg]
        \in \mathbb{R}^{3N_h\times N}
\end{equation*}
and the reduced system is obtained by a Galerkin projection of the full-order operators onto the reduced space\footnote{Remark that, to avoid notational heaviness, an Euclidean projection has been reported in equation \eqref{projection of the full-order problem}. In practice, a projection with respect to the $L^2(\Omega)$ scalar product is employed.}:
\begin{equation}\label{projection of the full-order problem}
    V^T M_h(\mub)V \boldsymbol{a}(\mub) = V^T f_h(\mub) \Longrightarrow M_N(\mub) \boldsymbol{a}(\mub) = f_N(\mub),
\end{equation}
where $\boldsymbol{a}(\mub) = (a_1(\mub), \dots, a_N(\mub)) \in \mathbb{R}^N$ is the reduced coefficients vector. These coefficients serve as coordinates for the online linear combination to reconstruct the high-dimensional fields
\begin{equation*}
    \boldsymbol{u}_h(\mub) \simeq \boldsymbol{u}_N(\mub) \coloneqq V \boldsymbol{a}(\mub) = \sum_{k=1}^N a_k(\mub)\boldsymbol{\varphi}_k.
\end{equation*}
In general, the full-order part of model reduction, involving snapshots' computation and POD-Galerkin projection, is known as \textit{offline} phase. Conversely, fields' reduced reconstruction for unknown parameters is addressed as \textit{online} phase. A schematic overview of this offline-online decoupling procedure is depicted in Figure \ref{ROM_schema}.

In the following, the \textit{lift function} approach \cite{star2021noveliterativepenaltymethod, Stabile2017CAIM} is adopted to impose non-homogeneous boundary conditions on the reduced velocity fields. This method consists of expressing reduced solutions as variations with respect to a known reference field $\boldsymbol{\zeta}$, which satisfies the non-homogeneous boundary constraints. The full-order velocity snapshots $\boldsymbol{u}_h(\mub_i)$ are first homogenized by subtracting the lift function:
\begin{equation*}
    \boldsymbol{u}'_h(\mub_i) \coloneqq \boldsymbol{u}_h(\mub_i) - \boldsymbol{\zeta}.
\end{equation*}
POD is then performed on these modified snapshots to yield a set of reduced modes $\{\boldsymbol{\varphi}'_k\}_{k=1}^N$ that naturally satisfy homogeneous boundary conditions. Finally, the lift function $\boldsymbol{\zeta}$ is reintroduced as a zeroth mode to reconstruct the reduced solution:
\begin{equation}\label{lift function approach}
    \boldsymbol{u}_N(\mub) = a_0(\mub)\boldsymbol{\zeta} + \sum_{k=1}^N a_k(\mub)\boldsymbol{\varphi}'_k.
\end{equation}
To guarantee that the non-homogeneous boundary conditions are exactly satisfied, the reduced algebraic system \eqref{projection of the full-order problem} is properly modified to enforce the constraint on the zeroth coefficient $a_0(\mub)$.\\
As with the full-order formulation, the resulting reduced-order problem can be solved using different numerical strategies.
\subsection{Reduced coupled approach}\label{coupled model} 
By discretizing the operators involved in Equation \eqref{N-S continuos matrix form} \cite{Stabile2017CAIM, stabile2018finite}, the full-order problem can be expressed as:
\begin{equation}\label{FOM saddle-point}
    \begin{bmatrix}
        A_h(\ub_h, \boldsymbol{\mu}) & B_h(\boldsymbol{\mu}) \\
        P_h(\boldsymbol{\mu}) & 0
    \end{bmatrix}
    \begin{bmatrix}
        \boldsymbol{u}_h(\boldsymbol{\mu}) \\
        p_h(\boldsymbol{\mu})
    \end{bmatrix}
    =
    \begin{bmatrix}
        f_h(\boldsymbol{\mu}) \\
        g_h(\boldsymbol{\mu})
    \end{bmatrix}.
\end{equation}
After projecting these operators onto the reduced spaces $\mathcal{V} \coloneqq \mathrm{span}\{\boldsymbol{\zeta}, \boldsymbol{\varphi}^{\boldsymbol{u}}_1, \dots, \boldsymbol{\varphi}^{\boldsymbol{u}}_{N_{\boldsymbol{u}}}\}$ and $\mathcal{Q} \coloneqq \mathrm{span}\{\varphi^p_1, \dots, \varphi^p_{N_p}\}$, Equation \eqref{FOM saddle-point} reads:
\begin{equation}\label{almost saddle point ROM}
    \begin{bmatrix}
        A_N(\boldsymbol{a}, \boldsymbol{\mu}) & B_N(\boldsymbol{\mu}) \\
        P_N(\boldsymbol{\mu}) & 0
    \end{bmatrix}
    \begin{bmatrix}
        \boldsymbol{a}(\mub)\\
        \boldsymbol{b}(\mub)
    \end{bmatrix}
    =
    \begin{bmatrix}
        f_N(\boldsymbol{\mu}) \\
        g_N(\boldsymbol{\mu})
    \end{bmatrix},
\end{equation}
where the reduced convective-diffusive operator is defined via the trilinear form $(A_N(\boldsymbol{a}))_{ij} = \sum_{k=1}^{N_{\boldsymbol{u}}} (A_N)_{ijk} a_k$, and the respective reduced components are given by:
\begin{equation*}
\begin{split}
    & (A_N)_{ijk} = \langle\varphib^{\ub}_i, A_h(\varphib^{\ub}_k)\varphi^{\ub}_j\rangle_{L^2(\Omega)} \simeq \langle \boldsymbol{\varphi}^{\boldsymbol{u}}_i, \nabla \cdot (\boldsymbol{\varphi}^{\boldsymbol{u}}_j \otimes \boldsymbol{\varphi}^{\boldsymbol{u}}_k) - \nu\Delta \boldsymbol{\varphi}^{\boldsymbol{u}}_j\rangle_{L^2(\Omega)}, \\
    & (B_N)_{ij} = \langle \boldsymbol{\varphi}^{\boldsymbol{u}}_i, B_h \varphi^p_j \rangle_{L^2(\Omega)} \simeq \langle \boldsymbol{\varphi}^{\boldsymbol{u}}_i, \nabla \varphi_j^p \rangle_{L^2(\Omega)}, \\
    & (P_N)_{ij} = \langle \varphi^p_i, P_h \boldsymbol{\varphi}^{\boldsymbol{u}}_j \rangle_{L^2(\Omega)} \simeq \langle \varphi^p_i, \nabla \cdot \boldsymbol{\varphi}^{\boldsymbol{u}}_j \rangle_{L^2(\Omega)}.
\end{split}
\end{equation*}
Using the divergence theorem, equation \eqref{almost saddle point ROM} can be rewritten in a symmetric saddle-point form:
\begin{equation}\label{saddle point ROM}
    \begin{bmatrix}
        A_N(\boldsymbol{a}, \boldsymbol{\mu}) & B_N(\boldsymbol{\mu}) \\
        B_N^{T}(\boldsymbol{\mu}) & 0
    \end{bmatrix}
    \begin{bmatrix}
        \boldsymbol{a}(\mub)\\
        \boldsymbol{b}(\mub)
    \end{bmatrix}
    =
    \begin{bmatrix}
        f_N(\boldsymbol{\mu}) \\
        \overline{g}_N(\boldsymbol{\mu})
    \end{bmatrix},
\end{equation}
where $\boldsymbol{a}(\mub) \in \mathbb{R}^{N_{\boldsymbol{u}}+1}$ and $\boldsymbol{b}(\mub) \in \mathbb{R}^{N_p}$ denote the reduced velocity and pressure coefficient vectors, respectively.

It is well established that such a system is well-posed if and only if it satisfies the Ladyzhenskaya–Babuška–Brezzi (LBB) inf-sup condition, stating that there exists a parameter-independent constant $\beta > 0$ such that:
\begin{equation}\label{inf-sup condition}
    \beta(\boldsymbol{\mu}) = \inf_{p \in \mathcal{Q}} \sup_{\boldsymbol{v} \in \mathcal{V}} \dfrac{\langle B_N(\boldsymbol{\mu}) \boldsymbol{v}, p\rangle_{L^2(\Omega)}}{\|\boldsymbol{v}\| \|p\|} \geq \beta > 0 \:, \quad \forall \boldsymbol{\mu} \in \mathcal{P}.
\end{equation}
In general, standard POD reduced spaces fail to fulfill this constraint. Therefore, a common remedy \cite{ballarin2015supremizer, stabile2018finite} consists of enriching the velocity space $\mathcal{V}$ with properly chosen velocity fields $\boldsymbol{\eta}_i$, called \textit{supremizers}. These fields are obtained by solving the following auxiliary Poisson problem:
\begin{equation}\label{supremizer equation}
    \begin{dcases}
        \Delta \boldsymbol{\eta}_i = - \nabla p_i \: , & \text{in } \Omega \\
        \boldsymbol{\eta}_i = \boldsymbol{0} \: , & \text{on } \partial \Omega
    \end{dcases}
\end{equation}
where $p_i$ represents a prescribed pressure field. In the \textit{exact} supremizer enrichment procedure, pressure fields coincide with POD modes, i.e., $\{p_i\}_{i=1}^{N_p} = \{\varphi^p_i\}_{i=1}^{N_p}$. Conversely, in the \textit{approximate} approach, full-order pressure snapshots are used, $\{p_i\}_{i=1}^{n_s} = \{p_h(\mub_i)\}_{i=1}^{n_s}$, and a subsequent POD is performed on the resulting supremizers $\{\boldsymbol{\eta}_i\}_{i=1}^{n_s}$. Although the latter approach leverages pre-computed fields, making the online stage less computationally intensive, the former method is adopted in the present work due to its demonstrated theoretical accuracy \cite{ballarin2015supremizer}.\\
Consequently, the reduced system \eqref{saddle point ROM} is expanded to:
\begin{equation}\label{matrix ROM with supremizers}
    \begin{bmatrix}
        \tilde{A}_N(\tilde{\boldsymbol{a}}, \boldsymbol{\mu}) & \tilde{B}_N(\boldsymbol{\mu}) \\
        \tilde{B}_N^{T}(\boldsymbol{\mu}) & 0
    \end{bmatrix}
    \begin{bmatrix}
        \tilde{\boldsymbol{a}}(\mub)\\
        \boldsymbol{b}(\mub)
    \end{bmatrix}
    =
    \begin{bmatrix}
        \tilde{f}_N(\boldsymbol{\mu}) \\
        \tilde{g}_N(\boldsymbol{\mu})
    \end{bmatrix},
\end{equation}
where $\tilde{\boldsymbol{a}}(\mub) \in \mathbb{R}^{N_{\boldsymbol{u}}+1+N_{sup}}$ is the enriched velocity coefficients vector, used to define the reduced velocity:
\begin{equation*}
    \boldsymbol{u}_N(\mub) = \tilde{a}_0(\mub)\boldsymbol{\zeta} + \sum_{k=1}^{N_{\boldsymbol{u}}}\tilde{a}_k(\mub)\boldsymbol{\varphi}^{\boldsymbol{u}}_k + \sum_{j=1}^{N_{sup}}\tilde{a}_{N_{\ub} + j}(\mub)\boldsymbol{\eta}_j,
\end{equation*}
with $0 \leq N_s \leq N_p$ denoting the number of supremizers included in the formulation.\\
In contrast to works dealing with physical parameters, (e.g., \cite{stabile2018finite}), the reduced matrices cannot be pre-assembled during the offline phase, since the computational mesh deforms non-affinely. Indeed, the underlying discretization mesh must be reconstructed online to re-evaluate the discrete differential operators and, consequently, the entries of the reduced matrices. The non-linear reduced system is subsequently solved online using a hybrid non-linear Newton algorithm.
\subsection{Reduced SIMPLE algorithm}\label{simple model}
Another viable strategy consists of implementing a segregated approach directly at the reduced-order level \cite{stabile2020efficient, zancanaro2021hybrid, zancanaro2025segregated, Oberto2026, nkana2026}, thereby mimicking the full-order SIMPLE algorithm presented in Section \ref{FOM SIMPLE}. This mathematical formulation avoids the saddle-point structure of the monolithic problem, making the supremizer enrichment unnecessary.

Under the semi-implicit assumption, at the $k$-th iteration, the full-order system of equations \eqref{SIMPLE discretized N-S} can be written in a decoupled matrix form as:
\begin{equation}\label{FOM SIMPLE matrix form}
    \begin{dcases}
        M_{\boldsymbol{u}}(\boldsymbol{u}^{(k-1)}, \boldsymbol{\mu})\boldsymbol{u}^{(k)}_* = f_{\boldsymbol{u}}(p^{(k-1)}, \boldsymbol{\mu}) \\
        M_p (\boldsymbol{\mu})p^{(k)}_{cor} = f_p(\boldsymbol{u}^{(k)}_*, \mub)
    \end{dcases}
\end{equation}
After projecting these operators onto the reduced spaces $\mathcal{V} \coloneqq \mathrm{span}\{\boldsymbol{\zeta}, \boldsymbol{\varphi}^{\boldsymbol{u}}_1, \dots, \boldsymbol{\varphi}^{\boldsymbol{u}}_{N_{\boldsymbol{u}}}\}$ and $\mathcal{Q} \coloneqq \mathrm{span}\{\varphi^p_1, \dots, \varphi^p_{N_p}\}$, the reduced SIMPLE system reads:
\begin{equation}\label{SIMPLE reduced matrix form - with N}
    \begin{dcases}
        M_N^{\boldsymbol{u}}(\boldsymbol{a}^{(k-1)}, \boldsymbol{\mu}) \boldsymbol{a}^{(k)}_* = f^{\boldsymbol{u}}_N(\boldsymbol{b}^{(k-1)}, \boldsymbol{\mu}) \\
        M_N^p(\boldsymbol{\mu}) \boldsymbol{b}^{(k)}_{cor} = f^p_N(\boldsymbol{a}^{(k)}_*, \mub)
    \end{dcases}
\end{equation}
where the entries of the reduced matrices are computed as:
\begin{equation*}
    \begin{split}
        & \left(M_N^{\boldsymbol{u}}\right)_{ij} = \langle\boldsymbol{\varphi}^{\boldsymbol{u}}_i, M_{\boldsymbol{u}}(\boldsymbol{u}^{(k-1)}, \boldsymbol{\mu})\boldsymbol{\varphi}^{\boldsymbol{u}}_j\rangle_{L^2(\Omega)}, \\
        & \left(M_N^p\right)_{ij} = \langle\varphi^p_i, M_p(\boldsymbol{\mu})\varphi^p_j\rangle_{L^2(\Omega)}.
    \end{split}
\end{equation*}
The reduced algebraic system \eqref{SIMPLE reduced matrix form - with N} is solved sequentially within an online SIMPLE loop, employing reduced under-relaxation coefficients $\alpha_{\boldsymbol{u}}$ and $\alpha_p$ that can be calibrated independently from their full-order counterparts. At each iteration, the updated reduced coefficients $\boldsymbol{a}^{(k)}$ and $\boldsymbol{b}^{(k)}$ are obtained via the standard pressure-velocity correction steps, initialized with $\boldsymbol{a}^{(0)} = (1, 0, \dots, 0)^T \in \mathbb{R}^{N_{\boldsymbol{u}}+1}$ and $\boldsymbol{b}^{(0)} = \boldsymbol{0} \in \mathbb{R}^{N_p}$.

The convergence of the iterative procedure is controlled by a prescribed maximum number of iterations $k_{max}$ and a tolerance criterion $tol$. The latter serves as an accuracy threshold for the absolute algebraic residuals, defined at the $k$-th iteration as:
\begin{equation}\label{reduced SIMPLE residuals}
    r^{(k)}_{\boldsymbol{u}} = \|M_N^{\boldsymbol{u}}(\boldsymbol{a}^{(k)}, \boldsymbol{\mu}) \boldsymbol{a}^{(k)} - f^{\boldsymbol{u}}_N(\boldsymbol{b}^{(k)}, \boldsymbol{\mu})\|_1, \qquad r^{(k)}_p = \|M_N^p(\boldsymbol{\mu}) \boldsymbol{b}^{(k)} - f^p_N(\boldsymbol{a}^{(k)}, \mub)\|_1.
\end{equation}
To ensure a parameter-independent evaluation, the normalized residuals are considered:
\begin{equation}\label{reduced SIMPLE normalized residuals}
    \overline{r}^{(k)}_{\boldsymbol{u}} = \dfrac{r^{(k)}_{\boldsymbol{u}}}{\|f^{\boldsymbol{u}}_N(\boldsymbol{b}^{(k)}, \boldsymbol{\mu})\|_1} \:, \quad \overline{r}^{(k)}_p = \dfrac{r^{(k)}_p}{\|f^p_N(\boldsymbol{a}^{(k)}, \mub)\|_1}.
\end{equation}
Additionally, to monitor stagnation or sudden divergence within the online segregated loop, the absolute residual jumps are evaluated:
\begin{equation}\label{reduced SIMPLE residual jumps}
    j^{(k)}_{\boldsymbol{u}} = \left| r^{(k)}_{\boldsymbol{u}} - r^{(k-1)}_{\boldsymbol{u}} \right|, \qquad j^{(k)}_p = \left| r^{(k)}_p - r^{(k-1)}_p \right|,
\end{equation}
and used as supplementary stopping criteria.
\section{Lid-driven cavity flow}\label{lid driven cavity}
The first considered benchmark test case is the well known lid-driven cavity flow \cite{schreiber1983driven}. The undeformed domain consists of a square cavity $\Omega$ of side $0.1$. The fluid flows from left to right, along the upper boundary, with velocity $U = (1, 0, 0)$ and kinematic viscosity $\nu = 0.001$. These parameters yield a Reynolds number of $Re = 100$, which is sufficiently small to ensure a steady-state laminar phenomenon \cite{Ghia1982-dn}.

\begin{figure}[!htb]
    \hspace{-1 cm}
    \begin{subfigure}{0.35\textwidth}
        \centering
        \includegraphics[width=\linewidth]{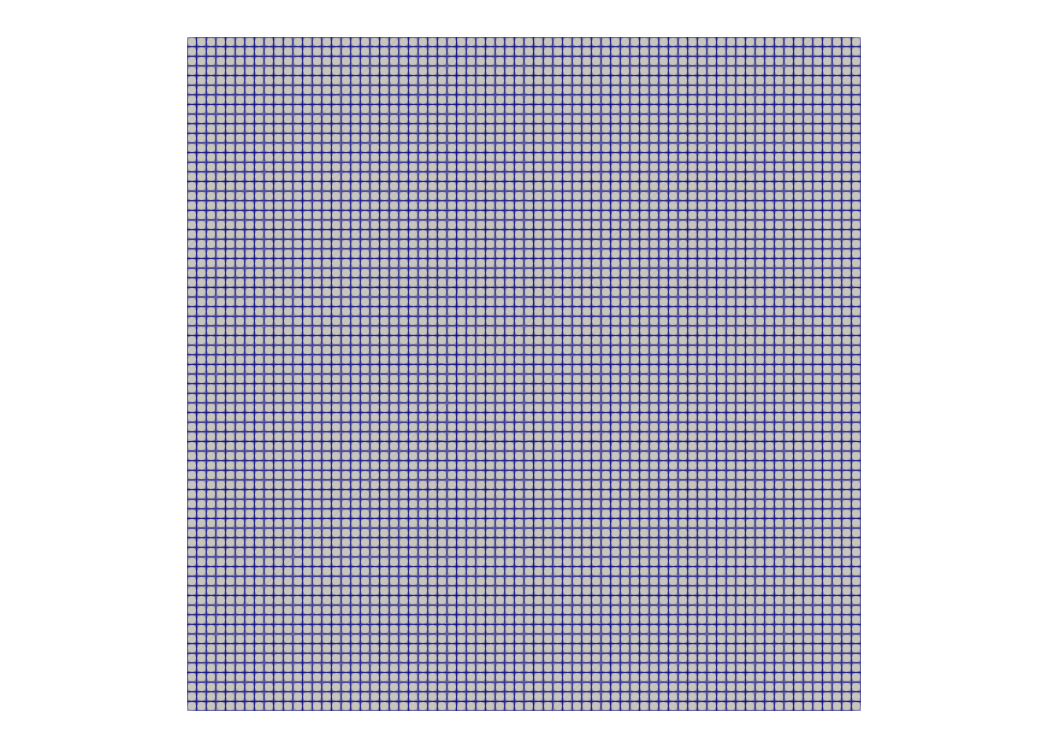}
    \end{subfigure}\hfill
    \begin{subfigure}{0.35\textwidth}
        \centering
        \includegraphics[width=\linewidth]{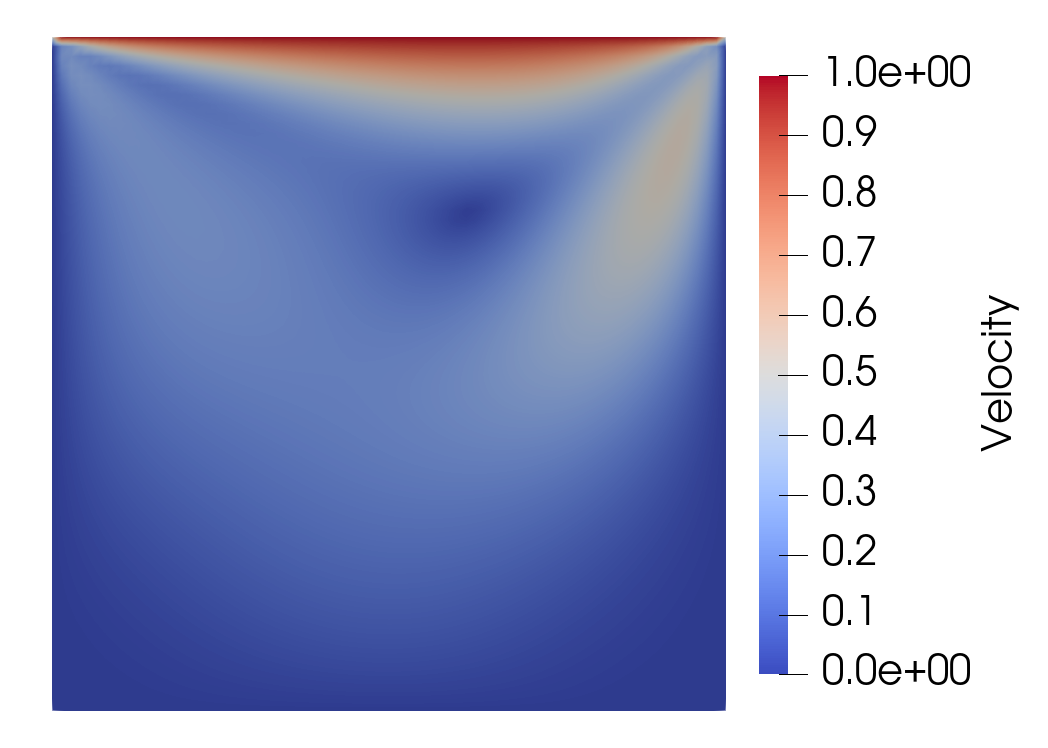}
    \end{subfigure}\hfill
    \begin{subfigure}{0.35\textwidth}
        \centering
        \includegraphics[width=\linewidth]{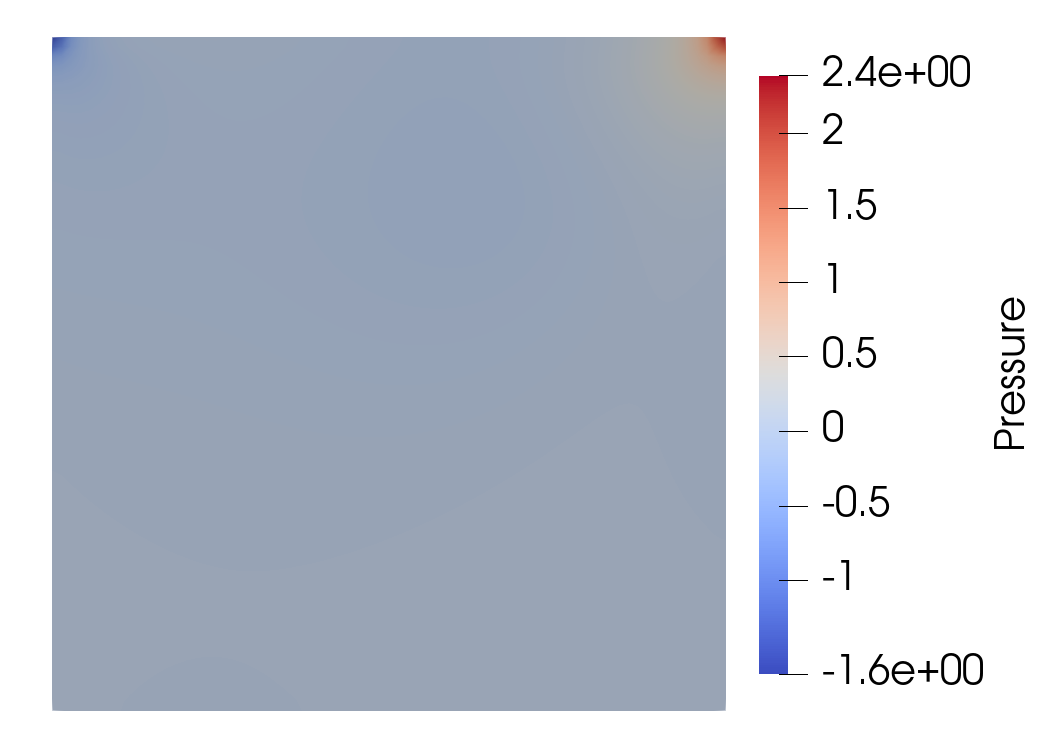}
    \end{subfigure}
    \caption{Undeformed case for the lid-driven cavity phenomenon.}
        \label{Cavity_reference_case}
\end{figure}

Full-order numerical simulations are performed using the SIMPLE algorithm implemented in OpenFOAM \cite{OpenFOAM}, employing a discretizing mesh composed of $70 \times 70$ uniform cells. Figure \ref{Cavity_reference_case} shows the discretizing grid along with the numerical solutions for the velocity and pressure fields.
\subsection{Offline stage}
\begin{figure}[!htb]
    \hspace{-1 cm}
    \begin{subfigure}{0.35\textwidth}
        \centering
        \includegraphics[width=\linewidth]{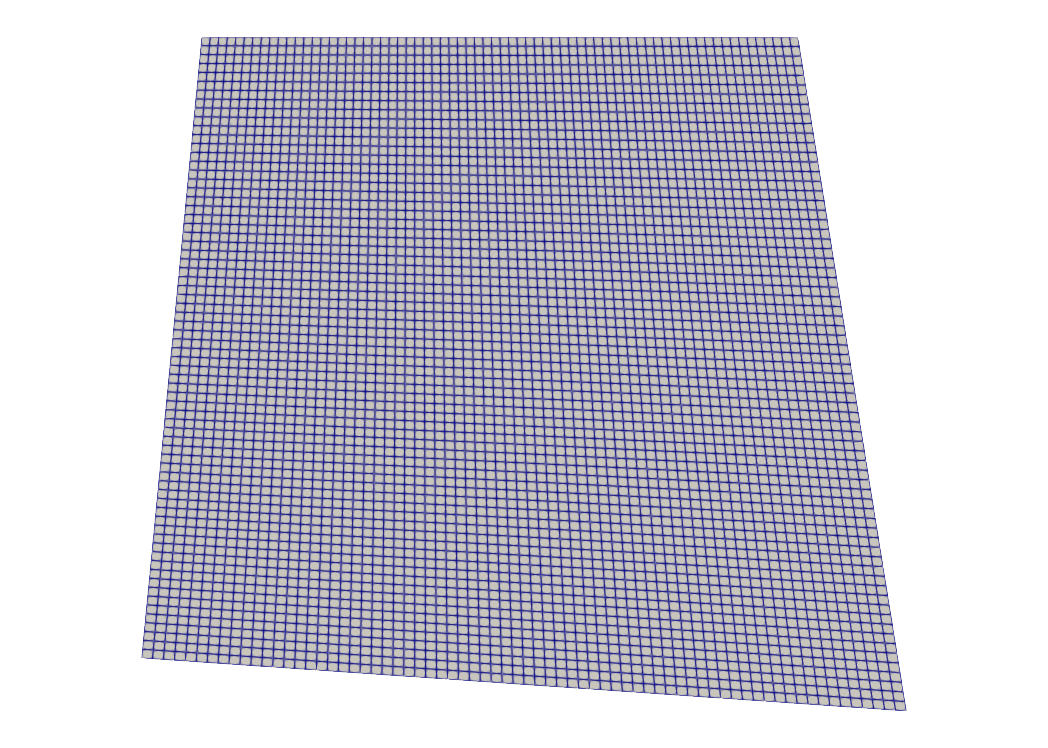}
    \end{subfigure}\hfill
    \begin{subfigure}{0.35\textwidth}
        \centering
        \includegraphics[width=\linewidth]{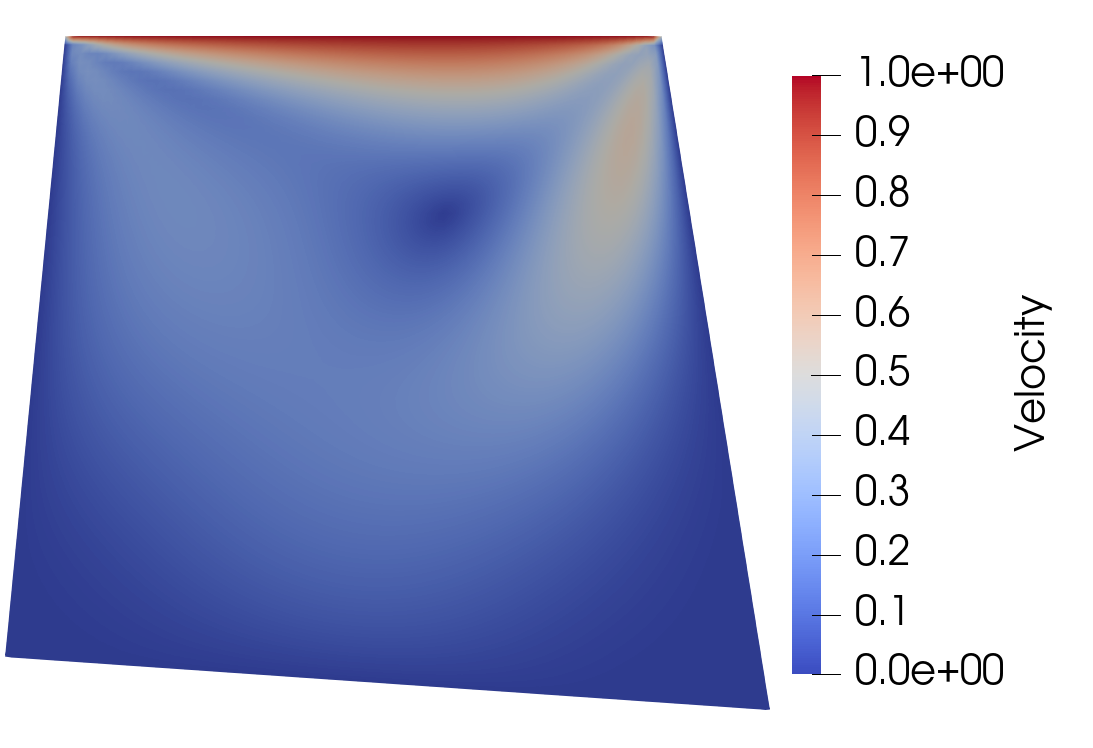}
    \end{subfigure}\hfill
    \begin{subfigure}{0.35\textwidth}
        \centering
        \includegraphics[width=\linewidth]{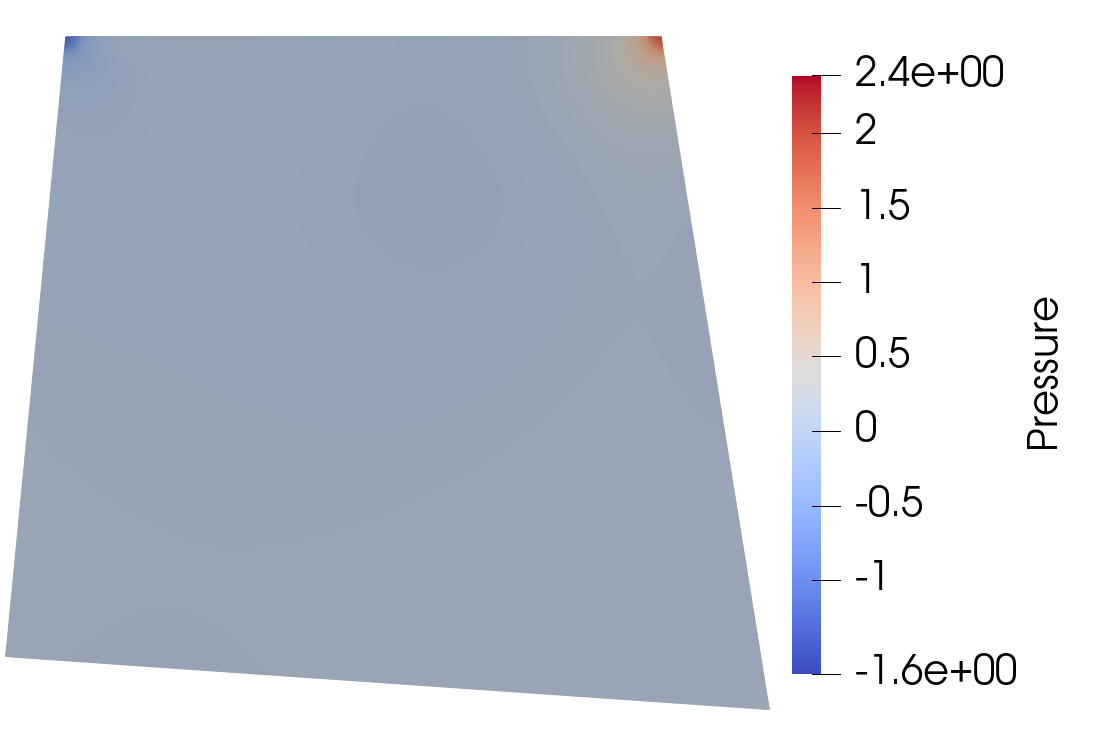}
    \end{subfigure}
    \caption{Deformed case for the lid-driven cavity phenomenon.}
        \label{Cavity_deformed_case}
\end{figure}

The domain is deformed moving cavity's lower vertices by random values ranging in $[-0.2, 0.2]$. Consequently, the left lower vertex, originally in $(0,0)$, will be located in $[-0.2, 0.2] \times [-0.2, 0.2]$, whereas the right one, originally in $(1,0)$, in $[0.8, 1.2]\times [-0.2, 0.2]$. Figure \ref{Cavity_deformed_case} shows the resulting deformed mesh, together with the corresponding full order velocity and pressure fields.

$n_s = 200$ deformations are randomly selected and the resulting full-order solutions serve as snapshots for the POD to define the reduced order models.

\begin{figure}[!htb]
    \hspace{-0.5 cm}
    \begin{subfigure}{0.25\textwidth}
        \centering
        \includegraphics[width=\linewidth]{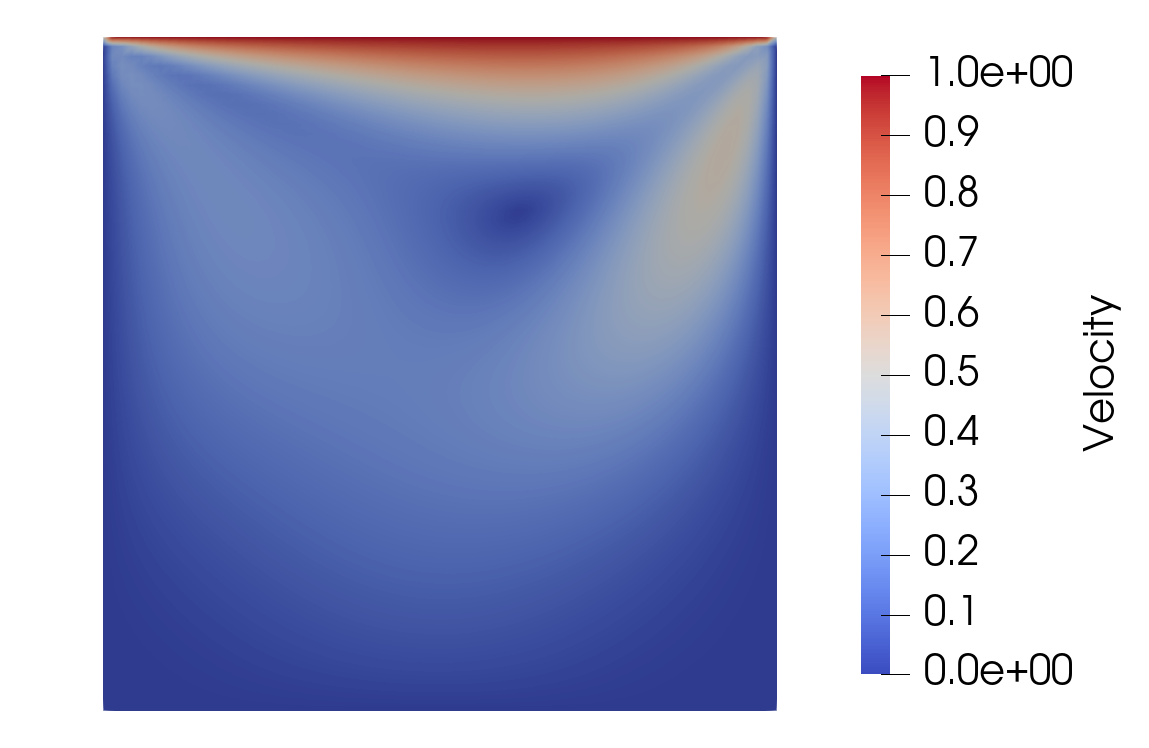}
    \end{subfigure}\hfill
    \begin{subfigure}{0.25\textwidth}
        \centering
        \includegraphics[width=\linewidth]{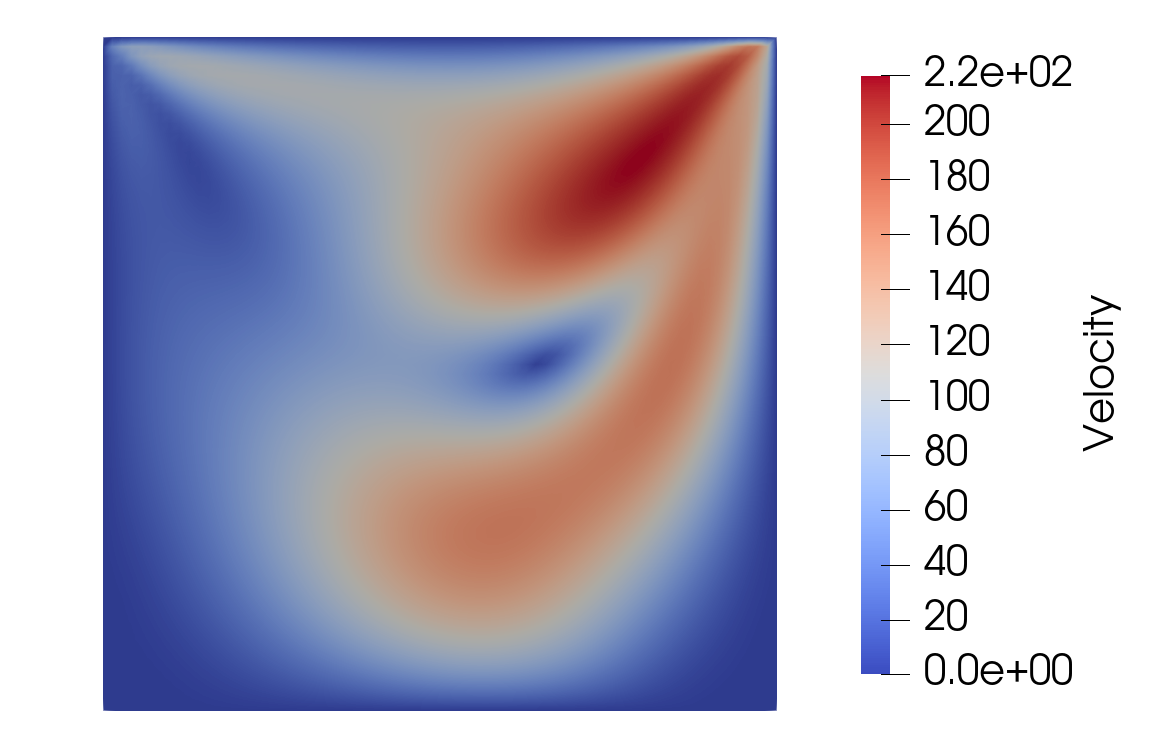}
    \end{subfigure}\hfill
    \begin{subfigure}{0.25\textwidth}
        \centering
        \includegraphics[width=\linewidth]{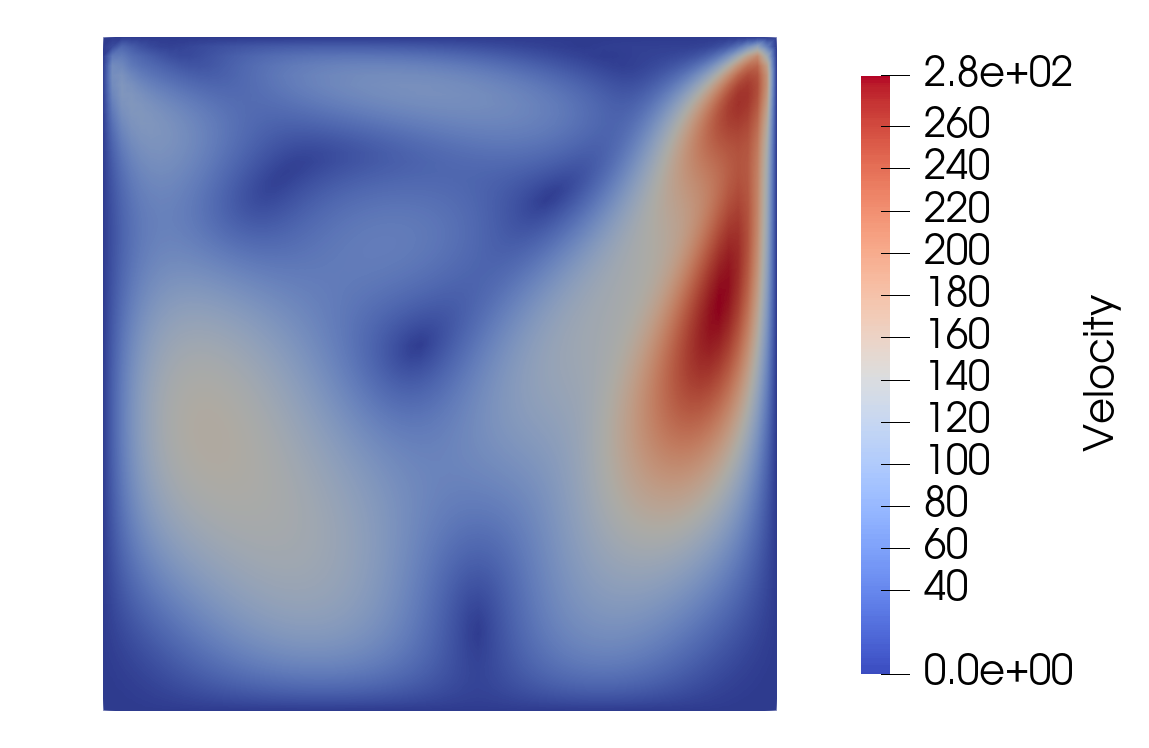}
    \end{subfigure}
    \begin{subfigure}{0.25\textwidth}
        \centering
        \includegraphics[width=\linewidth]{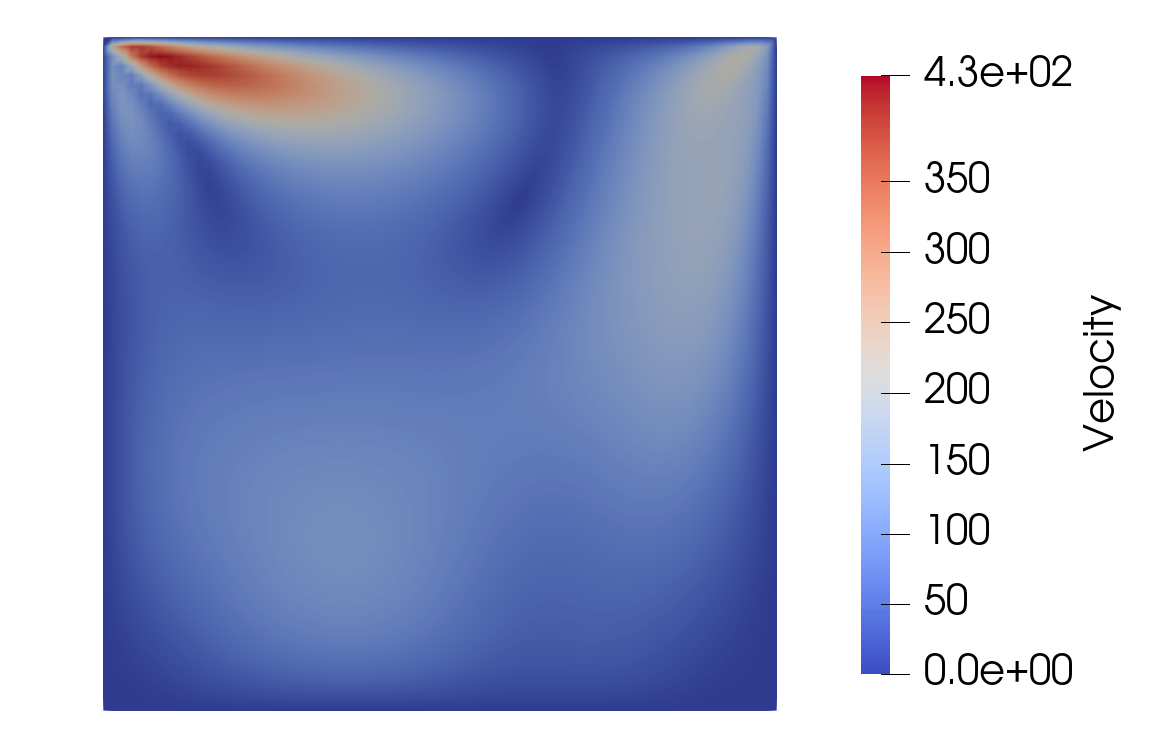}
    \end{subfigure}
    \medskip
    \hspace{-0.5cm}
    \begin{subfigure}{0.25\textwidth}
        \centering
        \includegraphics[width=\linewidth]{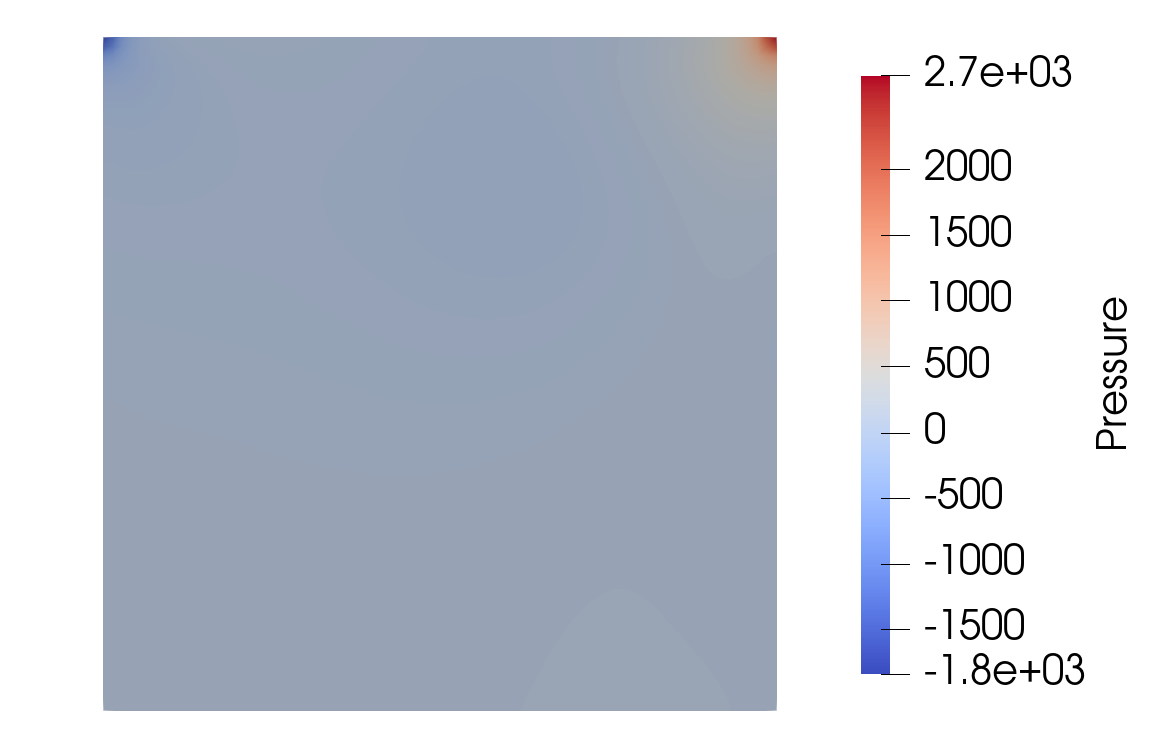}
    \end{subfigure}\hfill
    \begin{subfigure}{0.25\textwidth}
        \centering
        \includegraphics[width=\linewidth]{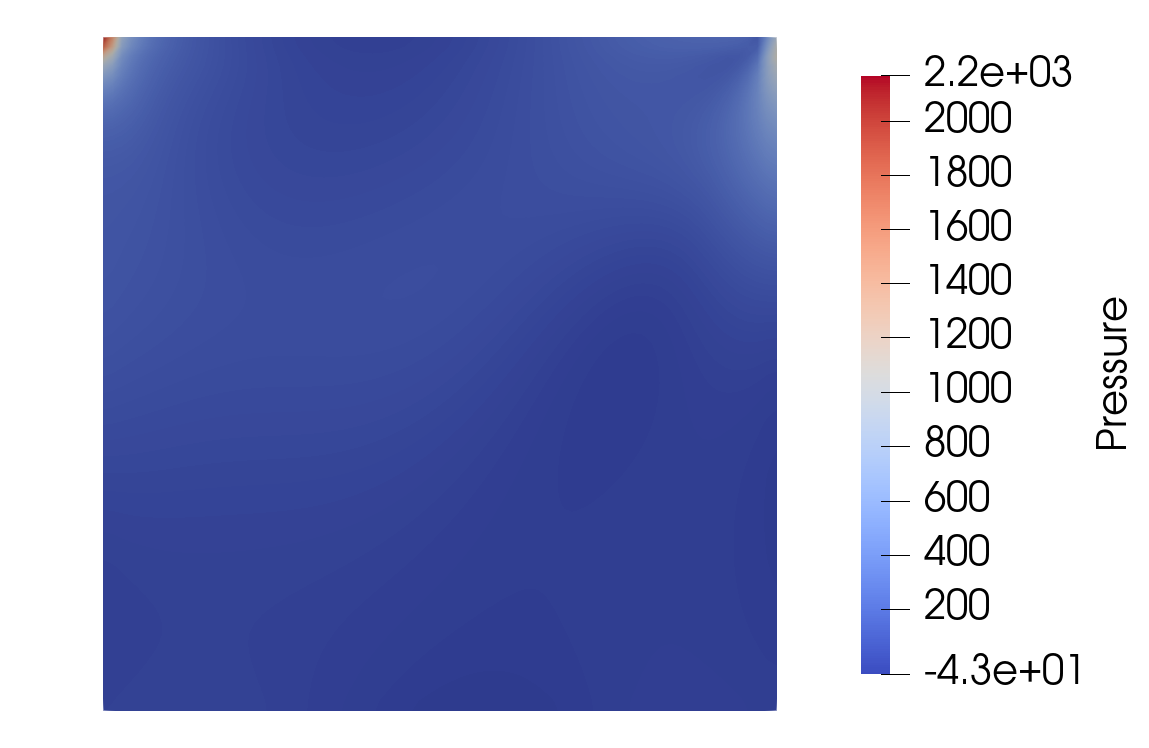}
    \end{subfigure}\hfill
    \begin{subfigure}{0.25\textwidth}
        \centering
        \includegraphics[width=\linewidth]{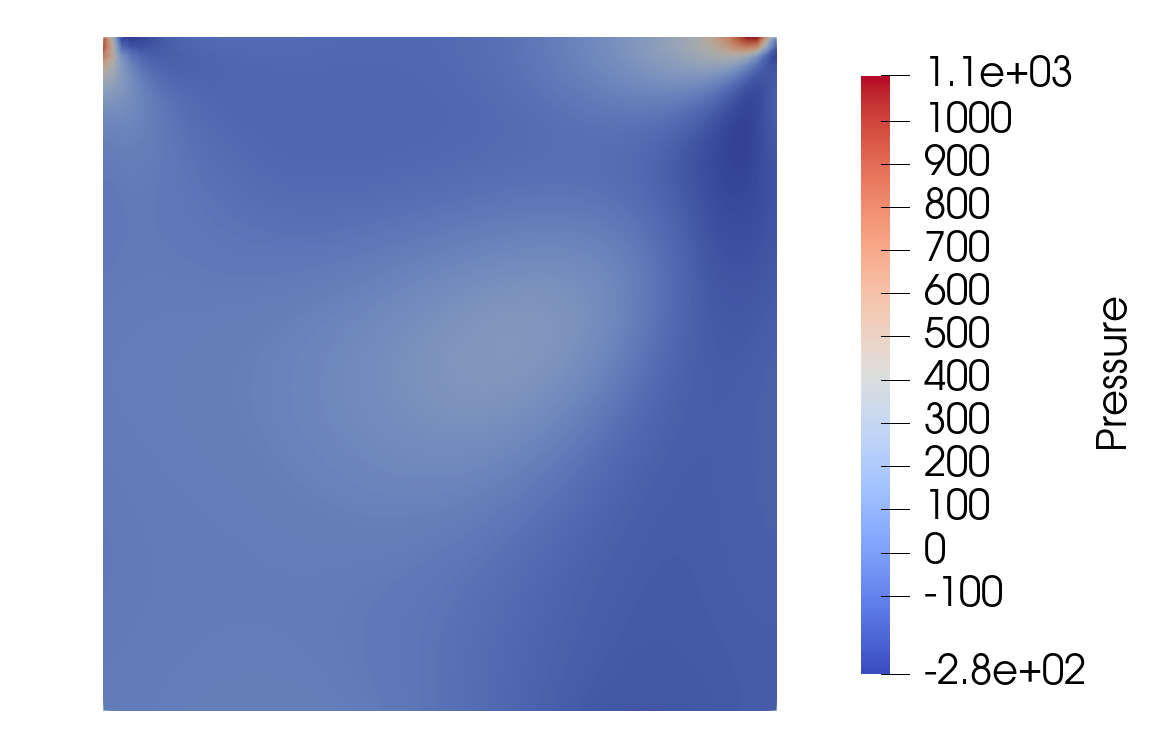}
    \end{subfigure}
    \begin{subfigure}{0.25\textwidth}
        \centering
        \includegraphics[width=\linewidth]{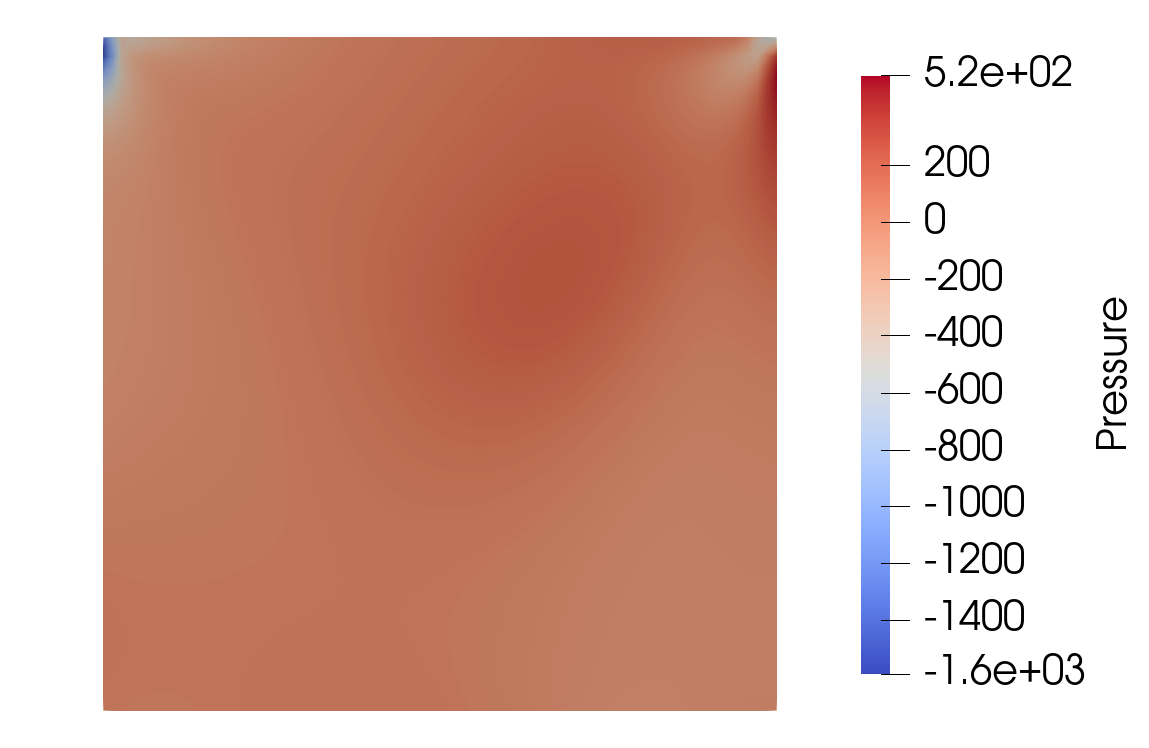}
    \end{subfigure}
    \caption{First four velocity (upper row) and pressure (lower row) modes in the lid-driven cavity case. Remark that for velocity we adopted a lift mode (the first on the left), coinciding with the solution of the undeformed case.}
    \label{POD cavity}
\end{figure}

Being $N_h = 4900$ the number of computational cells in the FV discretization, we define the snapshots matrices $S_p \in \mathbb{R}^{N_h \times n_s}$, $S_{\boldsymbol{u}} \in \mathbb{R}^{3N_h \times n_s}$ for velocity and pressure, respectively.\\
$S_p$ and $S_{\boldsymbol{u}}$ are processed through ITHACA-FV \footnote{https://ithaca-fv.github.io/ITHACA-FV/} (In real Time Highly Advanced Computational Applications for Finite Volumes) \cite{Stabile2017CAIM, stabile2018finite} to compute the reduced order bases with respect to the $L^2(\Omega)$ scalar product in the reference geometry.
Figure \ref{POD cavity} shows the first velocity and pressure modes in the reference undeformed domain. As expected, the lowest-order modes hold most of the information content coming from the collected snapshots. Conversely, increasing the number of modes introduces physically less relevant fields.

In the velocity case, we adopt the "lift function" method exposed in \ref{coupled model}. Consequently, the first mode in Figure \ref{POD cavity} corresponds to the solution in the undeformed case itself, which serves as the lift function.

The spectral analysis highlights that 15 and 6 modes, for velocity and pressure respectively, carry 99.99\% of the information content.
\subsection{Online stage}
\begin{figure}[!htb]
    \begin{subfigure}{0.48\textwidth}
        \centering
        \includegraphics[width=\linewidth]{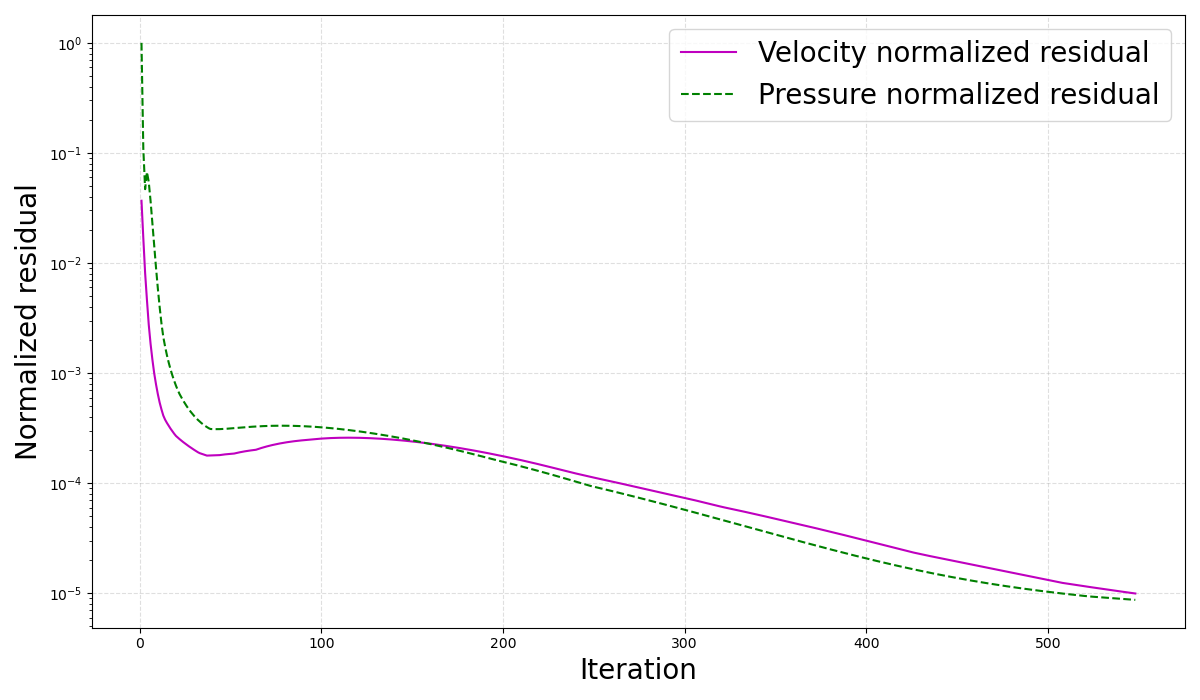}
        \subcaption{$\alpha_{\boldsymbol{u}} = 0.7$, $\alpha_p = 0.3$}
        \label{0.7-0.3-residuals}
    \end{subfigure}\hfill
    \begin{subfigure}{0.48\textwidth}
        \centering
        \includegraphics[width=\linewidth]{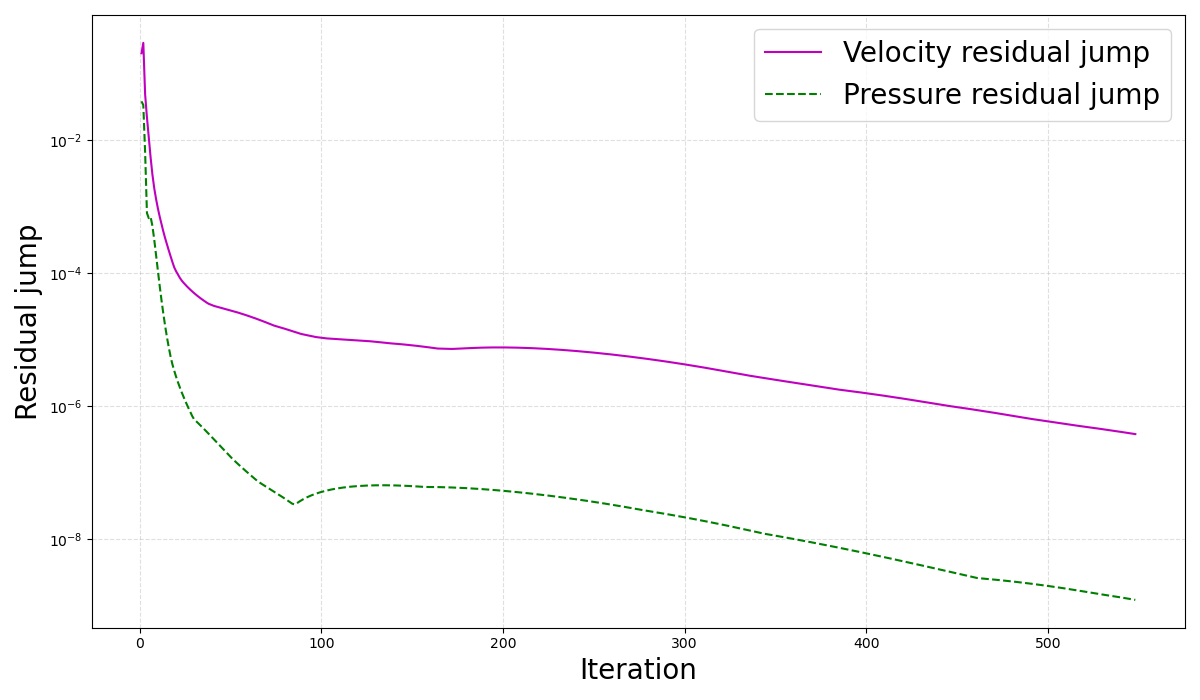}
        \subcaption{$\alpha_{\boldsymbol{u}} = 0.7$, $\alpha_p = 0.3$}
        \label{0.7-0.3-jumps}
    \end{subfigure}
    \medskip
    \begin{subfigure}{0.48\textwidth}
        \centering
        \includegraphics[width=\linewidth]{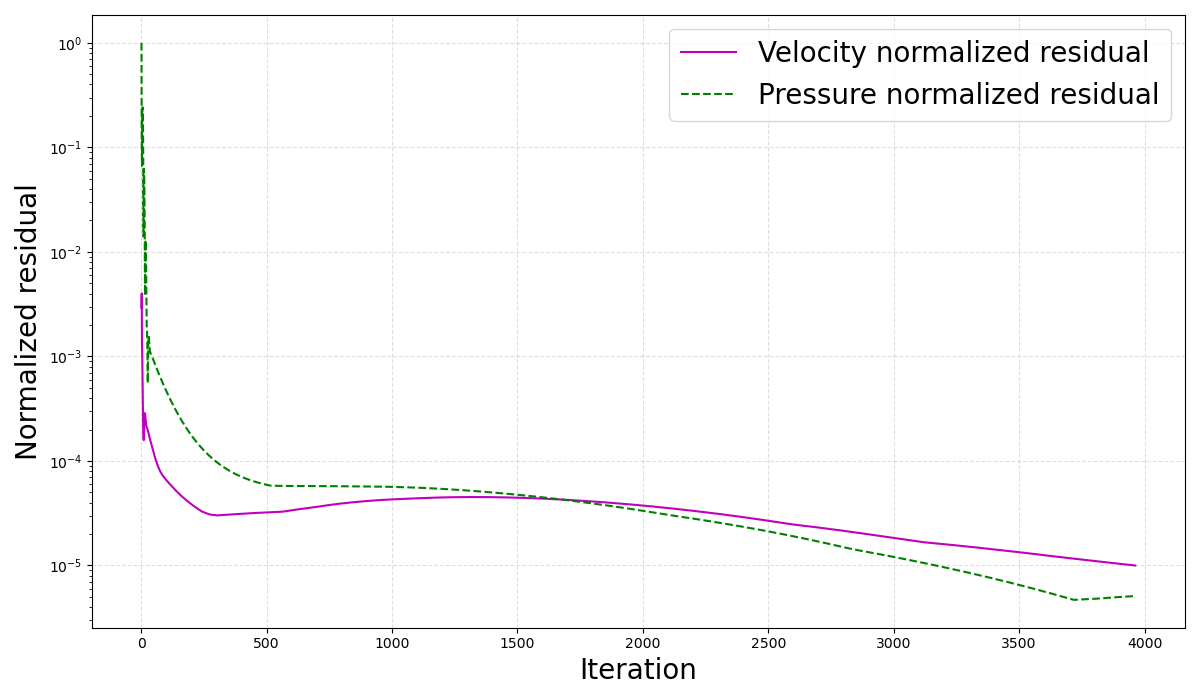}
        \subcaption{$\alpha_{\boldsymbol{u}} = 0.15$, $\alpha_p = 0.3$}
        \label{0.15-0.3-residuals}
    \end{subfigure}\hfill
    \begin{subfigure}{0.48\textwidth}
        \centering
        \includegraphics[width=\linewidth]{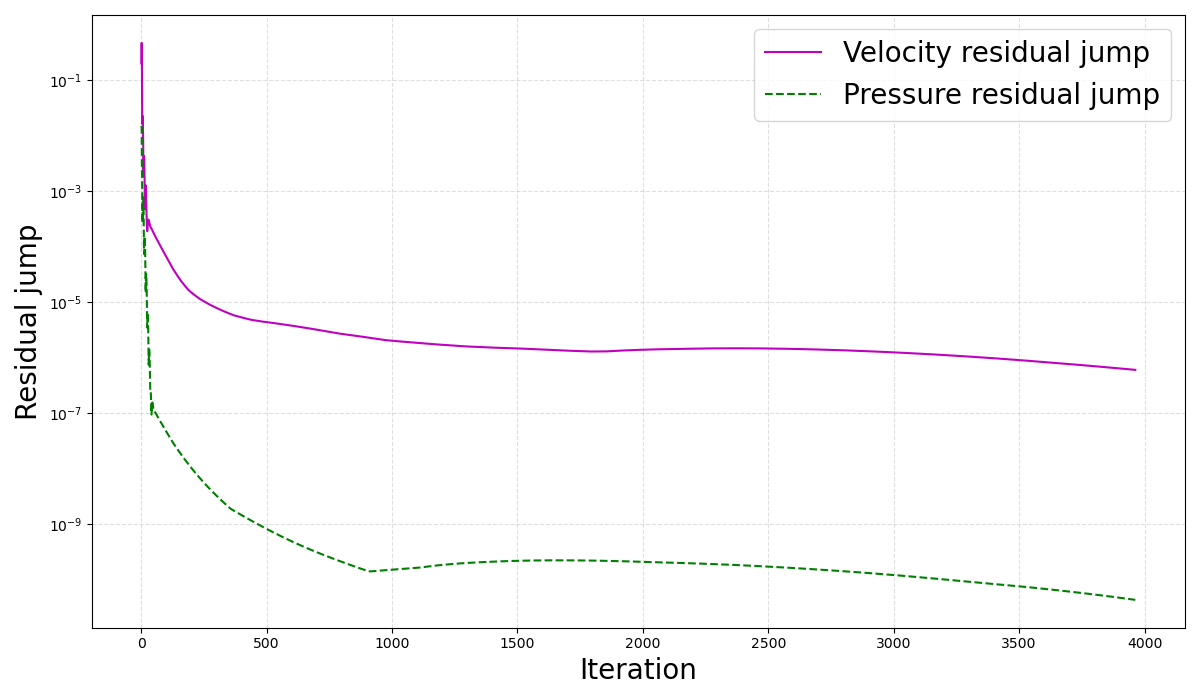}
        \subcaption{$\alpha_{\boldsymbol{u}} = 0.15$, $\alpha_p = 0.3$}
        \label{0.15-0.3-jumps}
    \end{subfigure}
    \caption{Under relaxation coefficients analysis for the reference case: Figures \ref{0.7-0.3-residuals} and \ref{0.7-0.3-jumps} refer to coefficients equal to 0.7 and 0.3 for velocity and pressure respectively, whereas Figures \ref{0.15-0.3-residuals} and \ref{0.15-0.3-jumps} stand for coefficients equal to 0.15 and 0.3. Simulations have been carried out with $N_{\boldsymbol{u}} = 15$ and $N_p = 6$.}
    \label{cavity-under_relax_coeff}
\end{figure}

Both coupled and SIMPLE ROM methods are considered for the online stage.\\
The SIMPLE algorithm is subject to numerical instabilities arising from the choice of the under-relaxation coefficients. Too large coefficients may lead to divergent schemes, whereas a too conservative choice could result in an excessively slow convergence.\\
Figure \ref{cavity-under_relax_coeff} shows how this choice affects the convergence rate. In this specific problem, the best option proves to be to mimic the structure of the full-order scheme, with $\alpha_{\boldsymbol{u}} = 0.7$ and $\alpha_p = 0.3$ for velocity and pressure respectively. Apart from guaranteeing the fastest convergence among the tested configurations, this choice ensures the cleanest comparison between the full and reduced order models. Contrary to \cite{stabile2018finite}, we did not include FOM snapshots coming from intermediate iterations of the SIMPLE procedure, because not needed to attain convergence of the SIMPLE ROM. This choice speedups and reduces memory needs for the offline stage. Additionally, we expect to extract more efficient modes, not affected by non-physical snapshots, from the POD procedure.

\begin{figure}[!htb]
    \begin{subfigure}{0.48\textwidth}
        \centering
        \includegraphics[width=\linewidth]{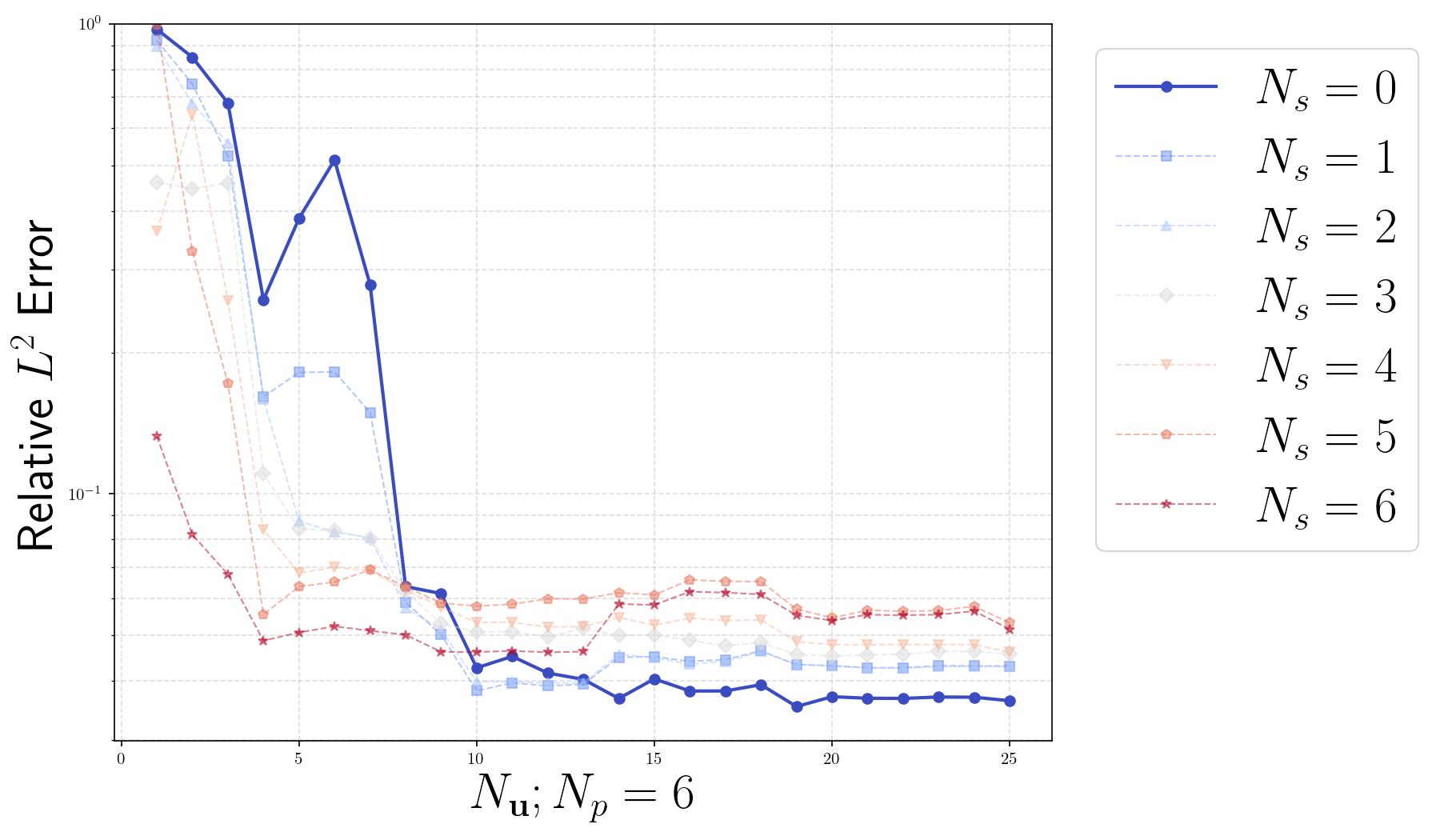}
        \caption{Relative velocity error for $N_p = 6$.}
        \label{sup_fixed_modes - velocity}
    \end{subfigure}\hfill
    \begin{subfigure}{0.48\textwidth}
        \centering
        \includegraphics[width=\linewidth]{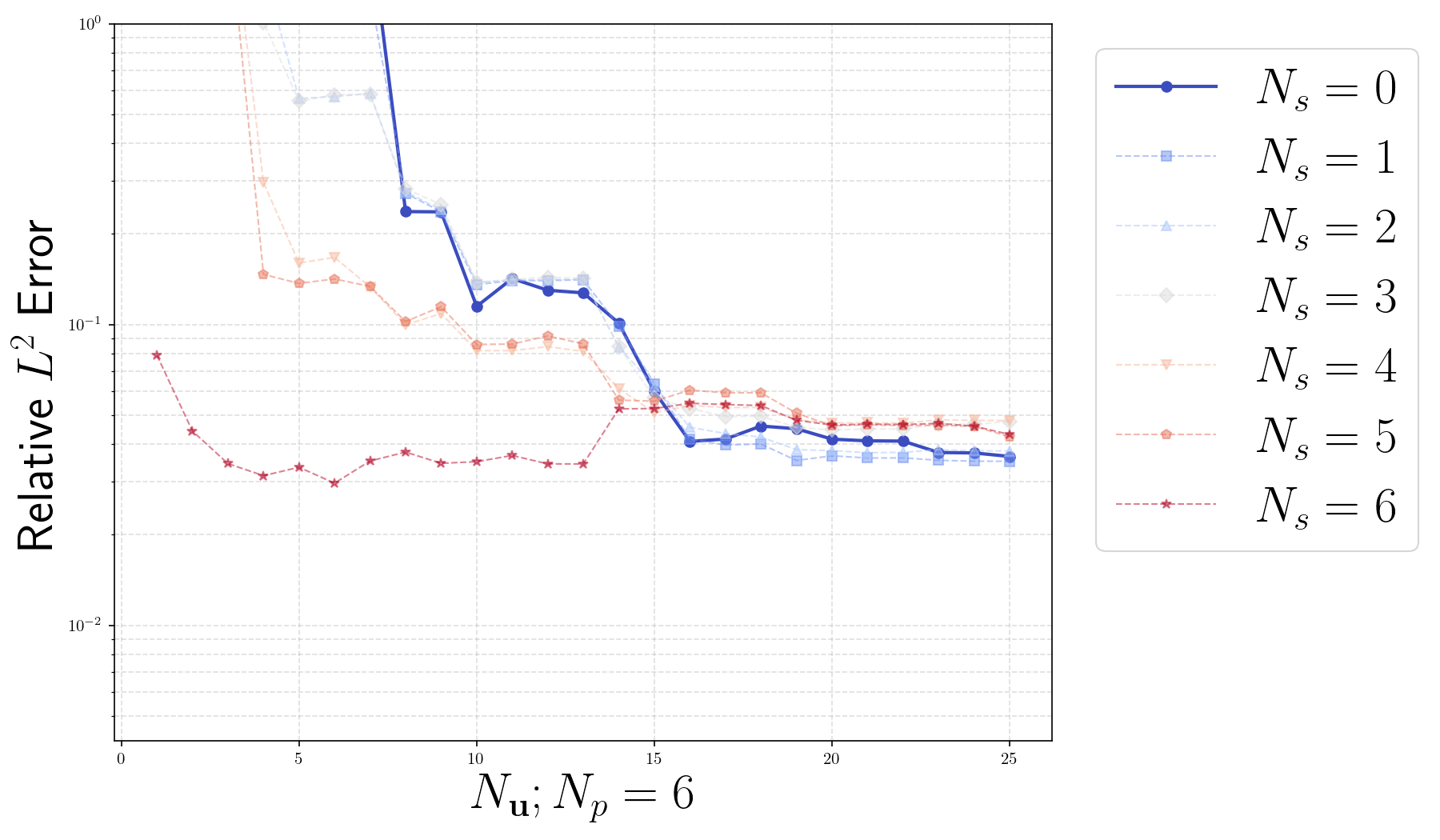}
        \caption{Relative pressure error for $N_p = 6$.}
        \label{sup_fixed_modes - pressure}
    \end{subfigure}
    \medskip
    \begin{subfigure}{0.48\textwidth}
        \centering
        \includegraphics[width=\linewidth]{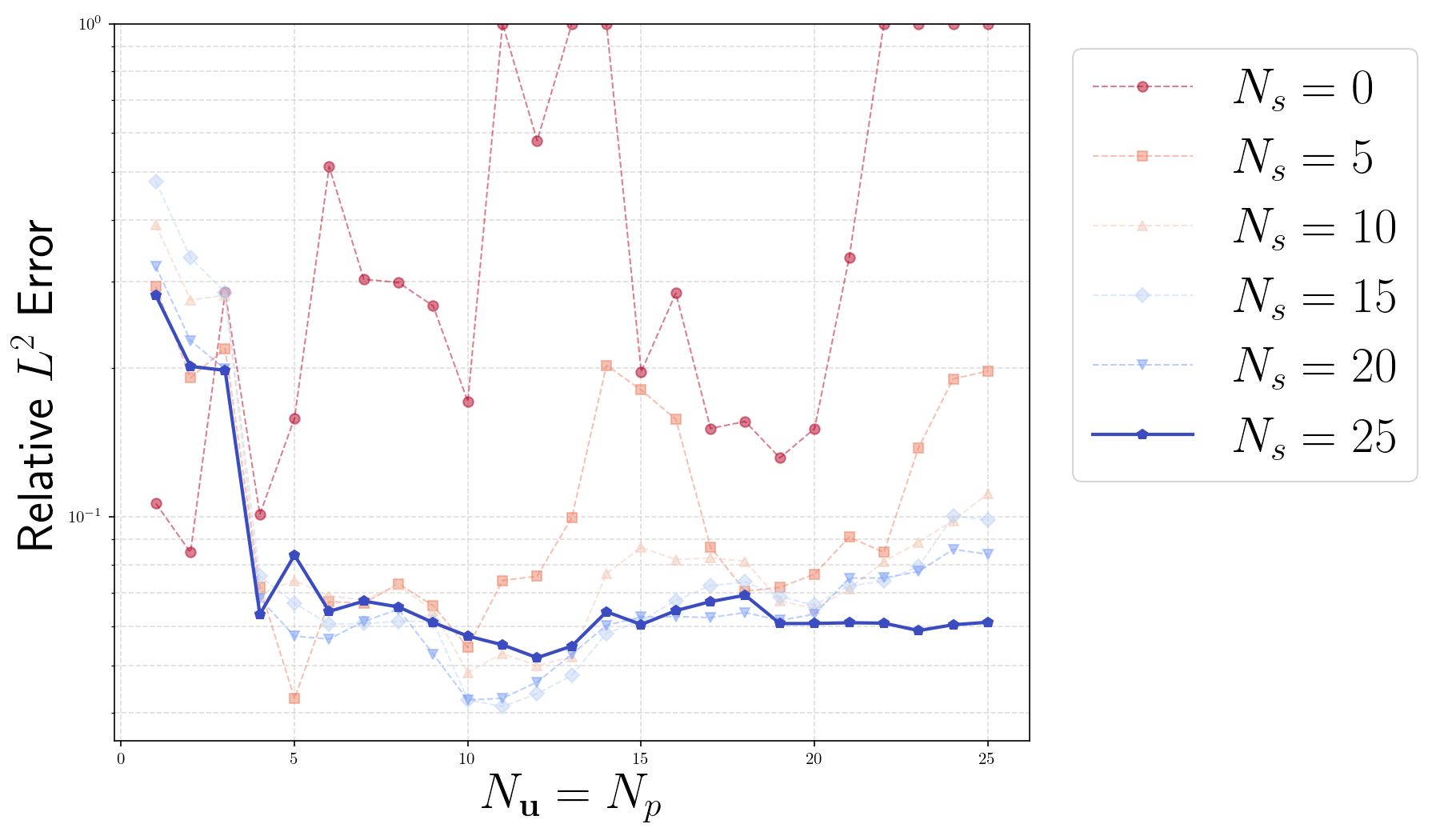}
        \caption{Relative velocity error for $N_{\boldsymbol{u}} = N_p$.}
        \label{sup_varying_modes - velocity}
    \end{subfigure}\hfill
    \begin{subfigure}{0.48\textwidth}
        \centering
        \includegraphics[width=\linewidth]{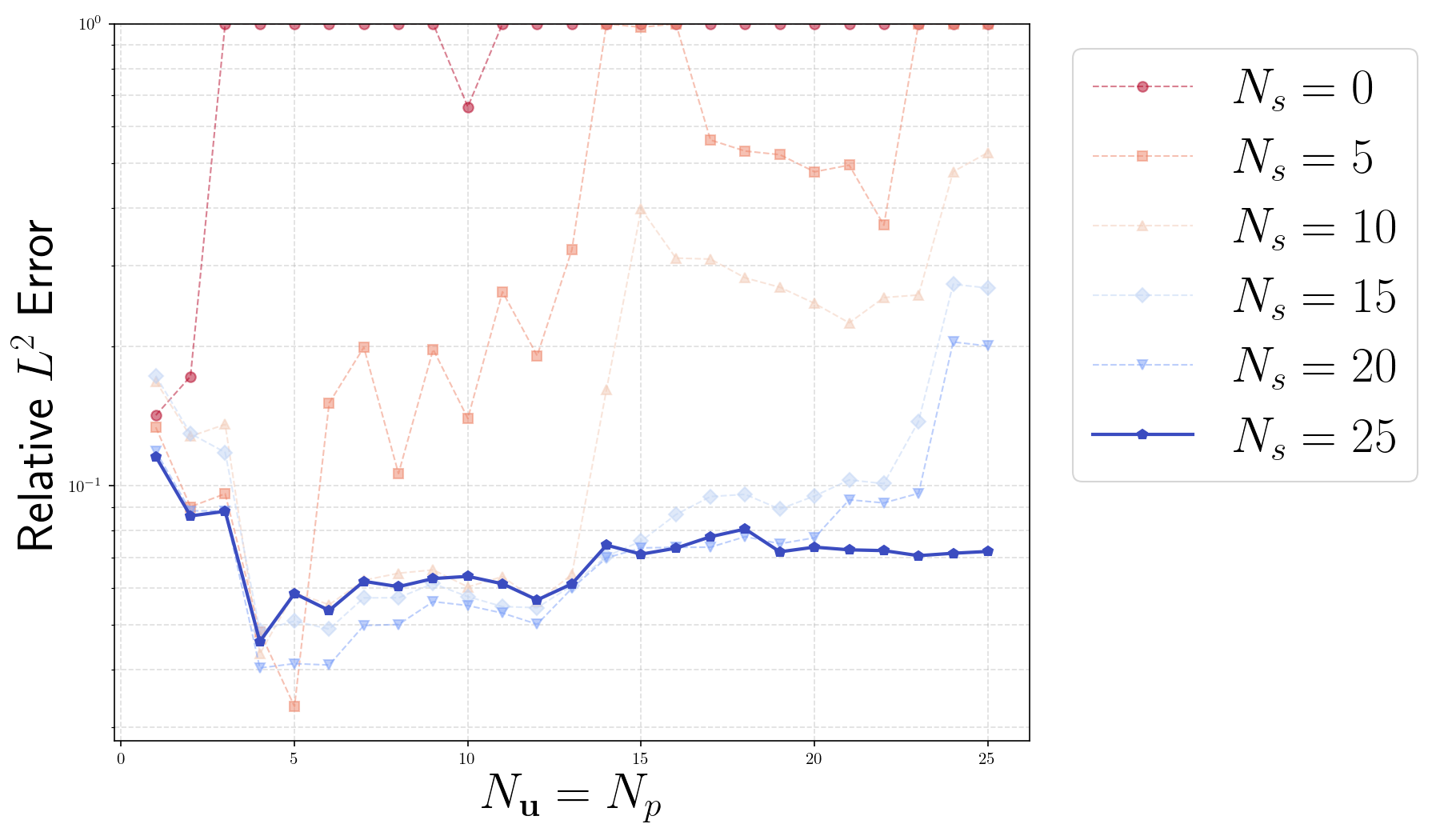}
        \caption{Relative pressure error for $N_{\boldsymbol{u}} = N_p$.}
        \label{sup_varying_modes - pressure}
    \end{subfigure}
    \caption{Analysis of the relative errors induced by the choice of the number of supremizers in the reduced coupled algorithm. Figures \ref{sup_fixed_modes - velocity} and \ref{sup_fixed_modes - pressure} present the results obtained by changing the number of velocity modes $N_{\boldsymbol{u}}$, while keeping $N_p = 6$. Figures \ref{sup_varying_modes - velocity} and \ref{sup_varying_modes - pressure} instead stand for the case in which the number of velocity and pressure modes equally vary.}
    \label{cavity_sup-analysis}
\end{figure}

Regarding the coupled approach, supremizers enrichment is needed due to its saddle-point structure to avoid spurious pressure solutions. Figure \ref{cavity_sup-analysis} shows the relative errors for different choices of the number of supremizers $N_s$ depending on the number of modes. Simulations are run in the undeformed geometry and two different frameworks are considered: in Figures \ref{sup_fixed_modes - velocity}, \ref{sup_fixed_modes - pressure}, the number of pressure modes $N_p$ is kept fixed to $6$, while in Figures \ref{sup_varying_modes - velocity} and \ref{sup_varying_modes - pressure} both $N_{\boldsymbol{u}}$ and $N_p$ vary with $N_{\boldsymbol{u}}=N_p$.\\
As expected, spurious pressure modes are obtained for the lowest $N_{\boldsymbol{u}}$ choices, increasing the relative error. Nevertheless, for $N_p = 6$, enlarging the reduced velocity space is sufficient to guarantee the stabilization of the saddle-point problem, even with $N_s = 0$. This is not the case for $N_p = N_{\boldsymbol{u}}$, where the increase of the reduced pressure space requires, for larger $N_{\boldsymbol{u}}=N_p$ values, the use of $N_s = 25$, the largest possible value in our analysis.

\begin{figure}[!htb]
    \centering
    \begin{minipage}[c]{0.32\textwidth}
        \centering
        \begin{subfigure}{\linewidth}
            \centering
            \includegraphics[width=\linewidth]{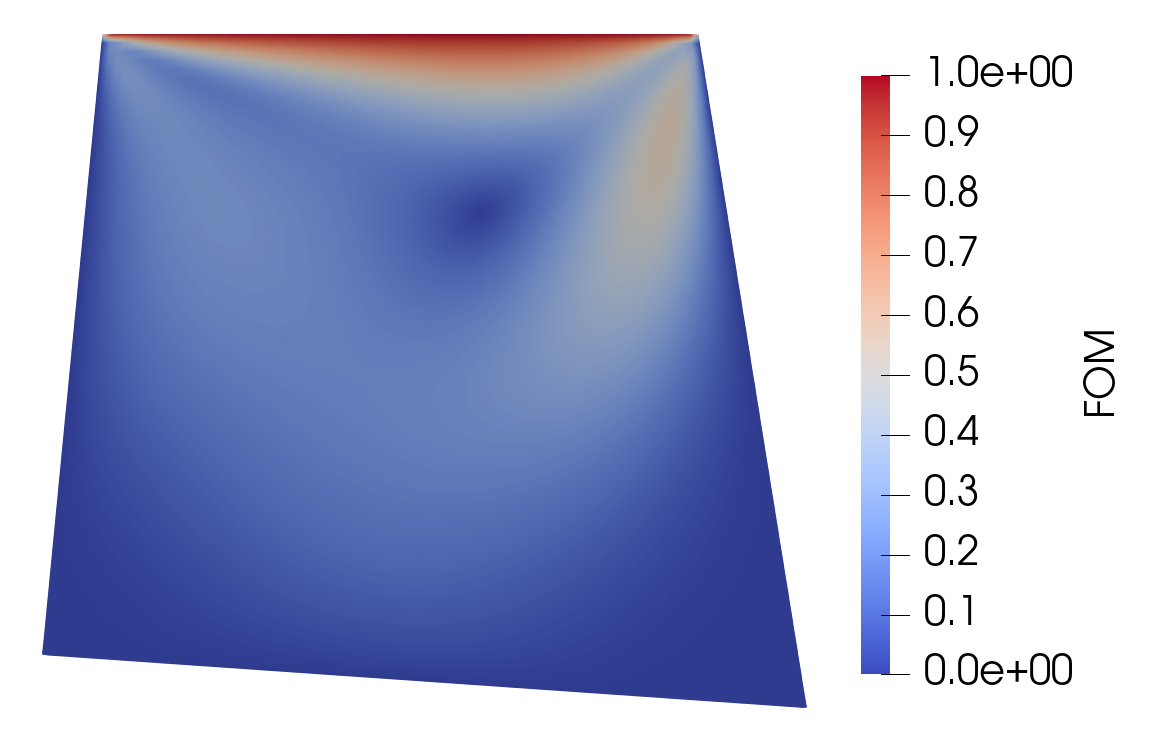}
        \end{subfigure}
    \end{minipage}
    \hfill
    \begin{minipage}[c]{0.65\textwidth}
        \begin{subfigure}{0.48\linewidth}
            \centering
            \includegraphics[width=\linewidth]{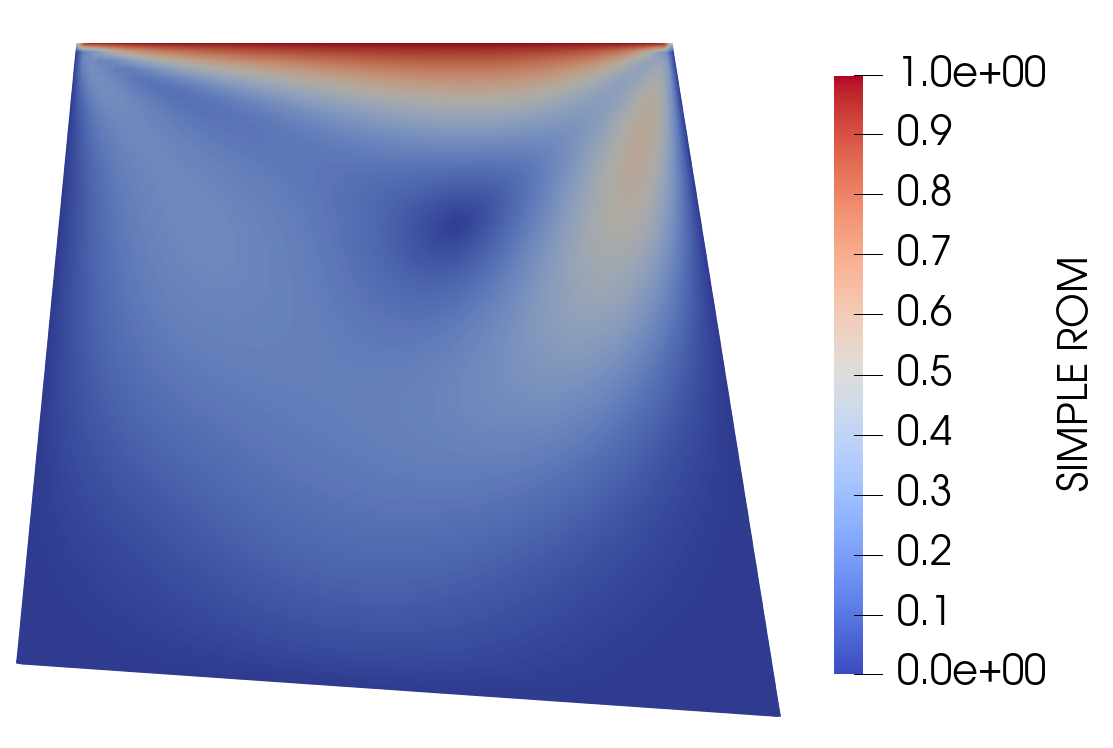}
        \end{subfigure}\hfill
        \begin{subfigure}{0.48\linewidth}
            \centering
            \includegraphics[width=\linewidth]{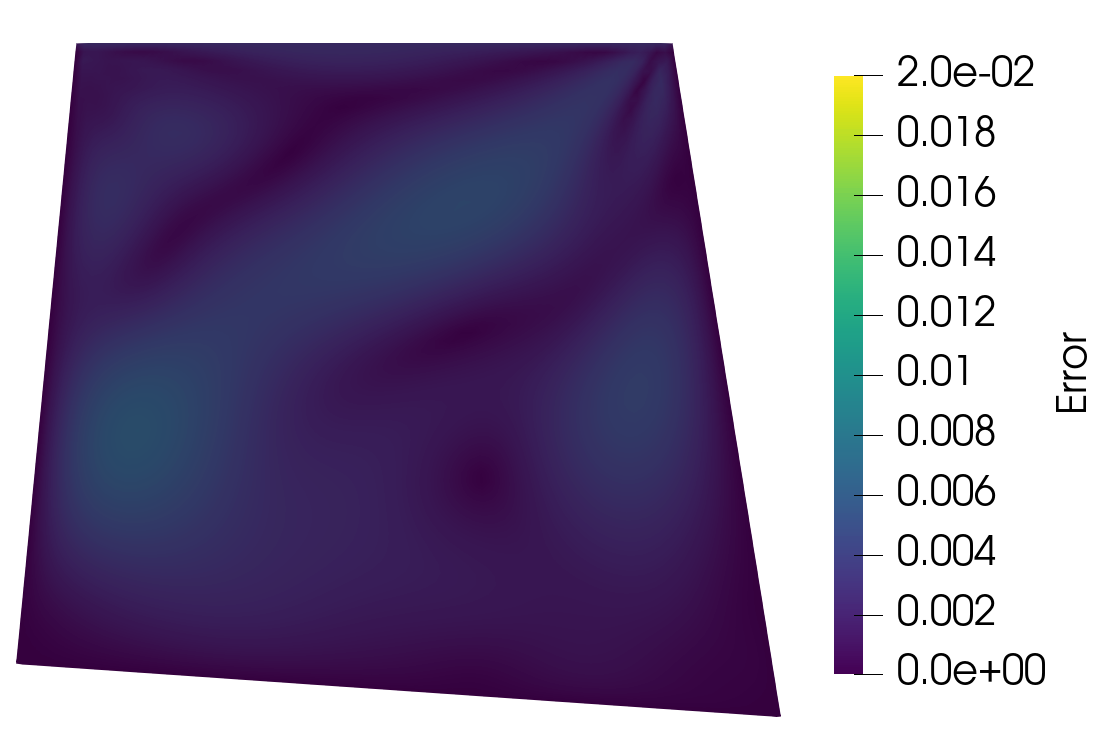}
        \end{subfigure}
        \\[2ex]
        \begin{subfigure}{0.48\linewidth}
            \centering
            \includegraphics[width=\linewidth]{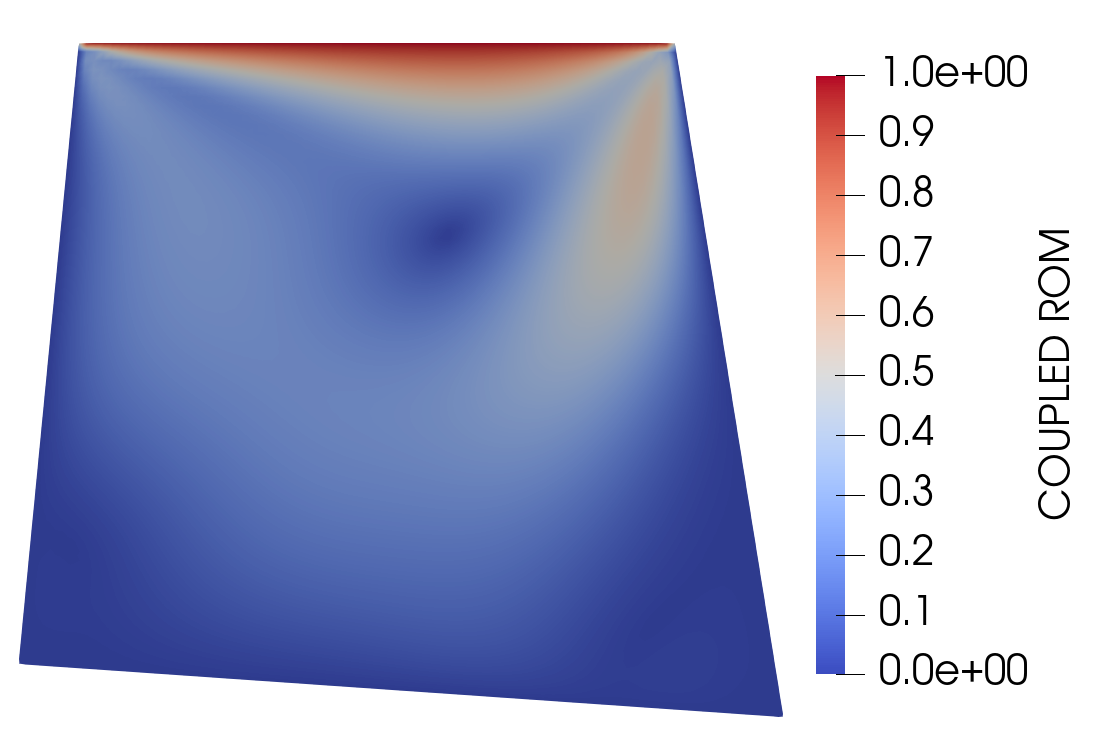}
        \end{subfigure}\hfill
        \begin{subfigure}{0.48\linewidth}
            \centering
            \includegraphics[width=\linewidth]{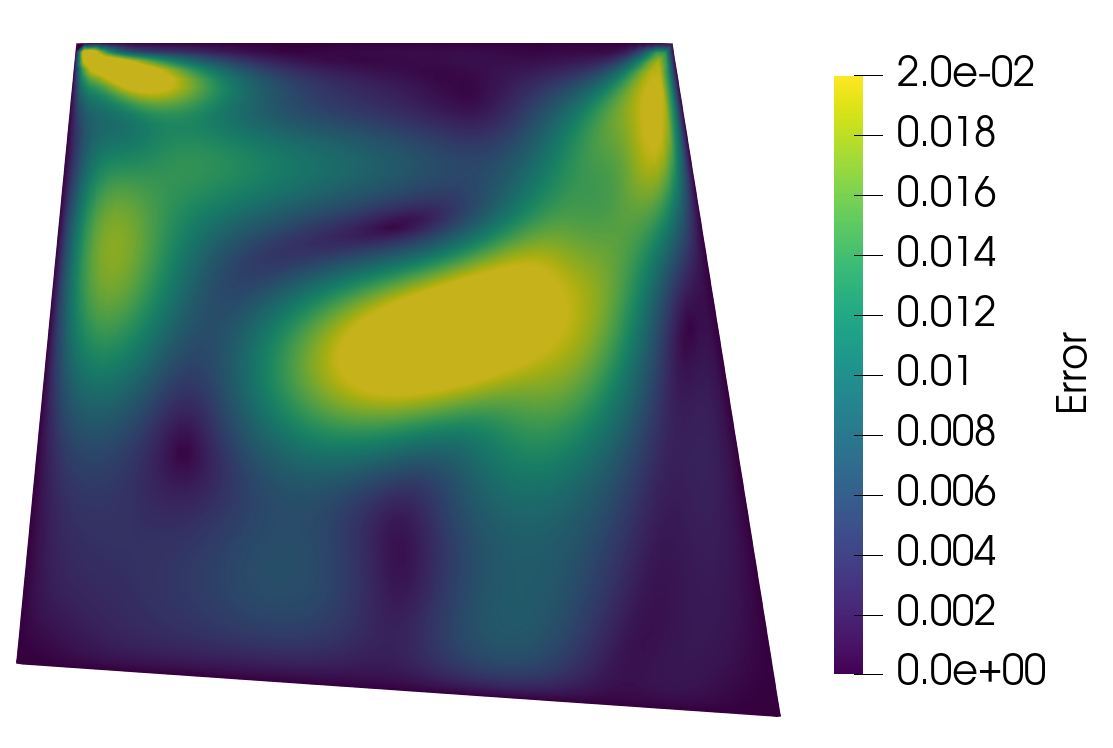}
        \end{subfigure}
    \end{minipage}

    \medskip

    \begin{minipage}[c]{0.32\textwidth}
        \centering
        \begin{subfigure}{\linewidth}
            \centering
            \includegraphics[width=\linewidth]{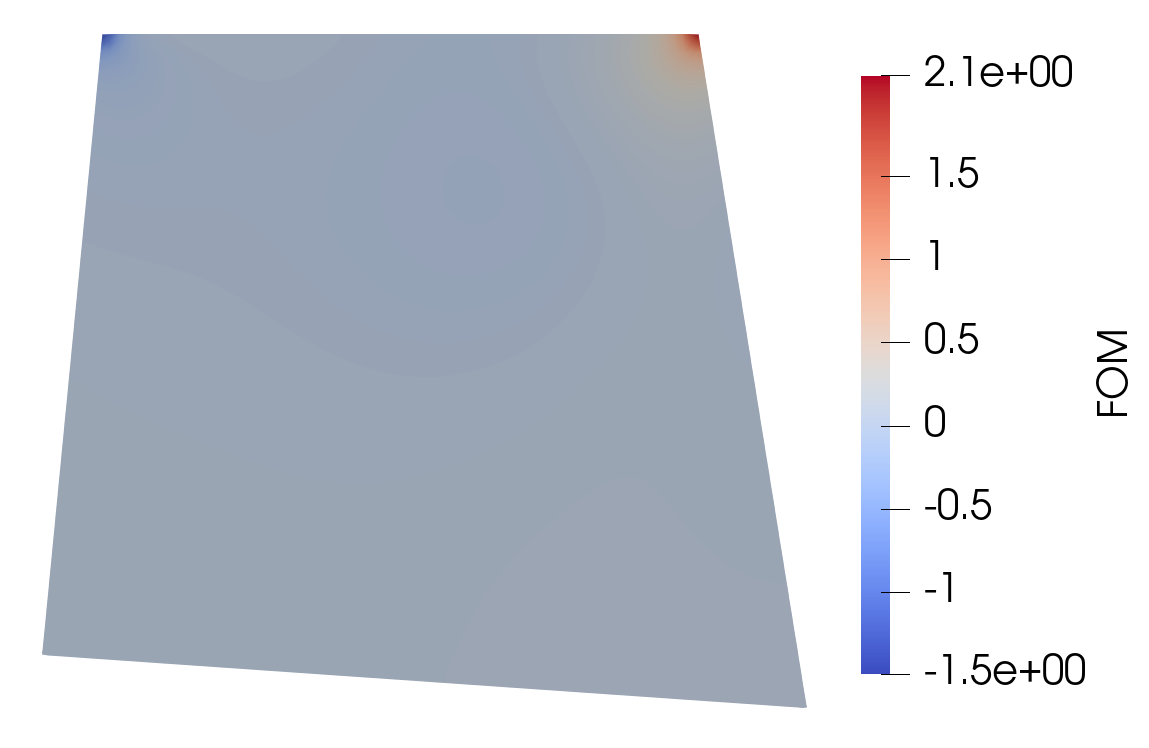}
        \end{subfigure}
    \end{minipage}
    \hfill
    \begin{minipage}[c]{0.65\textwidth}
        \begin{subfigure}{0.48\linewidth}
            \centering
            \includegraphics[width=\linewidth]{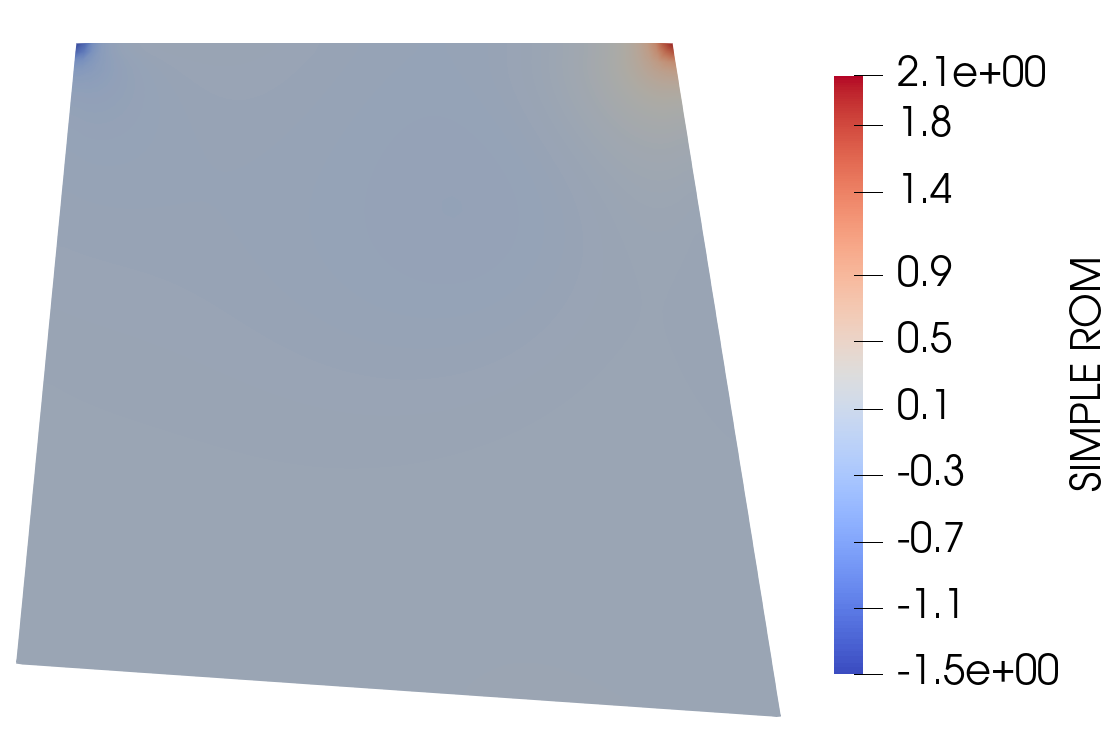}
        \end{subfigure}\hfill
        \begin{subfigure}{0.48\linewidth}
            \centering
            \includegraphics[width=\linewidth]{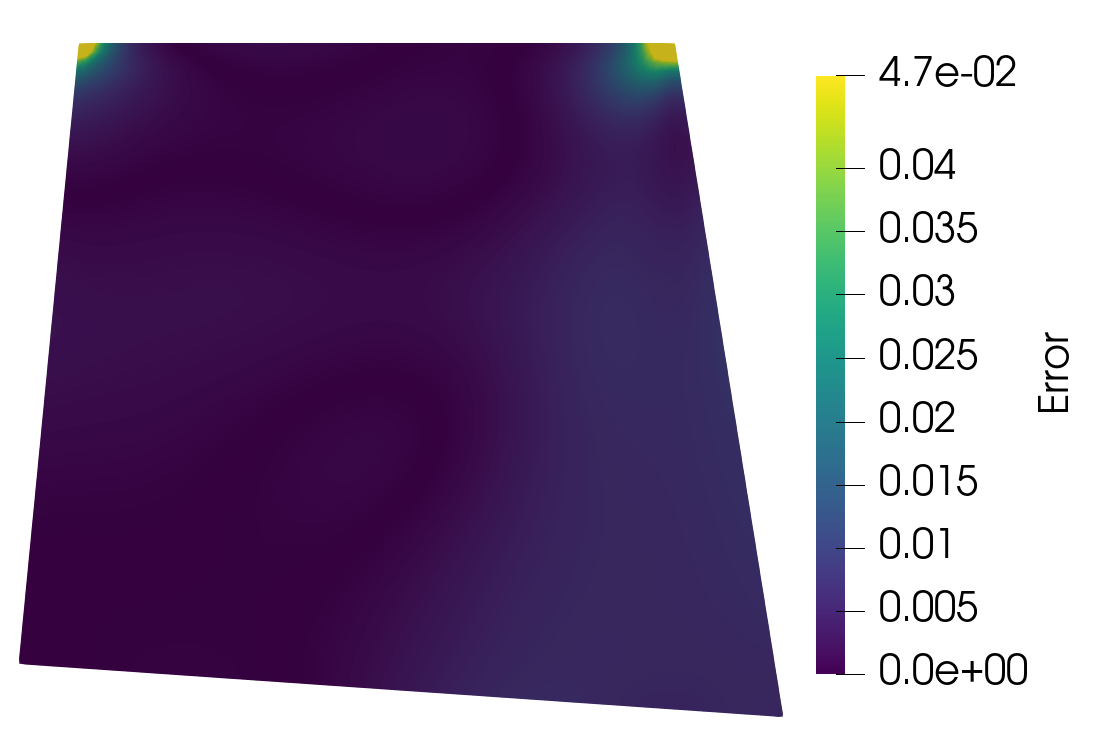}
        \end{subfigure}
        \\[2ex]
        \begin{subfigure}{0.48\linewidth}
            \centering
            \includegraphics[width=\linewidth]{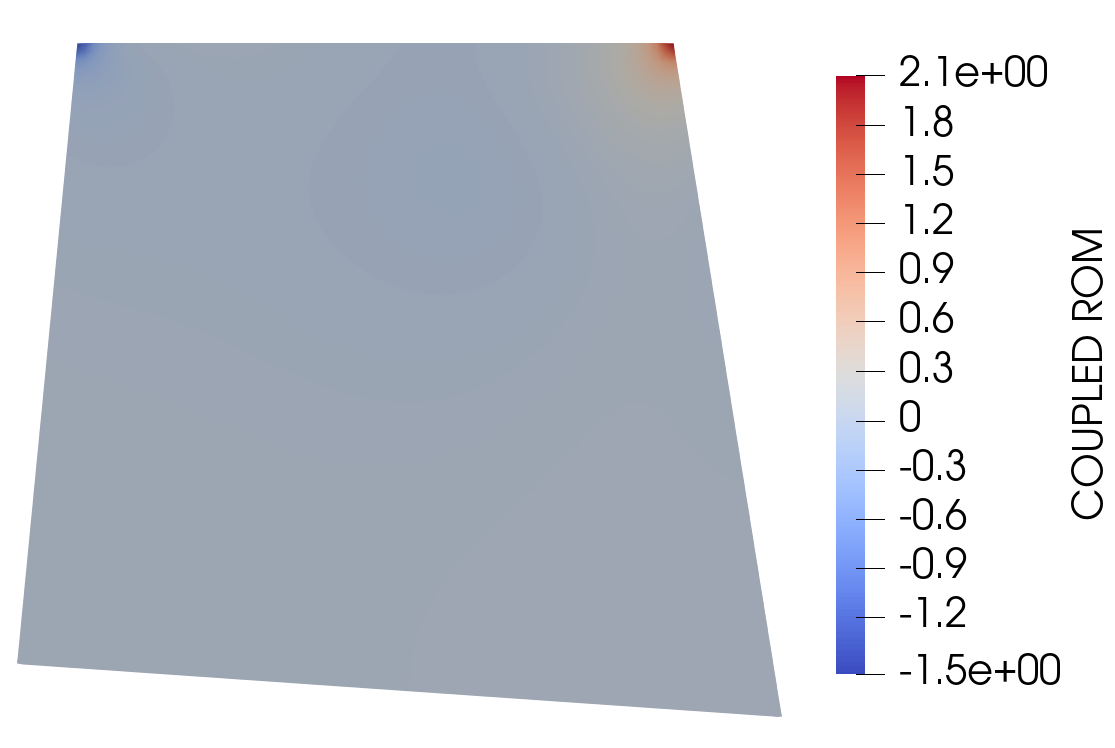}
        \end{subfigure}\hfill
        \begin{subfigure}{0.48\linewidth}
            \centering
            \includegraphics[width=\linewidth]{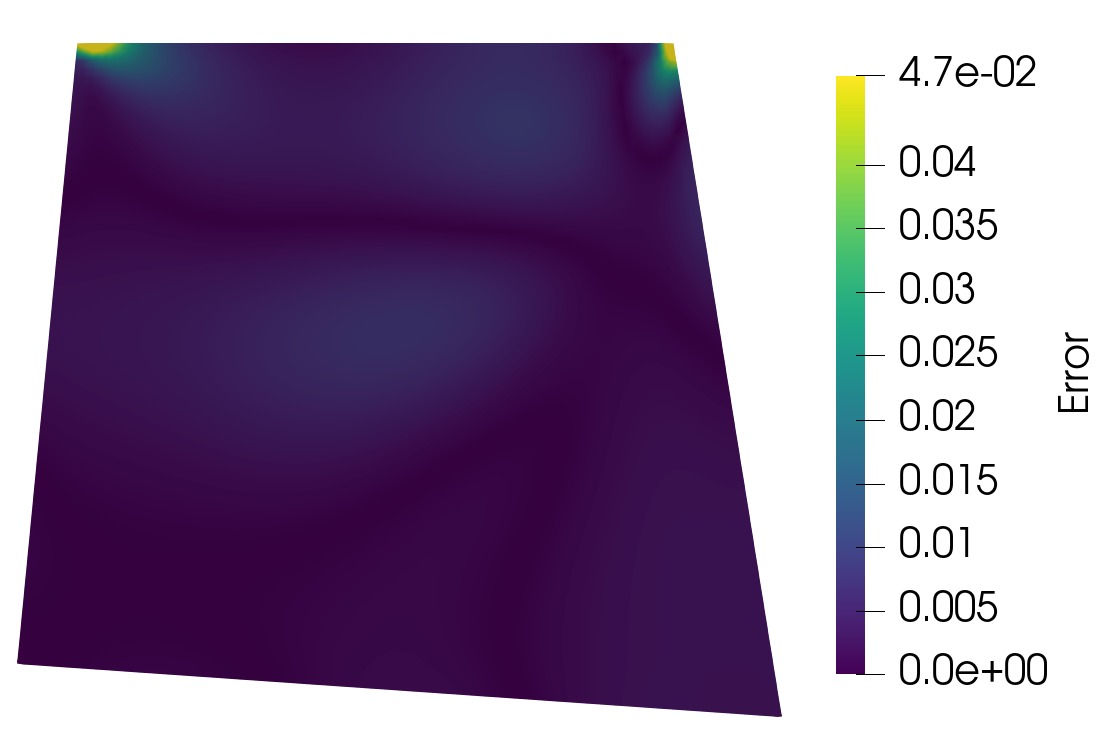}
        \end{subfigure}
    \end{minipage}
    \caption{Qualitative analysis of the absolute errors induced by the model order reduction: we compare the results obtained through the two reduced algorithms with those obtained at the full order level, both in terms of velocity and pressure (first and second set of figures respectively). Simulations have been carried out with $N_{\boldsymbol{u}} = 15$, $N_p = 6$ and $N_s = 0$.}
    \label{Cavity_absolute_error}
\end{figure}

To investigate the accuracy of both methods, we test them on a new deformed geometry. Figure \ref{Cavity_absolute_error} shows the predicted fields and absolute errors with respect to the full-order counterpart for one deformed case. Reduced solutions refer to $N_{\boldsymbol{u}} = 15$, $N_p = 6$ and $N_s = 0$. Both ROMs accurately approximate the full-order results, with comparable order of magnitude for the absolute error both in velocity and pressure. Nonetheless, the SIMPLE algorithm proves to be more accurate in the considered deformation. Specifically, relative errors are equal to 6.00\% (pressure) and 4.04\% (velocity) for the coupled algorithm, 6.53\% (pressure) and 1.19\% (velocity) for the SIMPLE one.

\begin{figure}[!htb]
    \begin{subfigure}[t]{0.16\textwidth}
        \vspace{0pt}
        \centering
        \includegraphics[width=\linewidth]{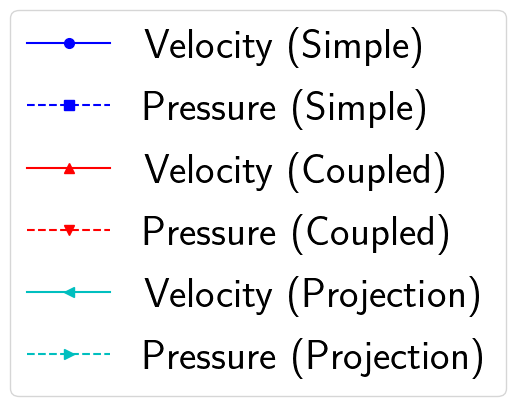}
    \end{subfigure}\hfill
    \begin{subfigure}[t]{0.41\textwidth}
        \vspace{0pt}
        \centering
        \includegraphics[width=\linewidth]{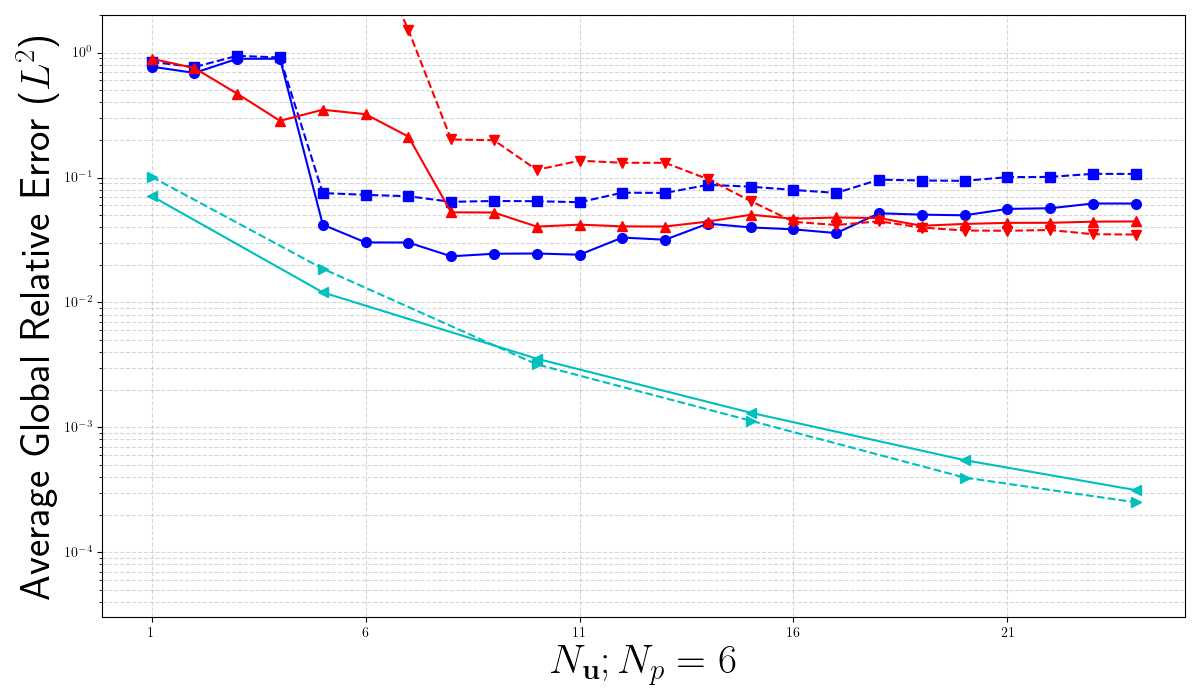}
        \caption{}
        \label{pres_fixed}
    \end{subfigure}\hfill
    \begin{subfigure}[t]{0.41\textwidth}
        \vspace{0pt}
        \centering
        \includegraphics[width=\linewidth]{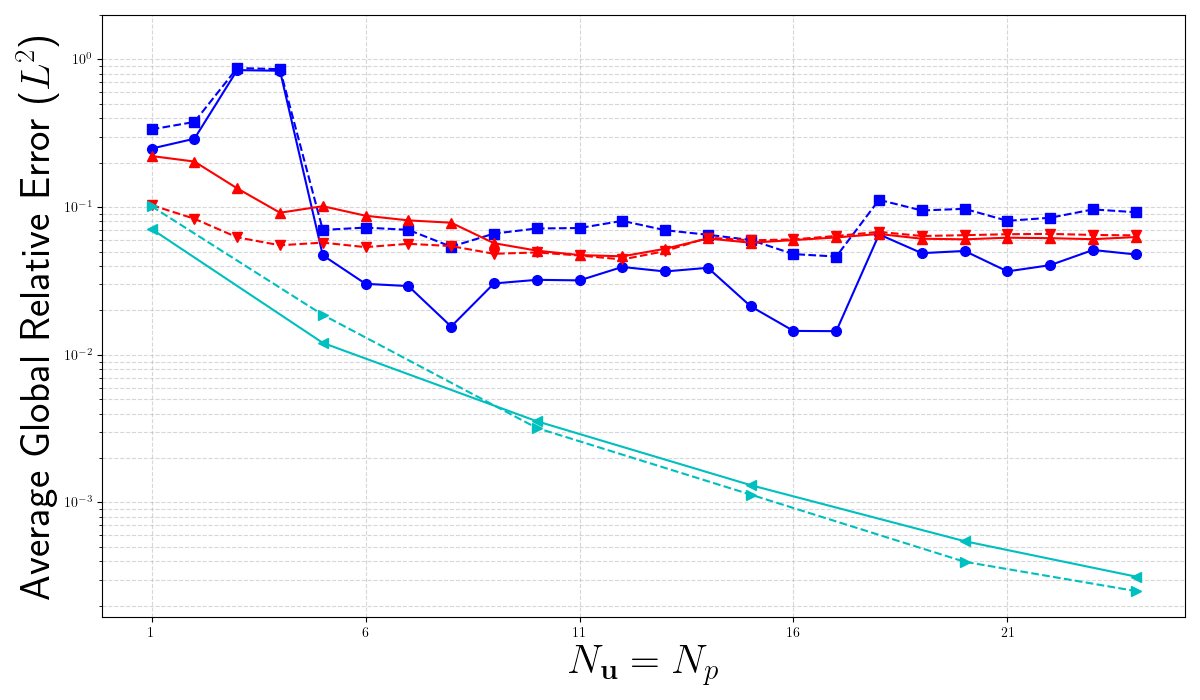}
        \caption{}
        \label{pres_varies}
    \end{subfigure}
    \caption{Comparison of the average relative errors induced by the reduced order models against the exact projection error. In Figure \ref{pres_fixed}, $N_p=6$ is fixed, whereas in Figure \ref{pres_varies} $N_{\boldsymbol{u}} = N_p$.}
    \label{relative_errors}
\end{figure}

To deepen our analysis, we test both ROMs on 10 new geometries and average the corresponding errors.
Figure \ref{pres_fixed} exhibits the average error when fixing $N_p = 6$ and $N_s = 0$. It appears that the SIMPLE algorithm tends to be more stable for $N_{\boldsymbol{u}} \leq 7$, due to the segregated nature of the problem, which circumvents handling a saddle-point problem. Additionally, the coupled approach exhibits in the same region completely out-of-range relative errors, caused by spurious pressure modes. Enlarging the reduced velocity space, namely for $8 \leq N_{\boldsymbol{u}} \leq 17$, overcomes the absence of supremizers and brings the errors to comparable values. Conversely, the SIMPLE algorithm struggles when handling a larger number of degrees of freedom, showing slightly worse performances for $N_{\boldsymbol{u}} \geq 18$.

In Figure \ref{pres_varies}, the case for $N_{\boldsymbol{u}} = N_p$ is presented. The coupled algorithm shows a more stable behaviour, enabled by the choice of a large number of supremizer $N_s = 25$, which stabilizes the problem even for lower dimensional spaces. The SIMPLE algorithm leads to errors within a comparable order of magnitude, yet with higher instability, probably due to the difficulty of handling some less relevant and noisier modes introduced by the expansion of the reduced spaces. This phenomenon goes under the name of \textit{saturation} \cite{quarteroni2015reduced, pinkus2012n}. Both methods reach relatively high plateaus with respect to exact projection errors, which are computed from the projection of the full-order solutions onto the reduced spaces. This discrepancy is mainly caused by two different factors: on the one hand, projecting the problem with respect to the undeformed $L^2(\Omega)$ scalar product implies some sort of inaccuracy, when dealing with other geometries; on the other hand, considering too many modes add more noise rather than information content to the model, in the aforementioned \textit{saturation} phenomenon. In addition, the linear reduced representation hypothesis introduces an intrinsic error when dealing with non-linear problems \cite{khamlich2025advanced}. A potential remedy is discussed in \cite{stabile2020efficient}, though investigating this methodology is outside the scope of this work.

Finally, from a computational point of view, both algorithms successfully reduce the complexity of the problem, decreasing the degrees of freedom from $4 \, N_h = 19600$ to $N = N_{\boldsymbol{u}} + N_p = 21$. Conversely, not implementing (D)EIM reduction \cite{stabile2020efficient}, we do not expect speed-ups in terms of computational time, thus we do not compare the algorithms under this aspect. Nevertheless, they present different behaviours when looking at the number of required non-linear iterations. Starting from a full-order SIMPLE algorithm reaching convergence in the order of $\mathcal{O}(100)$ iterations, the same order is preserved at the reduced segregated level. In contrast, the coupled algorithm reaches convergence in the order of $\mathcal{O}(10)$ non-linear iterations, showing faster convergence with respect to the reduced SIMPLE case.
\section{A gometrically deformed cylinder problem}\label{cylinder}
\begin{figure}[!htb]
    \begin{subfigure}{0.21\textwidth}
        \centering
        \includegraphics[width=\linewidth]{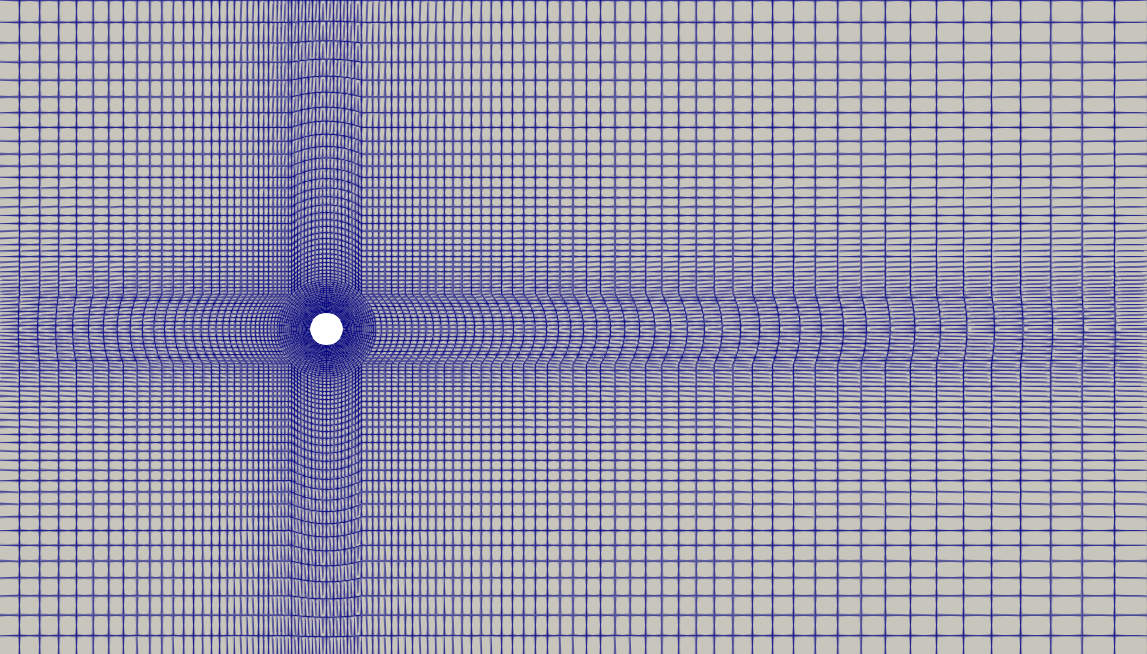}
    \end{subfigure}\hfill
    \begin{subfigure}{0.203\textwidth}
        \centering
        \includegraphics[width=\linewidth]{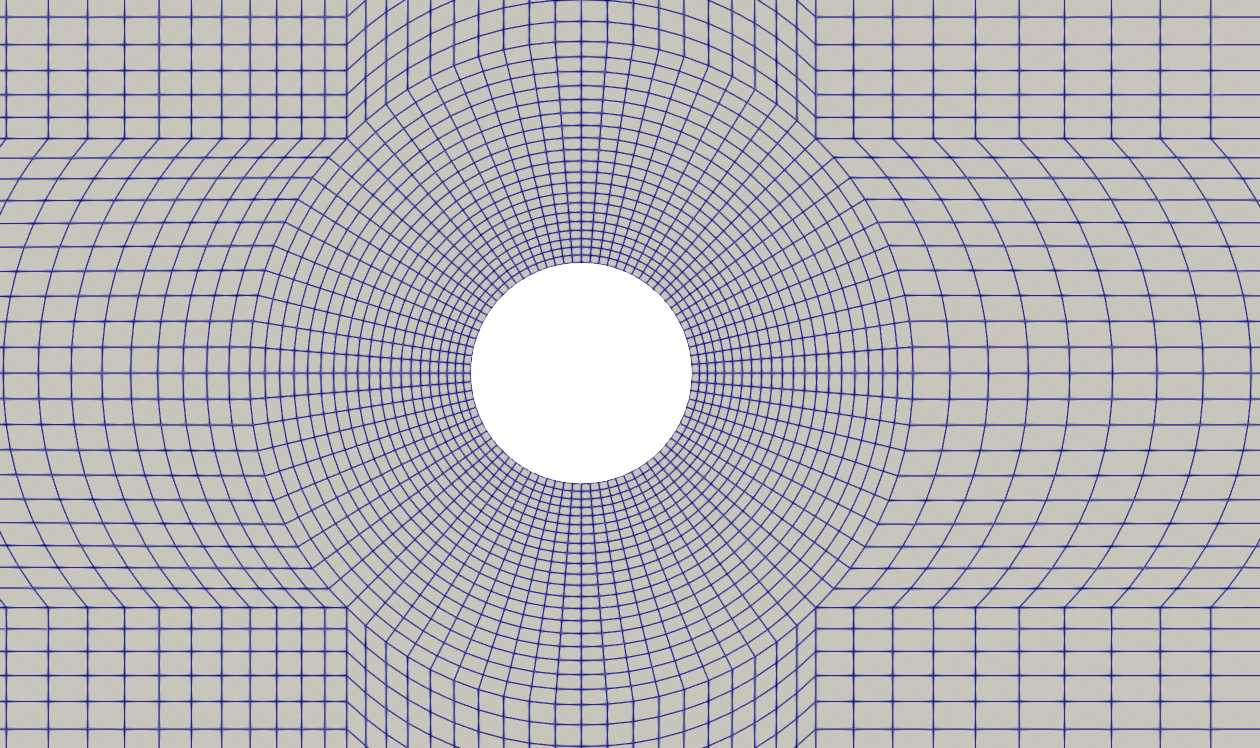}
    \end{subfigure}\hfill
    \begin{subfigure}{0.27\textwidth}
        \centering
        \includegraphics[width=\linewidth]{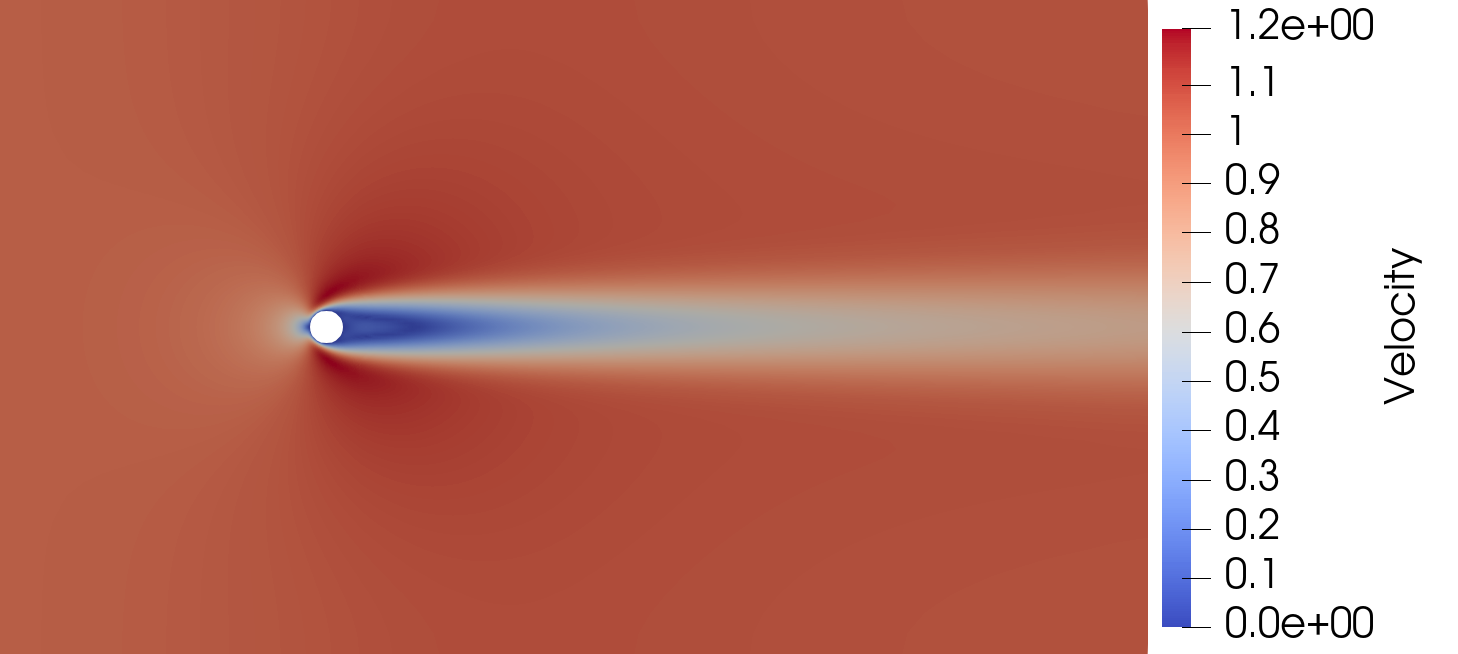}
    \end{subfigure}\hfill
    \begin{subfigure}{0.27\textwidth}
        \centering
        \includegraphics[width=\linewidth]{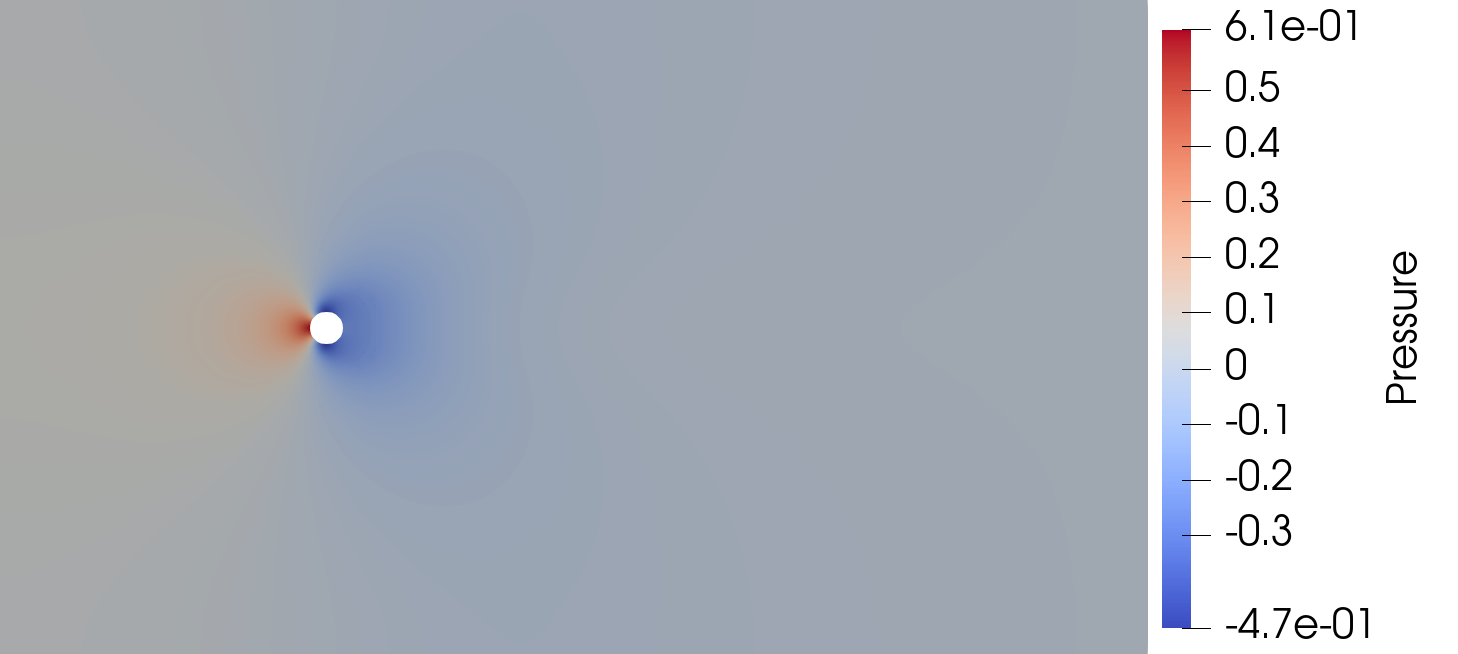}
    \end{subfigure}
    \caption{Cylinder reference case. From left to right: the discretizing mesh adopted for the finite volume method; zoom of the mesh around the obstacle; the full order velocity field; the full order pressure field.}
    \label{cylinder_reference}
\end{figure}

The second considered benchmark test case is the flow around a two-dimensional cylinder. The undeformed case consists of a circular cross-section cylinder of radius $r = 1$, immersed in a fluid flowing with inlet velocity $U = (1, 0, 0)$ and kinematic viscosity $\nu = 0.05$. These parameters yield a Reynolds number $Re = 40$, which is sufficiently small to ensure a steady-state phenomenon \cite{SEN_MITTAL_BISWAS_2009}. The computational domain, a two-dimensional grid of dimensions $70 \times 40$, is subdivided into 9200 computational cells. As shown in Figure \ref{cylinder_reference}, the mesh is refined closer to the obstacle.

A uniform velocity is prescribed at the inlet, the obstacle satisfies a no-slip condition, the upper and lower boundaries obeys a slip condition, whereas zero-gradient is imposed at the outlet. The pressure field is null at the outlet and satisfies a zero-gradient condition on the other boundaries.\\
Under these assumptions, full-order simulations are run employing the SIMPLE algorithm implemented in OpenFOAM \cite{OpenFOAM} and the resulting fields can be seen in Figure \ref{cylinder_reference}.

\begin{figure}[!htb]
    \begin{subfigure}{0.24\textwidth}
        \centering
        \includegraphics[width=\linewidth]{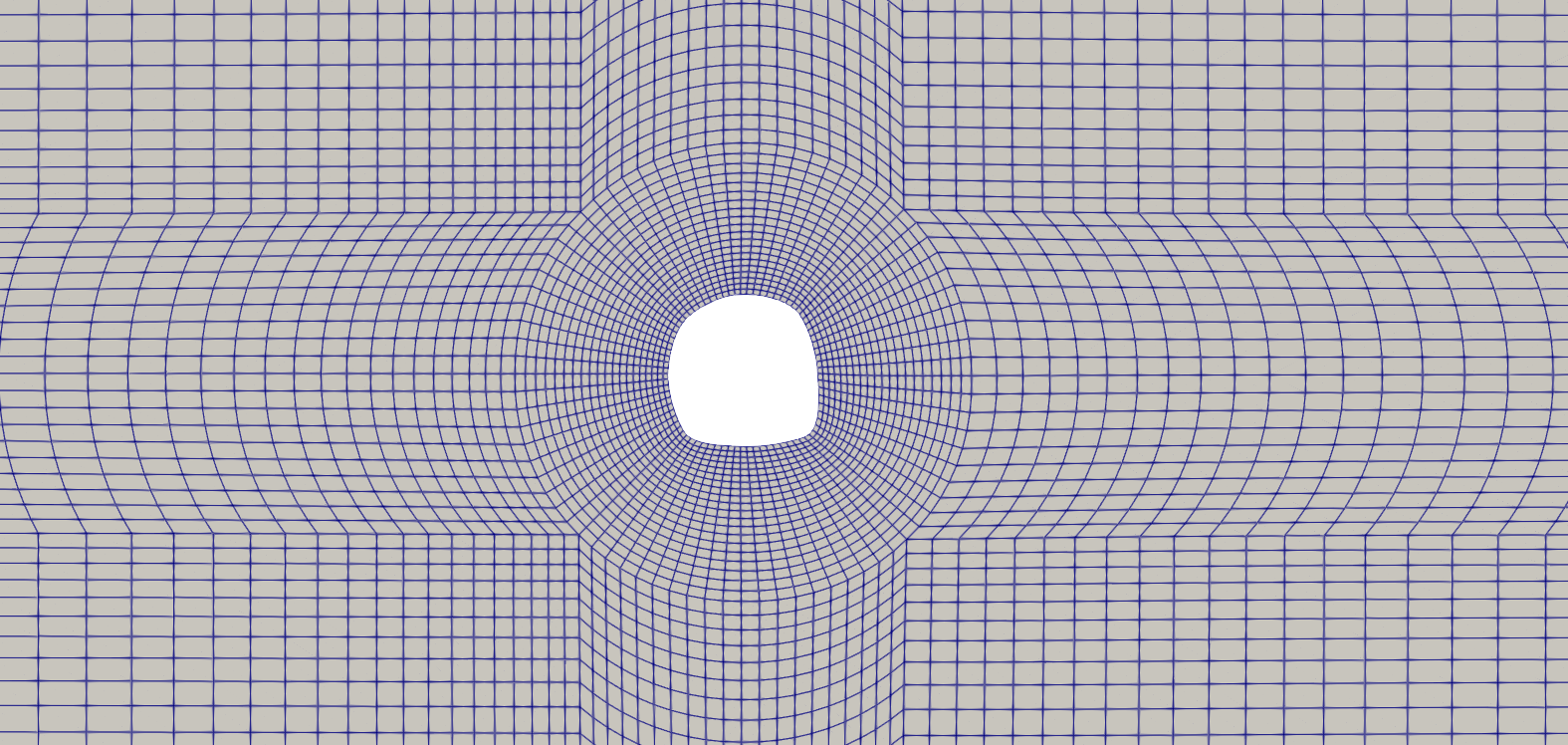}
    \end{subfigure}\hfill
    \begin{subfigure}{0.24\textwidth}
        \centering
        \includegraphics[width=\linewidth]{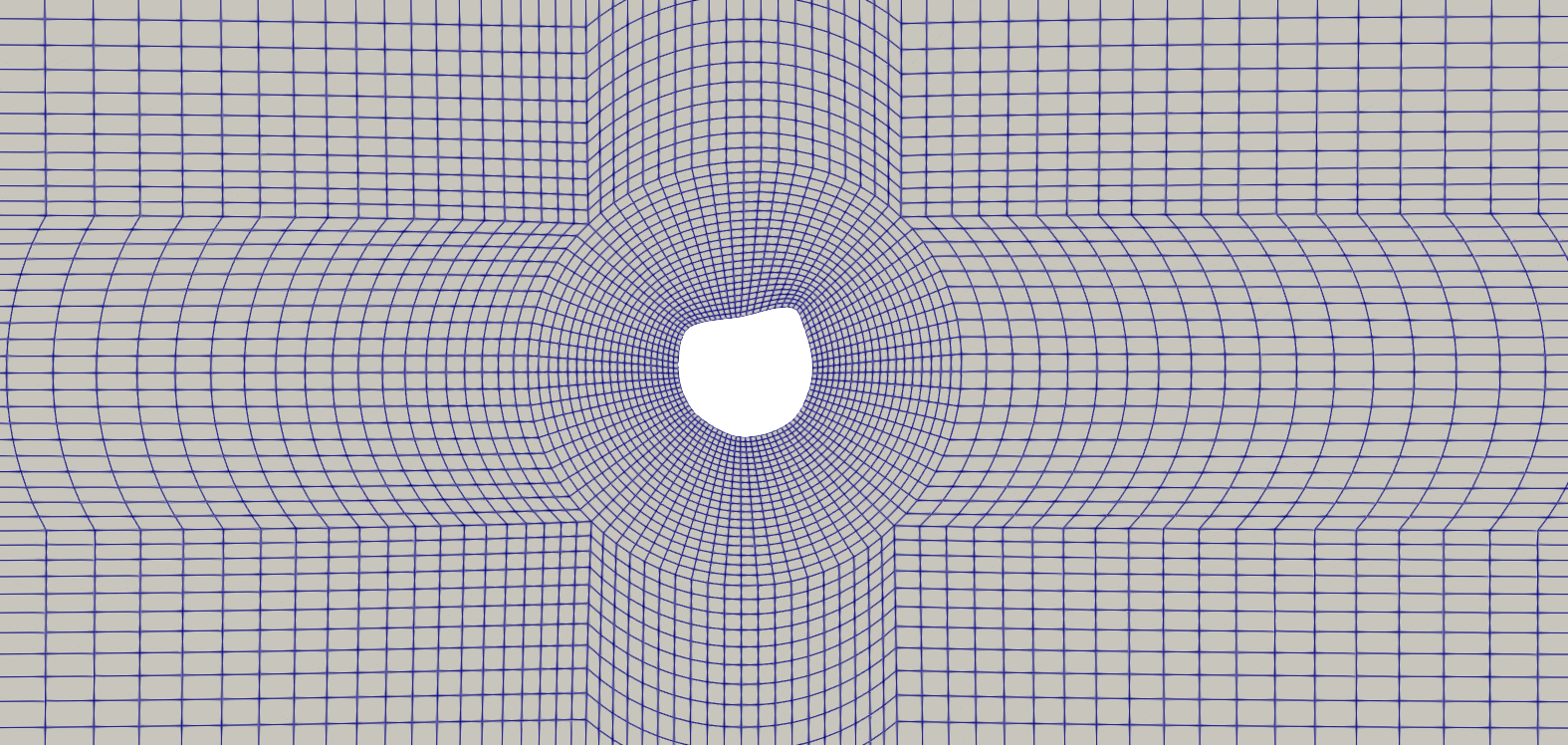}
    \end{subfigure}\hfill
    \begin{subfigure}{0.24\textwidth}
        \centering
        \includegraphics[width=\linewidth]{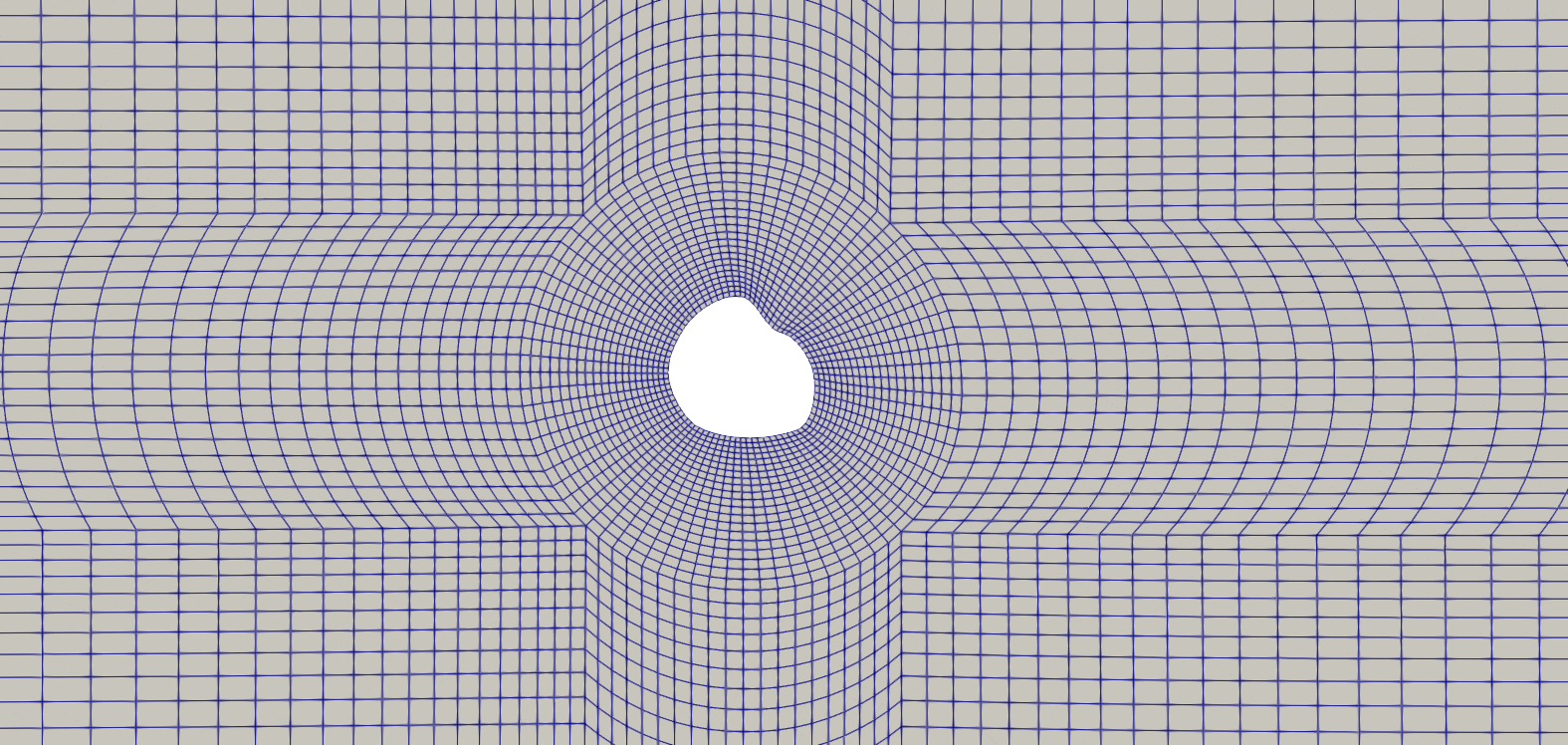}
    \end{subfigure}\hfill
    \begin{subfigure}{0.24\textwidth}
        \centering
        \includegraphics[width=\linewidth]{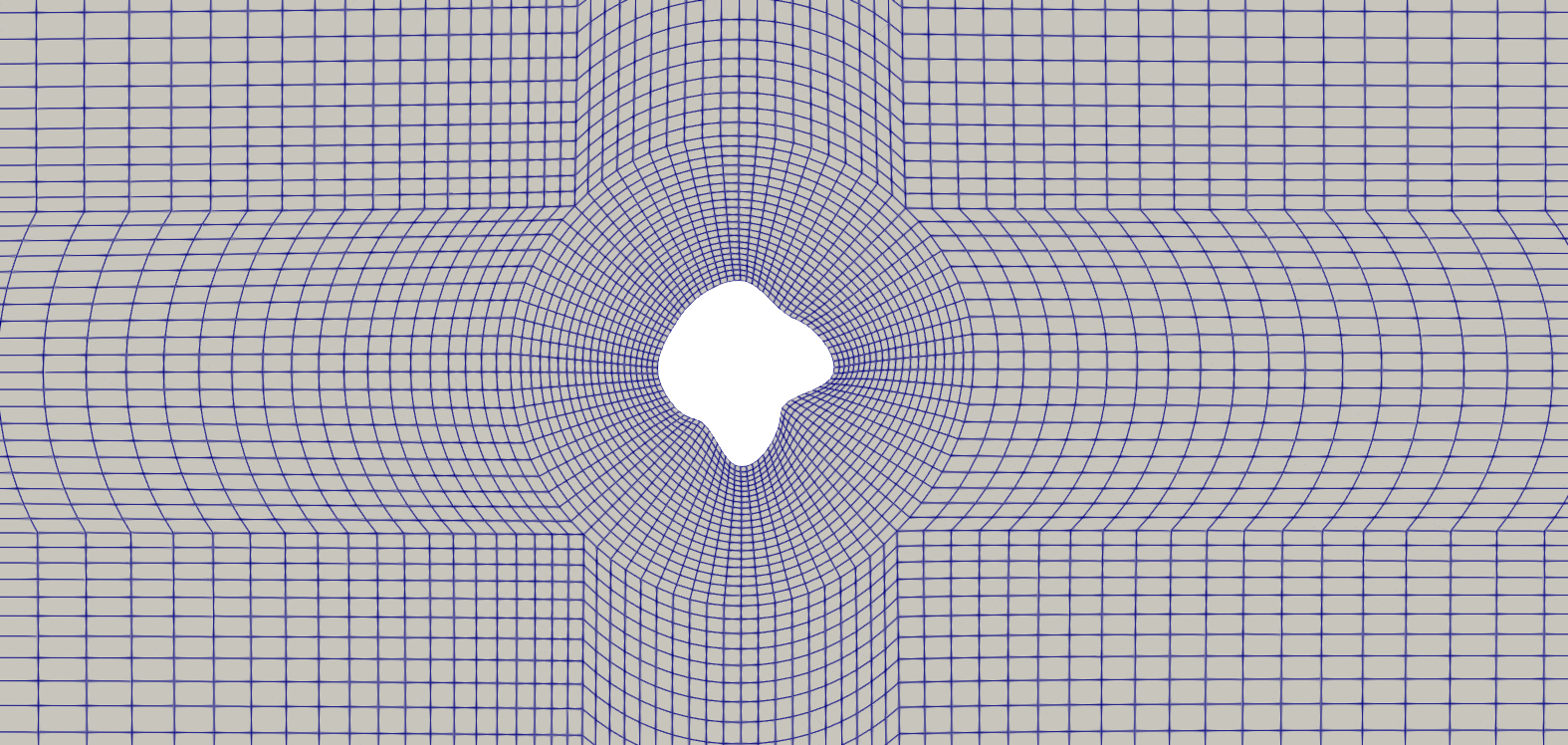}
    \end{subfigure}
    \caption{Examples of cylinder deformations with subsequent deformation of the mesh close to the obstacle.}
    \label{deformations' examples}
\end{figure}

Starting from the circular case, we generate deformed cylinders by randomly moving the 8 points of the circle intersecting the octants by random factors in $[0.75, 1.25]$. Subsequently, we link the new points through Bézier curves, following \cite{Viquerat_2020}. The subsequent mesh motion is handled via pyGeM \cite{TezzeleDemoMolaRozza2020PyGeM}, a Python library based on Radial Basis Functions (RBFs) interpolation, used to preserve the quality of the deformed mesh \cite{PhD_thesis,Bruno_2022}. For more details on the cylinder deformation and mesh motion strategies, we refer to Appendices \ref{Bezier} and \ref{RBF}, respectively. Figure \ref{deformations' examples} shows some deformations considered in the present analysis, along with the corresponding computational grids.
\subsection{Offline stage}
Full-order solution for 100 deformations and their symmetric with respect to the $x$-axis serve as snapshots for the POD. The reason for doubling the snapshots with this strategy is twofold: i) the dataset size is doubled without additional computational cost; ii) the resulting modes are either symmetric or with opposite sign with respect to the $x$-axis, making them not biased by the specific random deformations.

\begin{figure}[!htb]
    \begin{subfigure}{0.24\textwidth}
        \centering
        \includegraphics[width=\linewidth]{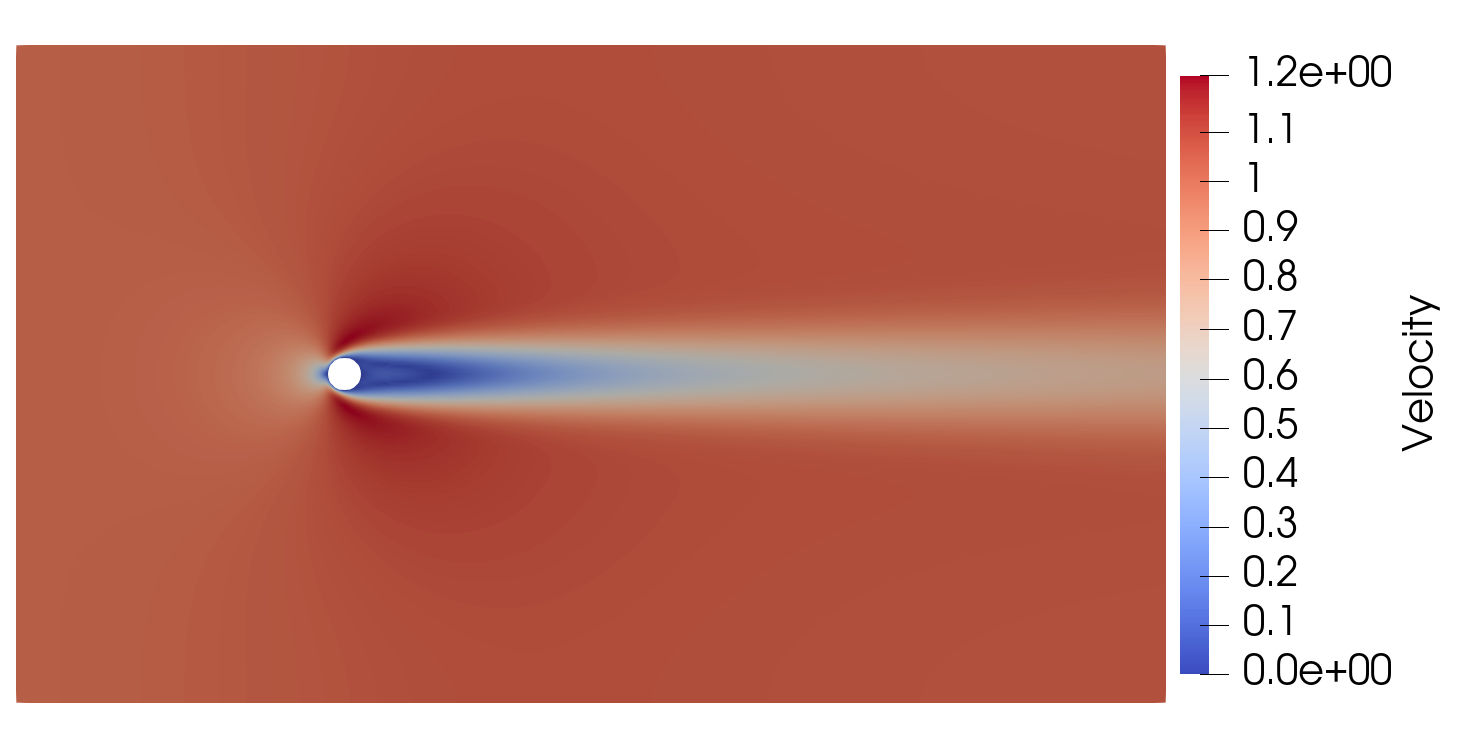}
    \end{subfigure}\hfill
    \begin{subfigure}{0.24\textwidth}
        \centering
        \includegraphics[width=\linewidth]{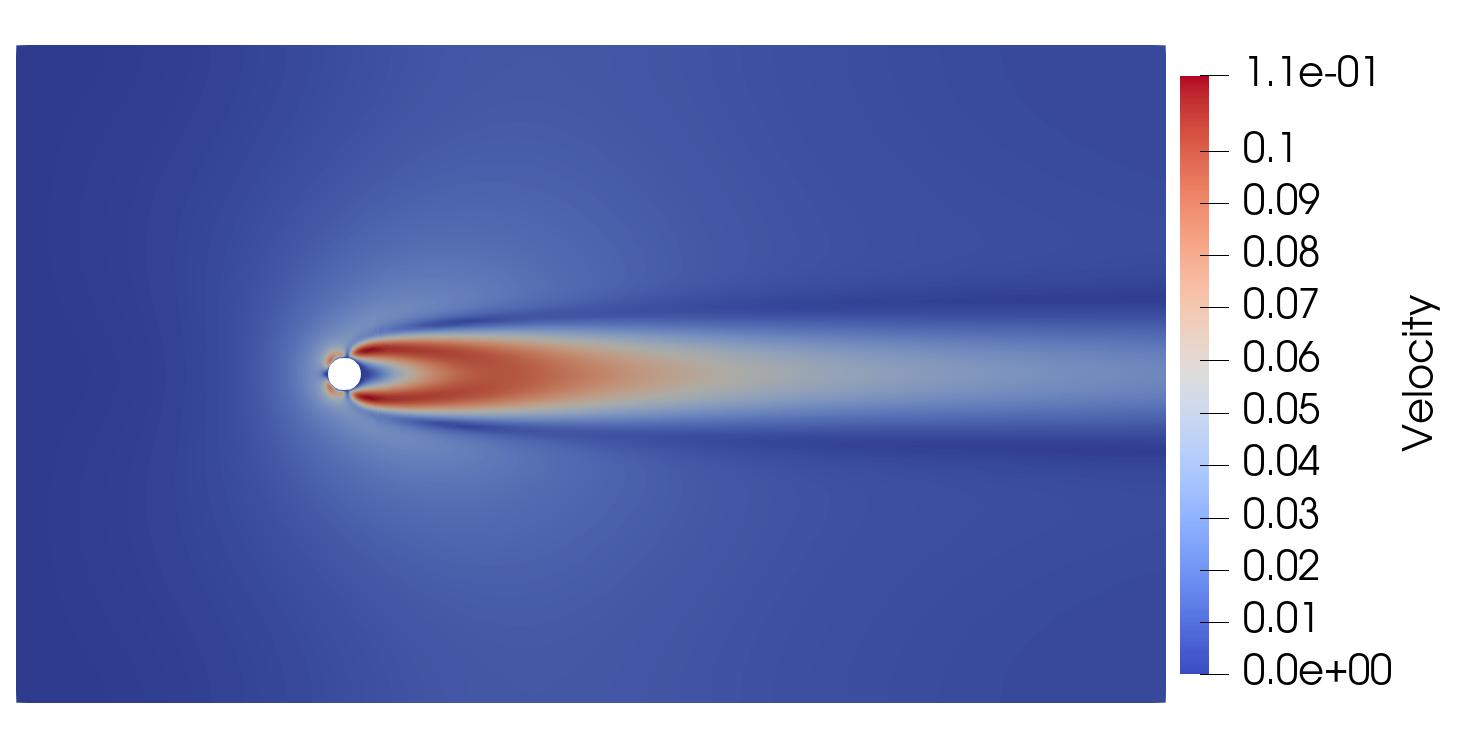}
    \end{subfigure}\hfill
    \begin{subfigure}{0.24\textwidth}
        \centering
        \includegraphics[width=\linewidth]{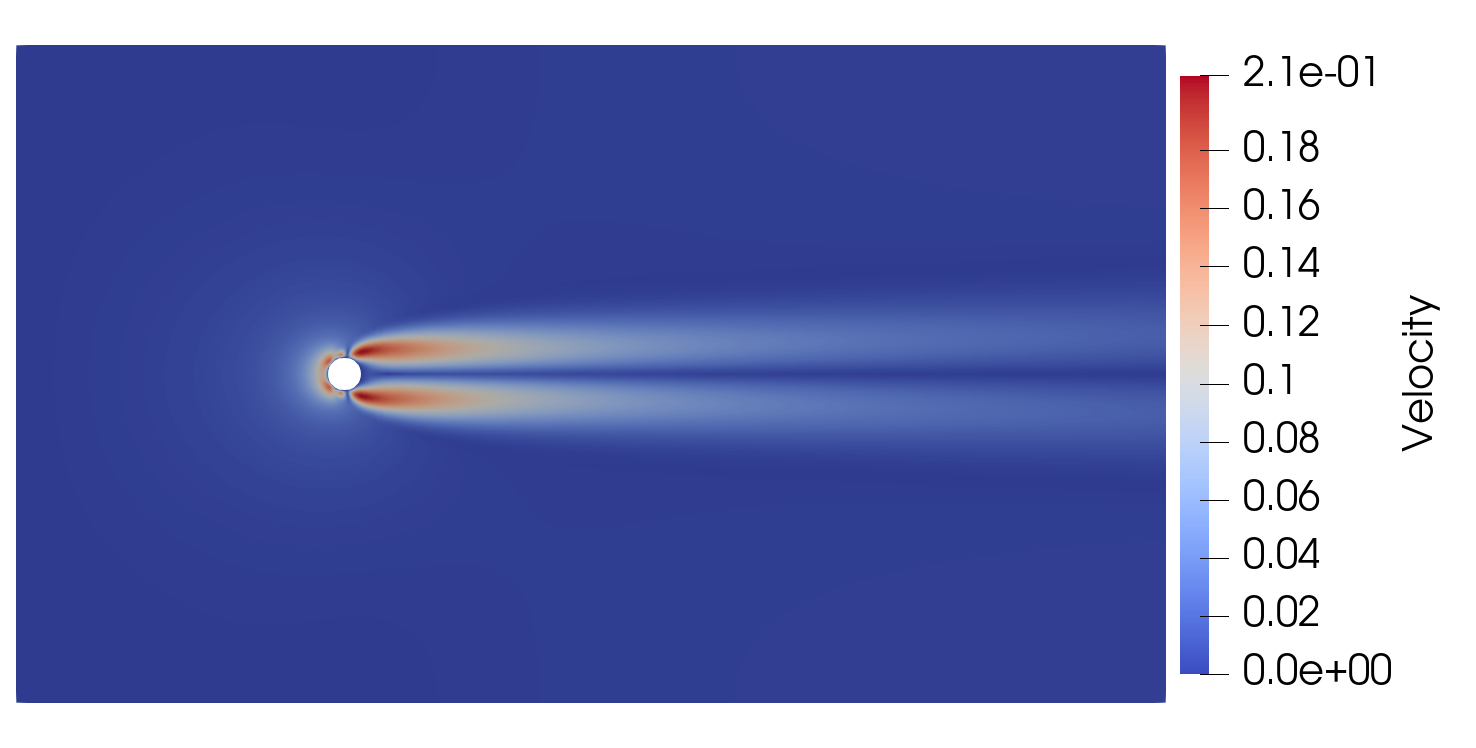}
    \end{subfigure}
    \begin{subfigure}{0.24\textwidth}
        \centering
        \includegraphics[width=\linewidth]{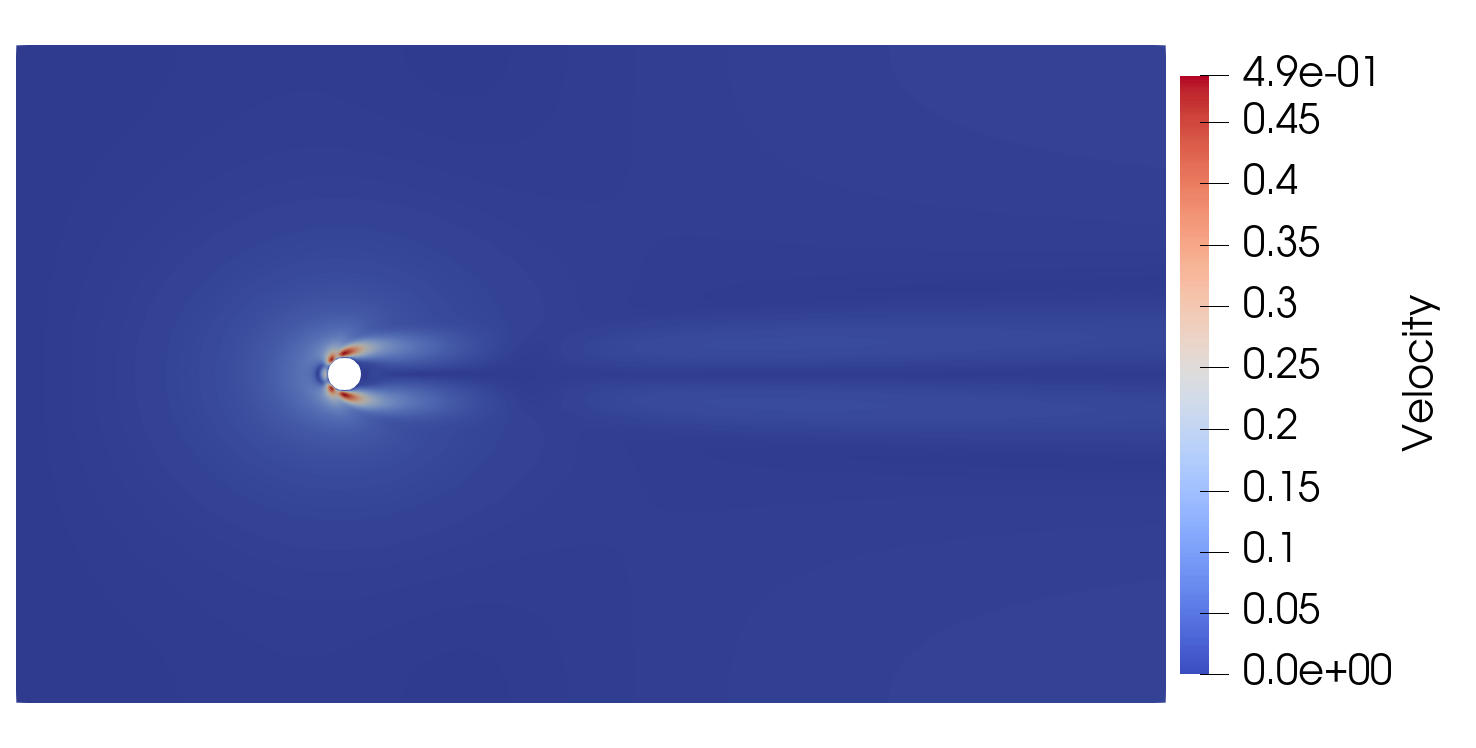}
    \end{subfigure}
    \medskip
    \begin{subfigure}{0.24\textwidth}
        \centering
        \includegraphics[width=\linewidth]{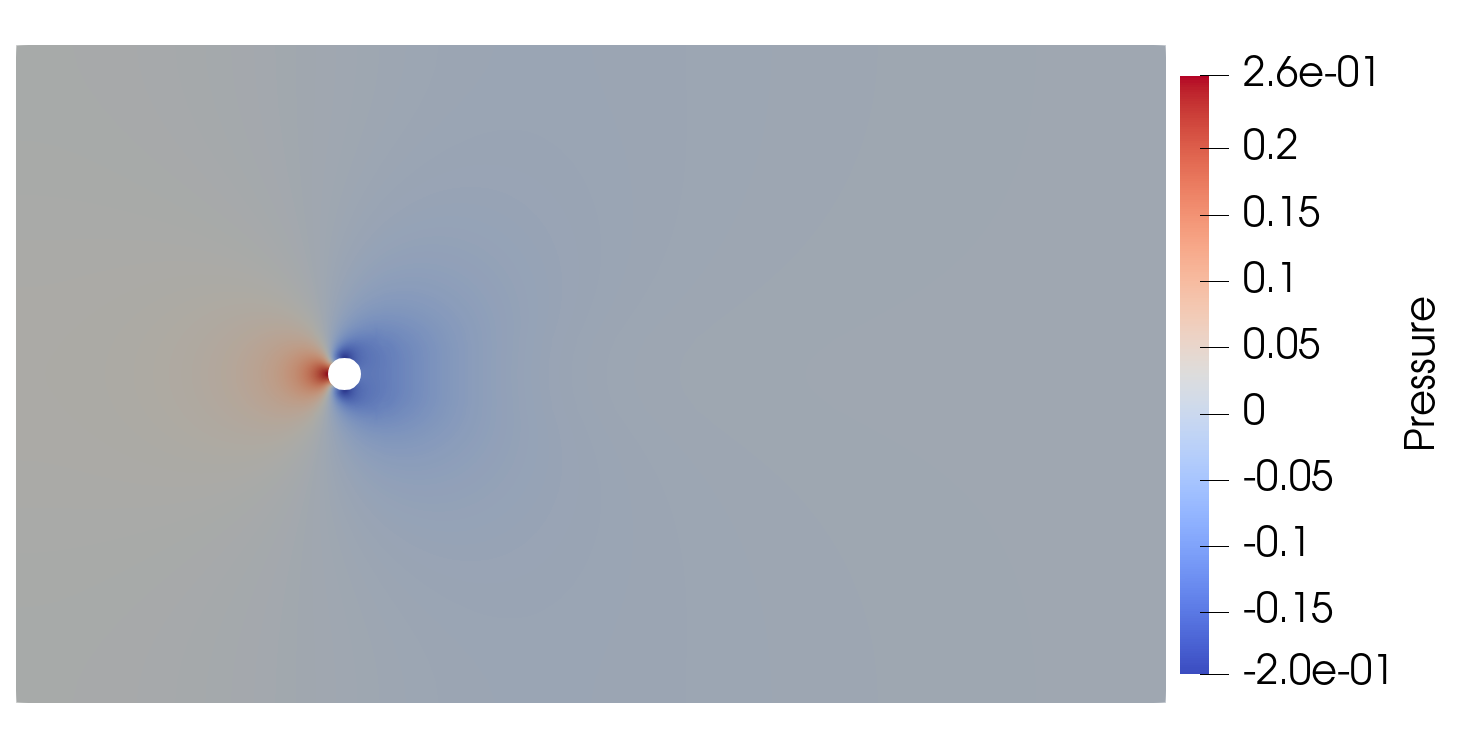}
    \end{subfigure}\hfill
    \begin{subfigure}{0.24\textwidth}
        \centering
        \includegraphics[width=\linewidth]{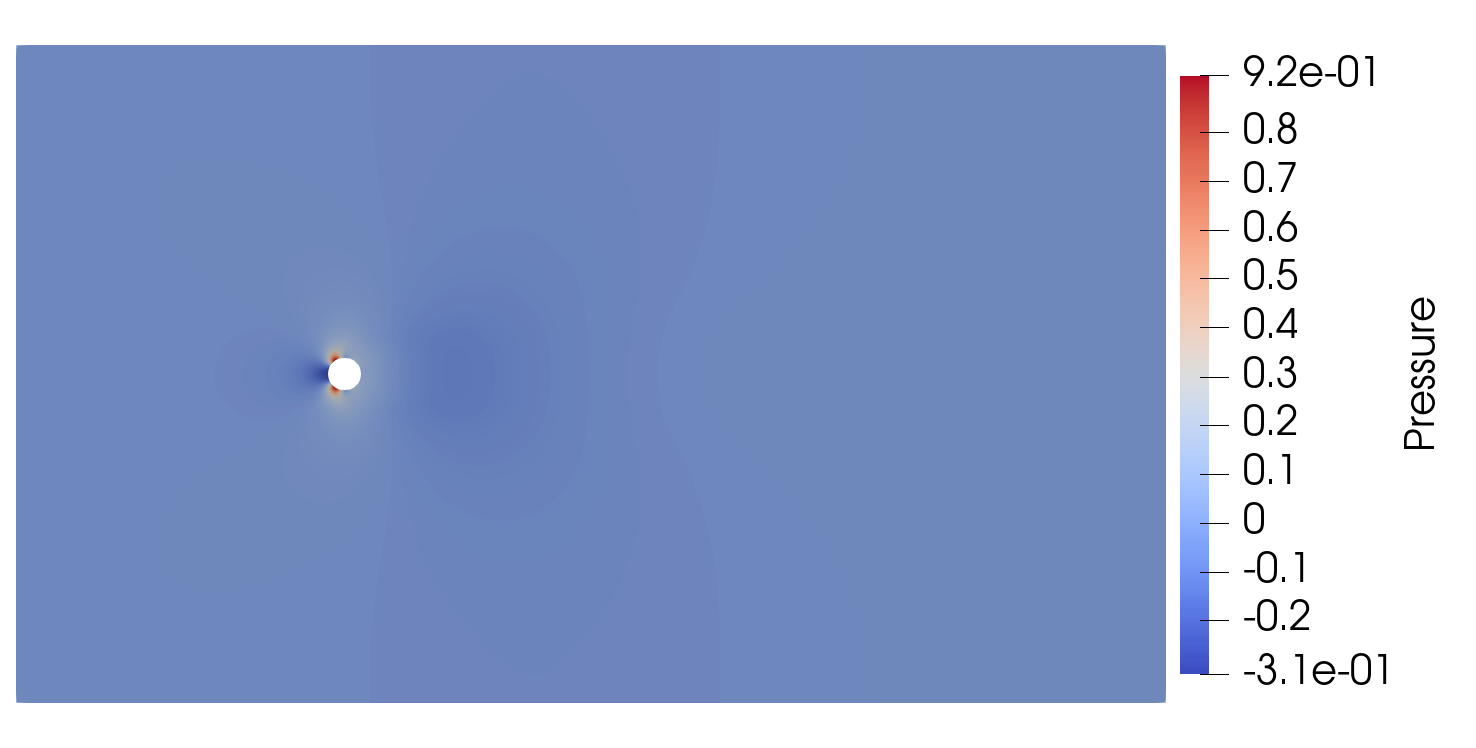}
    \end{subfigure}\hfill
    \begin{subfigure}{0.24\textwidth}
        \centering
        \includegraphics[width=\linewidth]{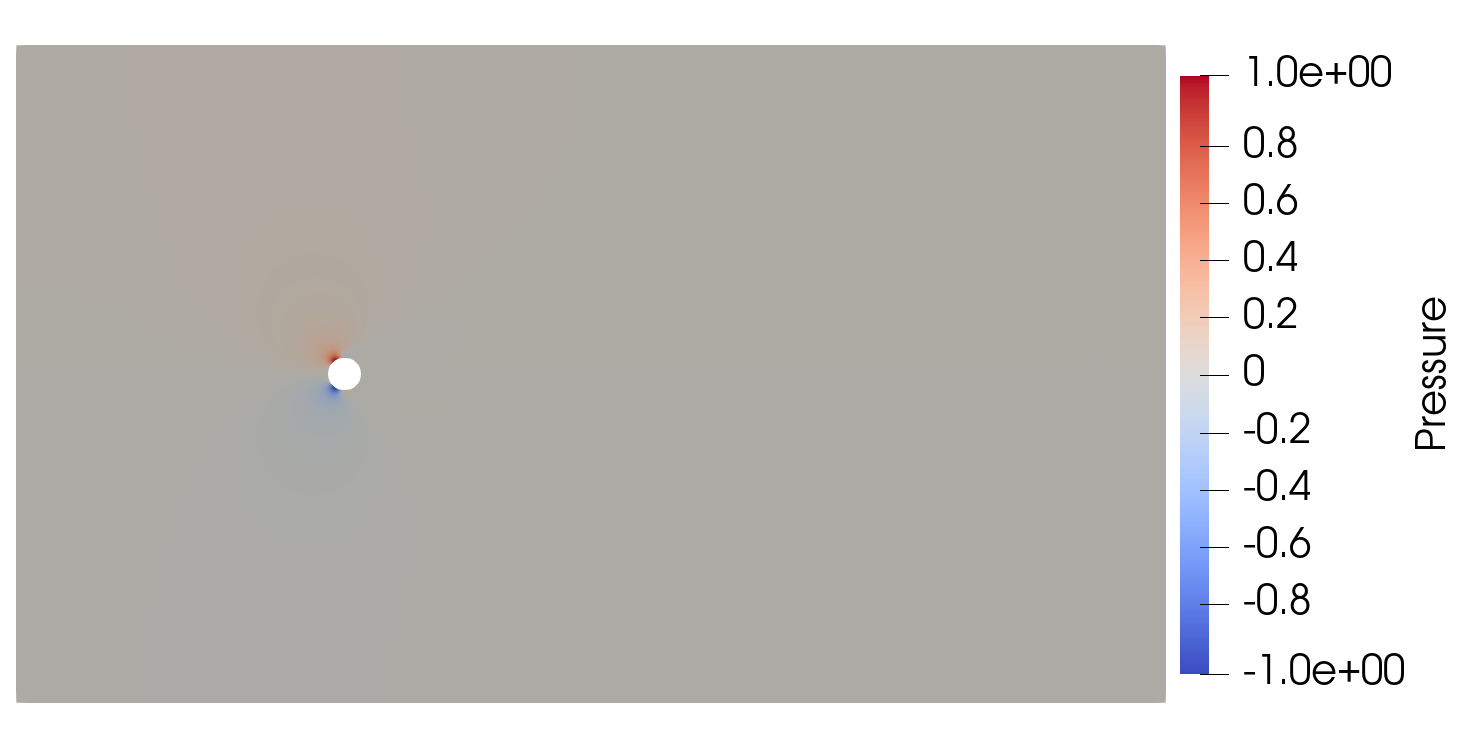}
    \end{subfigure}
    \begin{subfigure}{0.24\textwidth}
        \centering
        \includegraphics[width=\linewidth]{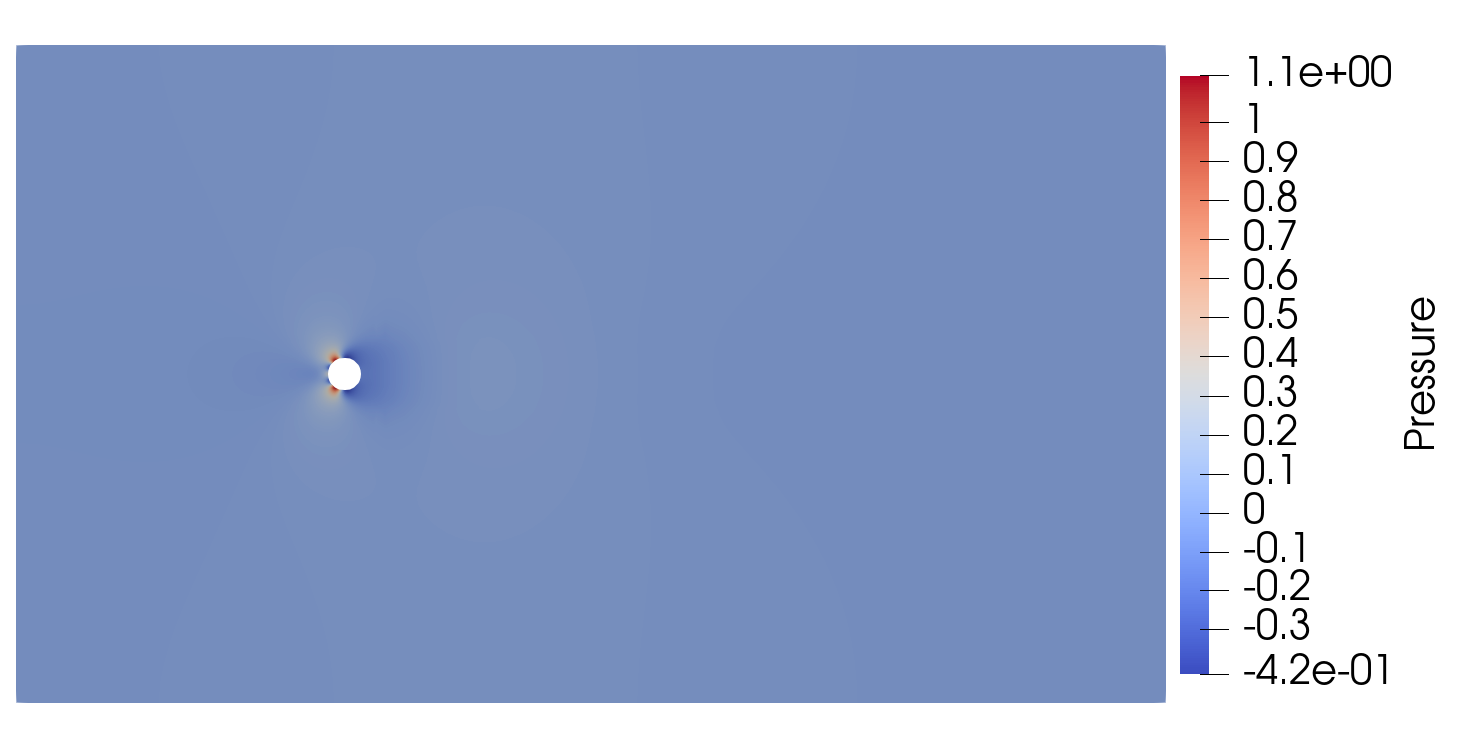}
    \end{subfigure}
    \caption{First four velocity (upper row) and pressure (lower row) modes in the cylinder case. Remark that for velocity we adopted a lift mode (the first on the left), coinciding with the solution of the reference case.}
    \label{POD cylinder}
\end{figure}

Being $N_h = 9200$ the number of computational cells in the FV discretization, we define the snapshots matrices $S_p \in \mathbb{R}^{N_h \times n_s}$, $S_{\boldsymbol{u}} \in \mathbb{R}^{3N_h \times n_s}$ for pressure and velocity, respectively.\\
$S_p$ and $S_{\boldsymbol{u}}$ are processed through ITHACA-FV \footnote{https://ithaca-fv.github.io/ITHACA-FV/} (In real Time Highly Advanced Computational Applications for Finite Volumes) \cite{Stabile2017CAIM, stabile2018finite} to compute the reduced order basis with respect to the $L^2(\Omega)$ scalar product in the undeformed geometry.\\
Figure \ref{POD cylinder} shows the first velocity and pressure modes in the circular obstacle configuration. As expected, the lowest-order modes hold most of the information content coming from the collected snapshots. Conversely, increasing the number of modes introduces physically less relevant fields.\\
In the velocity case, we adopt the "lift function" method exposed in \ref{coupled model}. Consequently, the first mode in Figure \ref{POD cylinder} corresponds to the solution in the undeformed case itself, which serves as the lift function.

\begin{figure}[!htb]
    \begin{minipage}{0.48\textwidth}
        \centering
        \includegraphics[width=\linewidth]{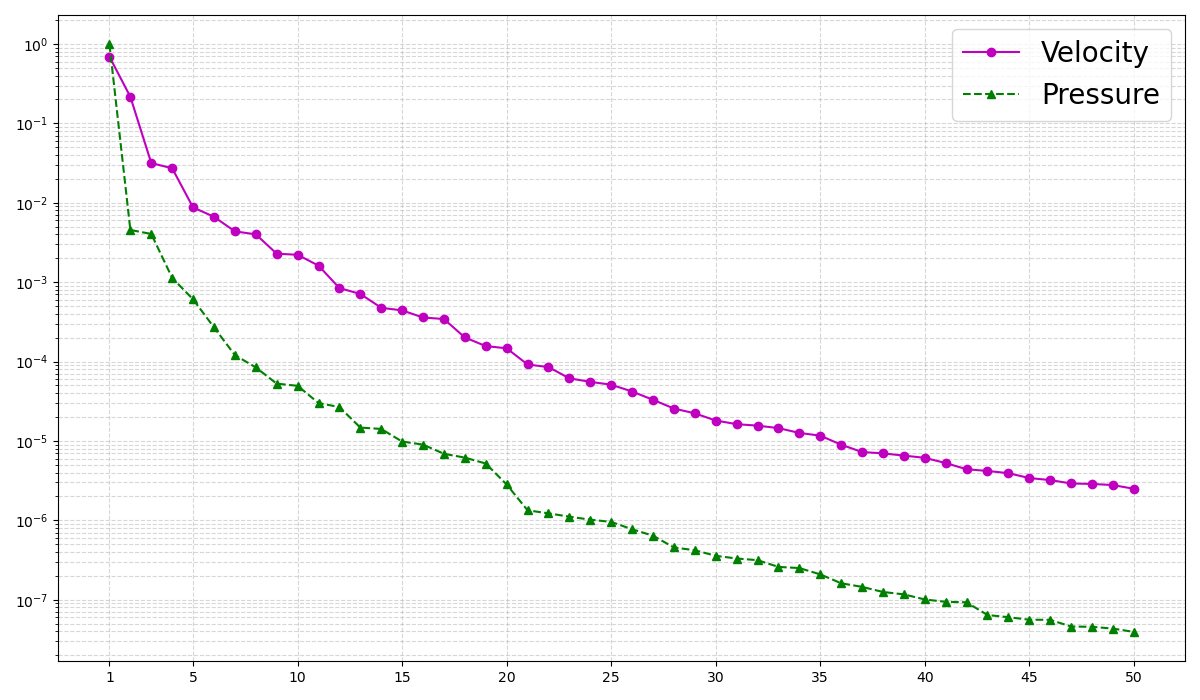}
        \caption{Decay in magnitude of the correlation matrices' eigenvalues, for both the velocity and pressure fields.}
        \label{eigenvalues_decay}
    \end{minipage}\hfill
    \begin{minipage}{0.48\textwidth}
    \centering
    \renewcommand{\arraystretch}{1.2}
    \begin{tabular}{|c|c|c|}
        \hline
        $N_{\boldsymbol{u}} = N_p$ & $I_{\boldsymbol{u}}(N_{\boldsymbol{u}})$ & $I_p(N_p)$\\
        \hline
        \hline
        1 & 0.690916 & 0.988951 \\
        \hline
        2 & 0.906952 & 0.993495 \\
        \hline
        $\dots$ & $\dots$ & $\dots$ \\
        \hline
        11 & 0.995669 & 0.999893 \\
        \hline
        12 & 0.996512 & 0.999920 \\
        \hline
        $\dots$ & $\dots$ & $\dots$ \\
        \hline
        34 & 0.999892 & 0.999998 \\
        \hline
        35 & 0.999904 & 0.999998 \\
        \hline
        $\dots$ & $\dots$ & $\dots$ \\
        \hline
    \end{tabular}
    \captionof{table}{Saturation of the relative information content due to the increasing of the considered number of modes.}
    \label{relative_information}
    \end{minipage}
    \label{eigenvalues_infos}
\end{figure}

The spectral analysis reported in Figure \ref{eigenvalues_decay} and Table \ref{relative_information} highlights that the number of eigenvalues needed to attain the 99.99\% of the information content for pressure and velocity is $N_{\boldsymbol{u}} = 35$ and $N_p = 12$ respectively.
\subsection{Online stage}
\begin{figure}[!htb]
    \begin{subfigure}{0.48\textwidth}
        \centering
        \includegraphics[width=\linewidth]{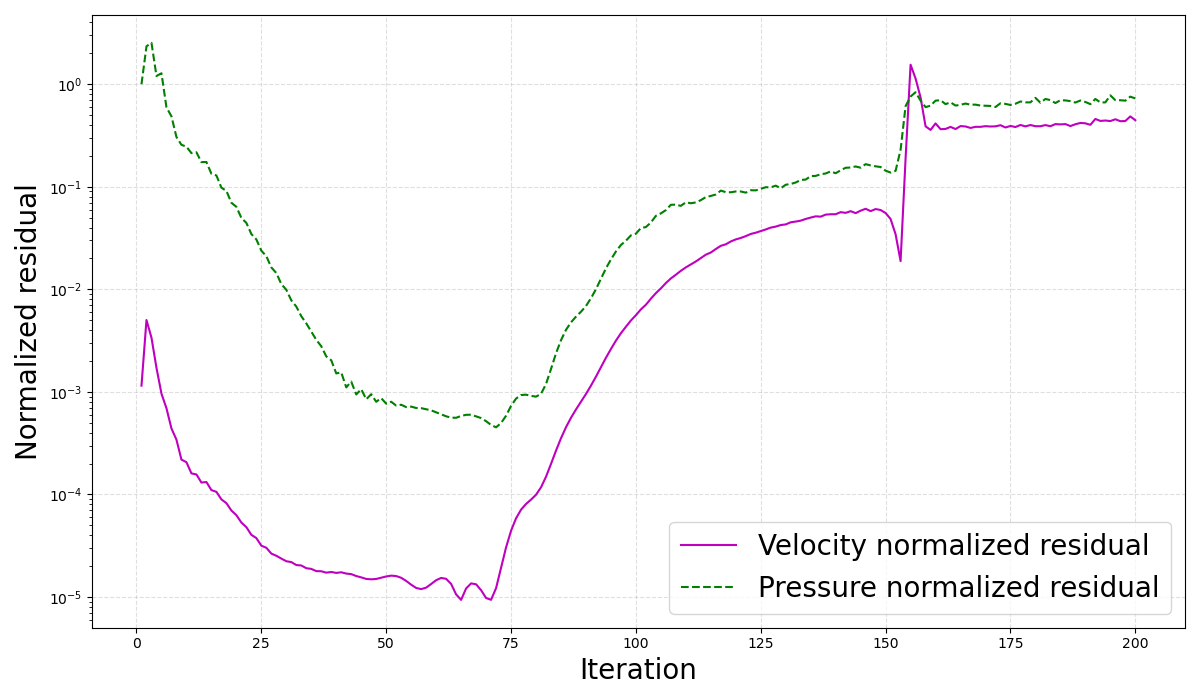}
        \subcaption{$\alpha_{\boldsymbol{u}} = 0.7$, $\alpha_p = 0.3$}
        \label{cylinder_0.7-0.3-residuals}
    \end{subfigure}\hfill
    \begin{subfigure}{0.48\textwidth}
        \centering
        \includegraphics[width=\linewidth]{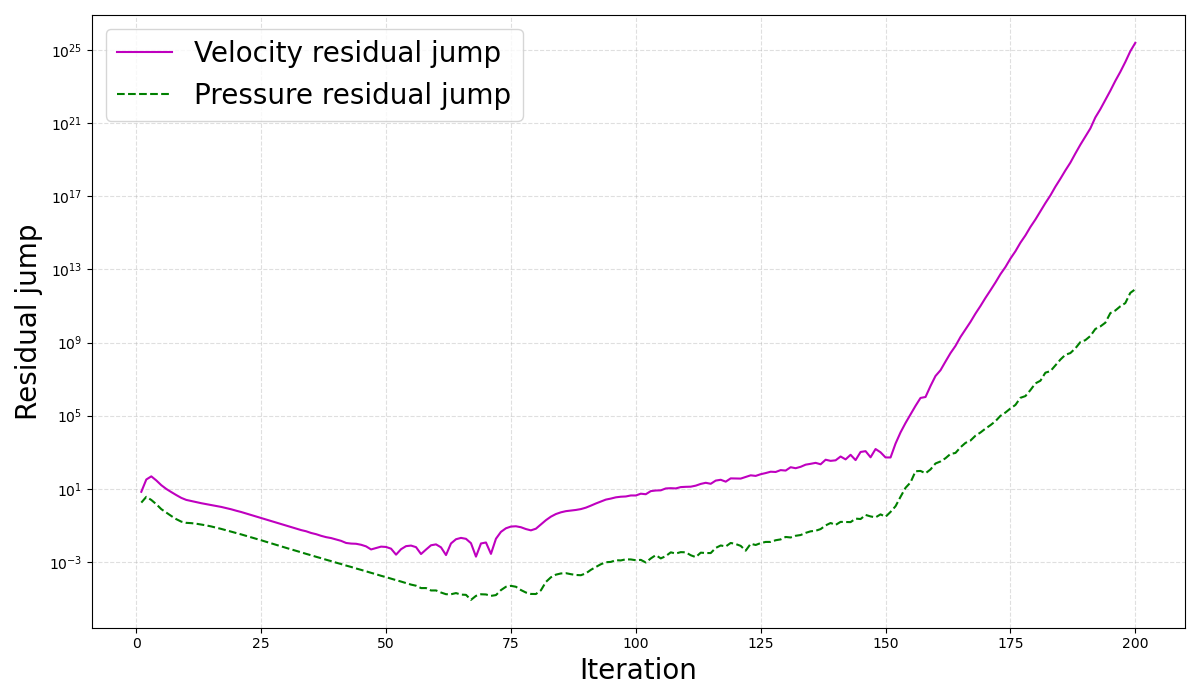}
        \subcaption{$\alpha_{\boldsymbol{u}} = 0.7$, $\alpha_p = 0.3$}
        \label{cylinder_0.7-0.3-jumps}
    \end{subfigure}
    \medskip
    \begin{subfigure}{0.48\textwidth}
        \centering
        \includegraphics[width=\linewidth]{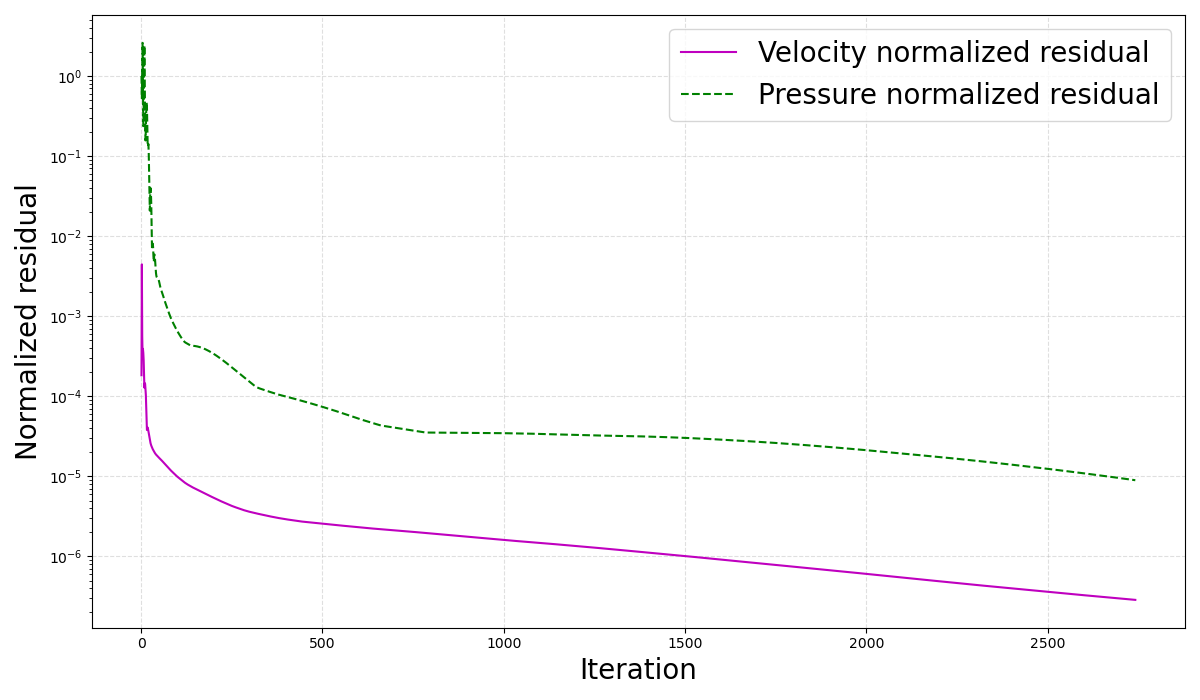}
        \subcaption{$\alpha_{\boldsymbol{u}} = 0.15$, $\alpha_p = 0.3$}
        \label{cylinder_0.15-0.3-residuals}
    \end{subfigure} \hfill
    \begin{subfigure}{0.48\textwidth}
        \centering
        \includegraphics[width=\linewidth]{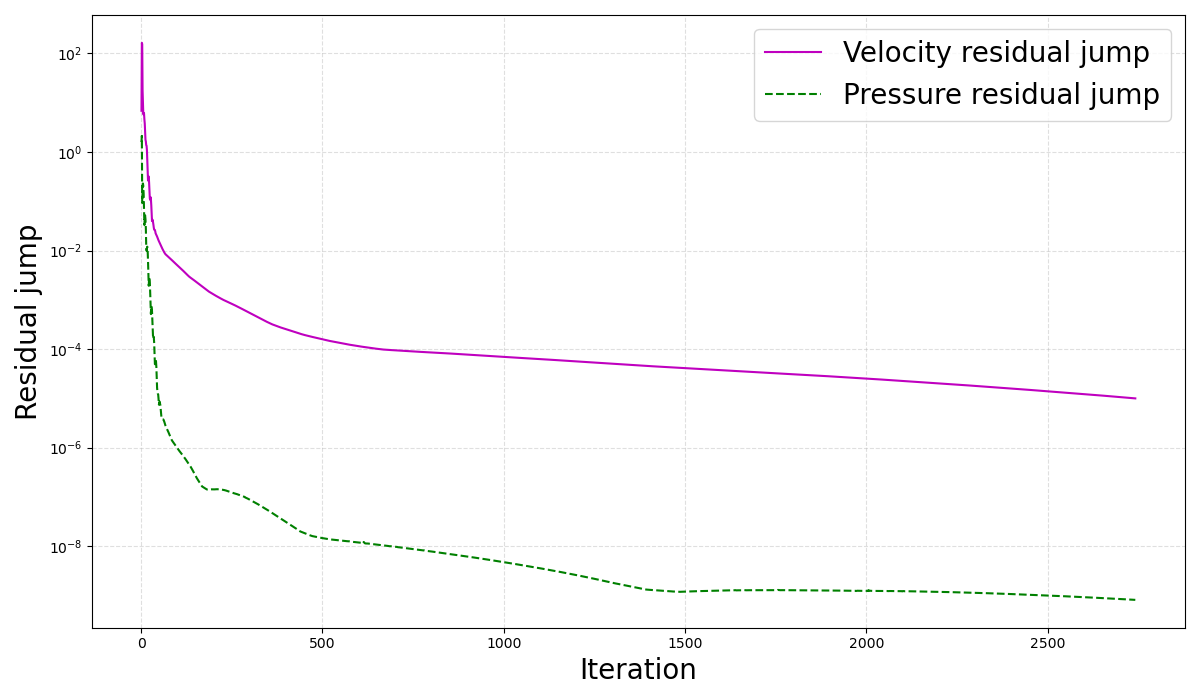}
        \subcaption{$\alpha_{\boldsymbol{u}} = 0.15$, $\alpha_p = 0.3$}
        \label{cylinder_0.15-0.3-jumps}
    \end{subfigure}
    \caption{Under relaxation coefficients analysis for the reference case: Figures \ref{cylinder_0.7-0.3-residuals} and \ref{cylinder_0.7-0.3-jumps} refer to coefficients equal to 0.7 and 0.3 for velocity and pressure respectively, whereas Figures \ref{cylinder_0.15-0.3-residuals} and \ref{cylinder_0.15-0.3-jumps} stand for coefficients equal to 0.15 and 0.3. Simulations have been carried out with $N_{\boldsymbol{u}} = 35$ and $N_p = 12$.}
    \label{cylinder-under_relax_coeff}
\end{figure}

Both coupled and SIMPLE ROM methods are considered for the online stage.\\
The SIMPLE algorithm is subject to numerical instabilities arising from the choice of the under-relaxation coefficients. Too large coefficients may lead to divergent schemes, whereas a too conservative choice could result in an excessively slow convergence.\\
Figure \ref{cylinder-under_relax_coeff} shows how this choice affects the convergence rate. With respect to the lid-driven cavity, the complexity of the phenomenon under investigation introduces higher numerical instabilities, making mimicking the full-order scheme impossible. These criticalities have to be mitigated through lower under-relaxation coefficients, ensuring scheme convergence, at the cost of compromising efficiency. Testing different configurations on the cylindrical reference case, the best option proves to be $\alpha_{\boldsymbol{u}} = 0.15$ and $\alpha_p = 0.3$.\\
After a trial-and-error analysis, contrary to \cite{stabile2018finite}, we observed that the inclusion of intermediate FOM snapshots did not guarantee ROM convergence when using the same under-relaxation factors. Thus, we did not include them, because not significant to attain convergence of the SIMPLE ROM. This choice speed-ups and reduces memory needs for the offline. Additionally, we expect to extract more efficient modes, not affected by non-physical snapshots, from the POD procedure.

\begin{figure}[!htb]
    \begin{subfigure}{0.48\textwidth}
        \centering
        \includegraphics[width=\linewidth]{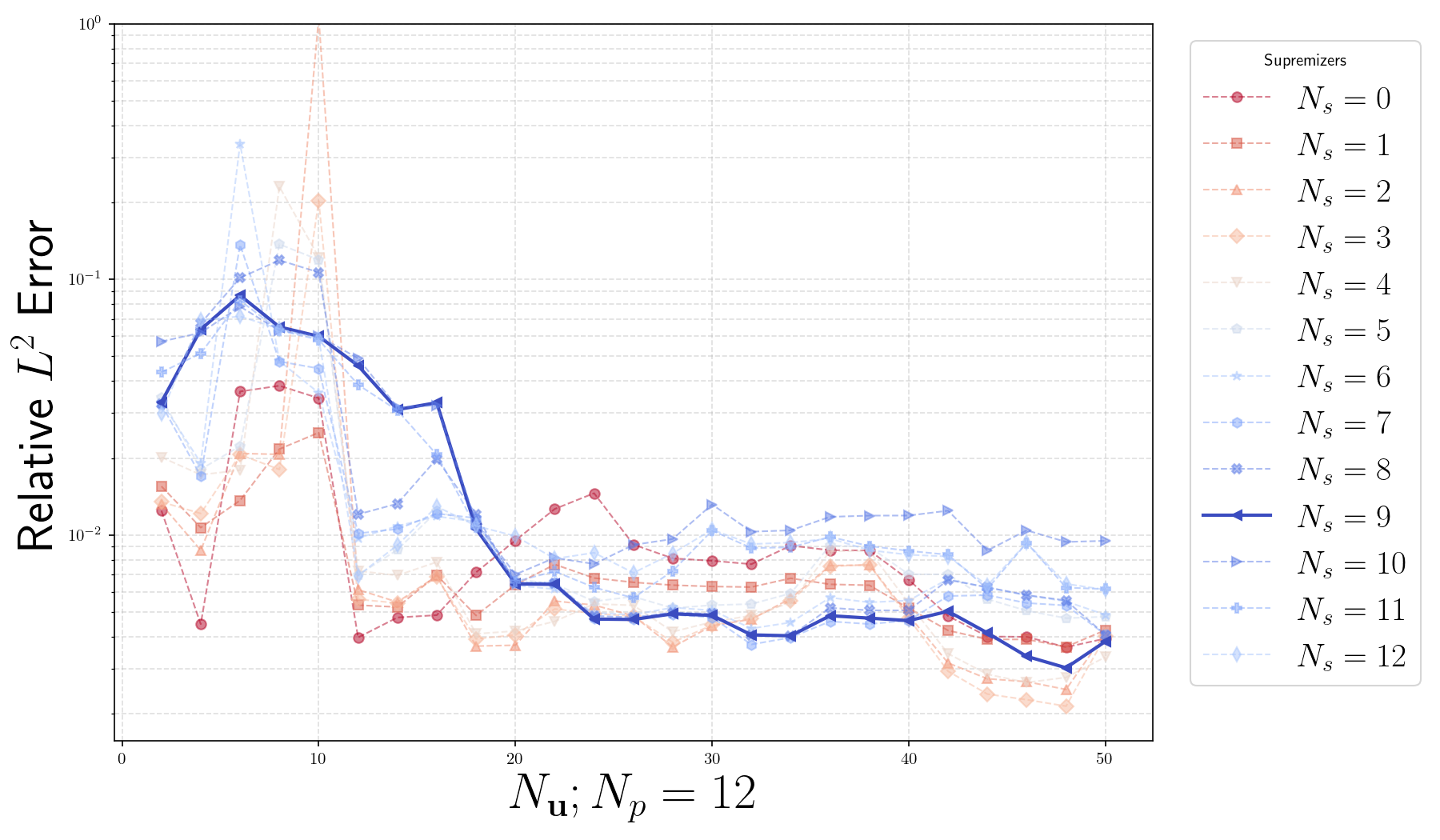}
        \caption{Relative velocity error for $N_p = 12$.}
        \label{cylinder - sup_fixed_modes - velocity}
    \end{subfigure}\hfill
    \begin{subfigure}{0.48\textwidth}
        \centering
        \includegraphics[width=\linewidth]{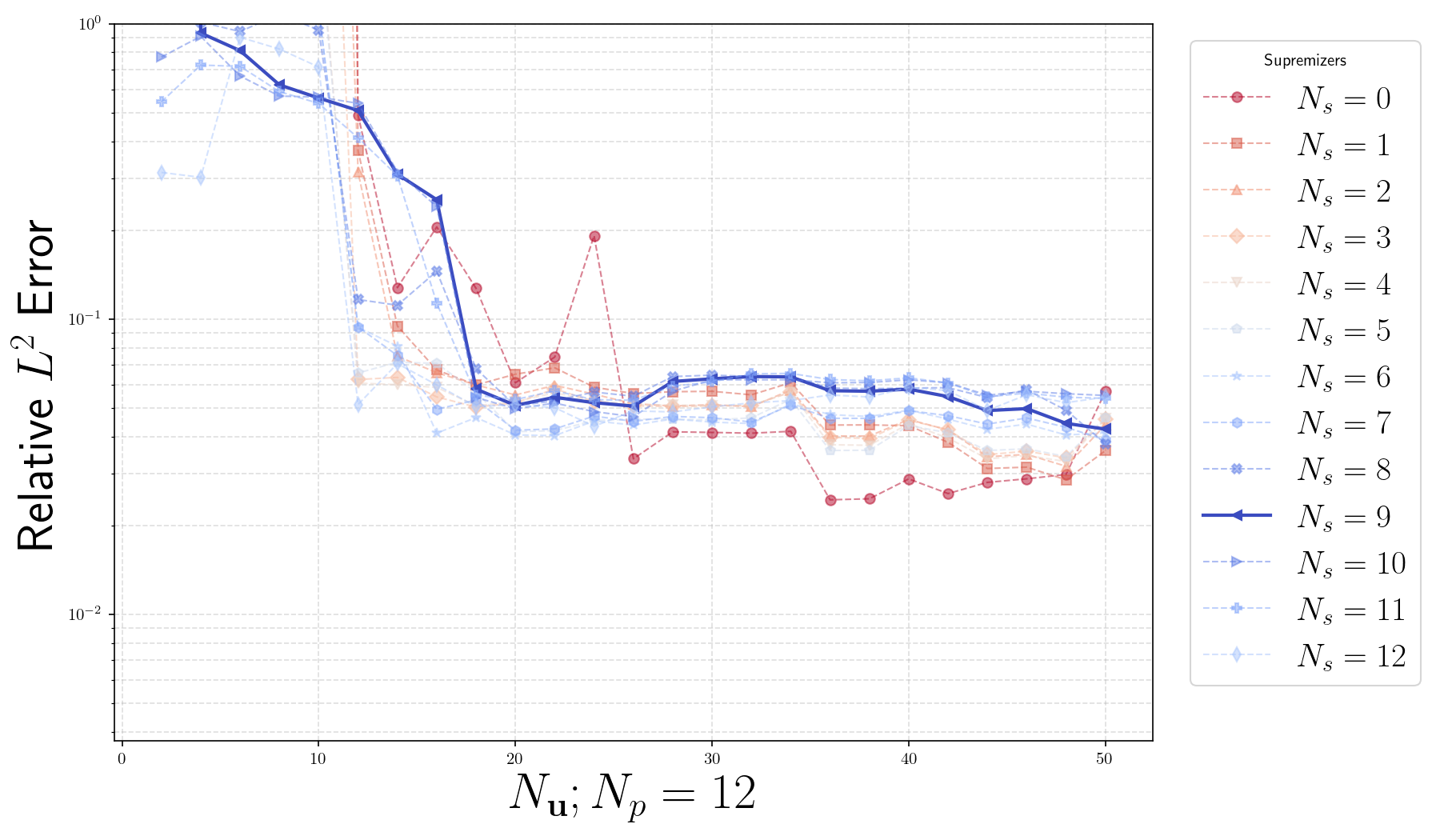}
        \caption{Relative pressure error for $N_p = 12$.}
        \label{cylinder - sup_fixed_modes - pressure}
    \end{subfigure}
    \medskip
    \begin{subfigure}{0.48\textwidth}
        \centering
        \includegraphics[width=\linewidth]{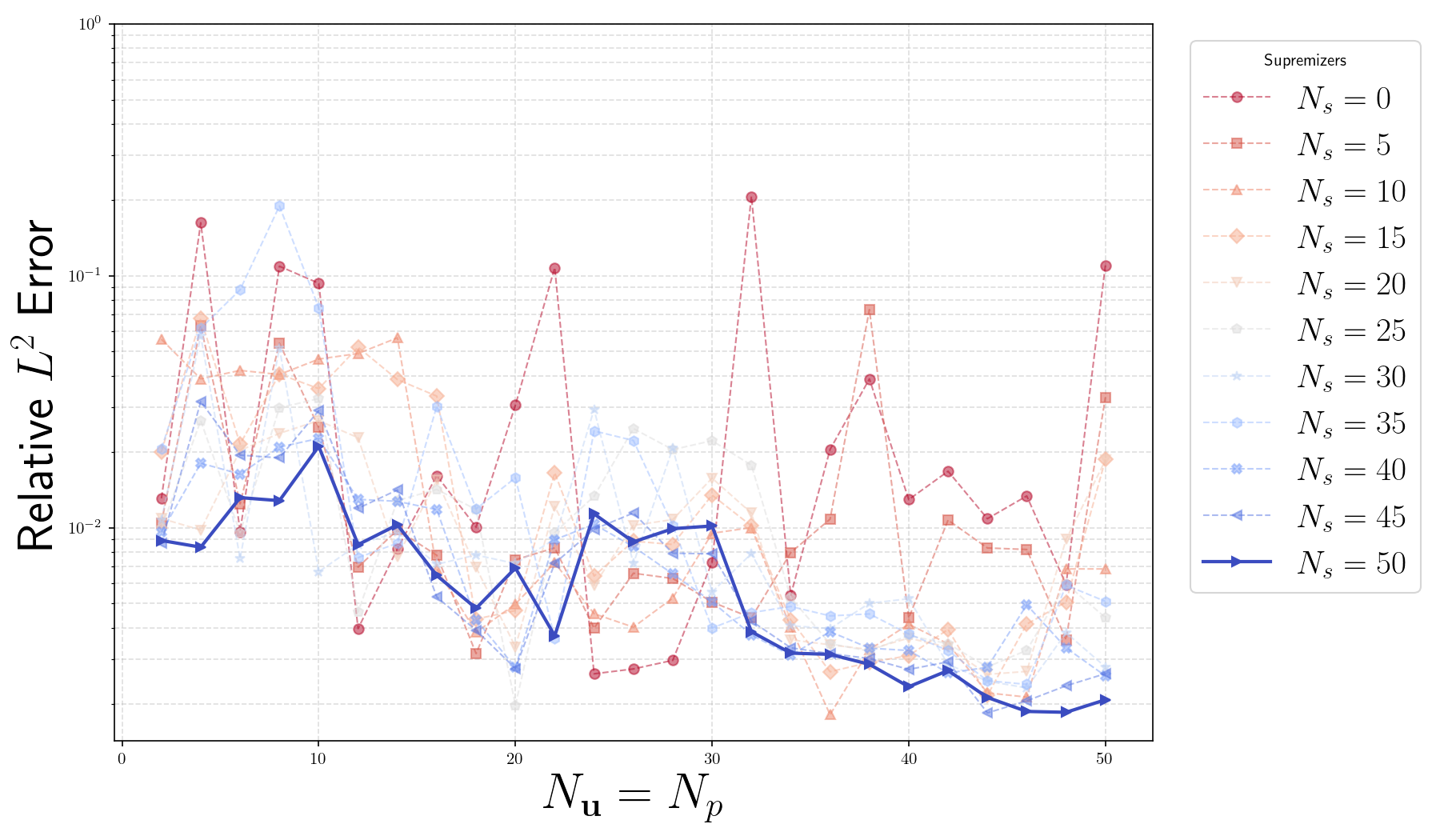}
        \caption{Relative velocity error for $N_{\boldsymbol{u}} = N_p$}
        \label{cylinder - sup_varying_modes - velocity}
    \end{subfigure}\hfill
    \begin{subfigure}{0.48\textwidth}
        \centering
        \includegraphics[width=\linewidth]{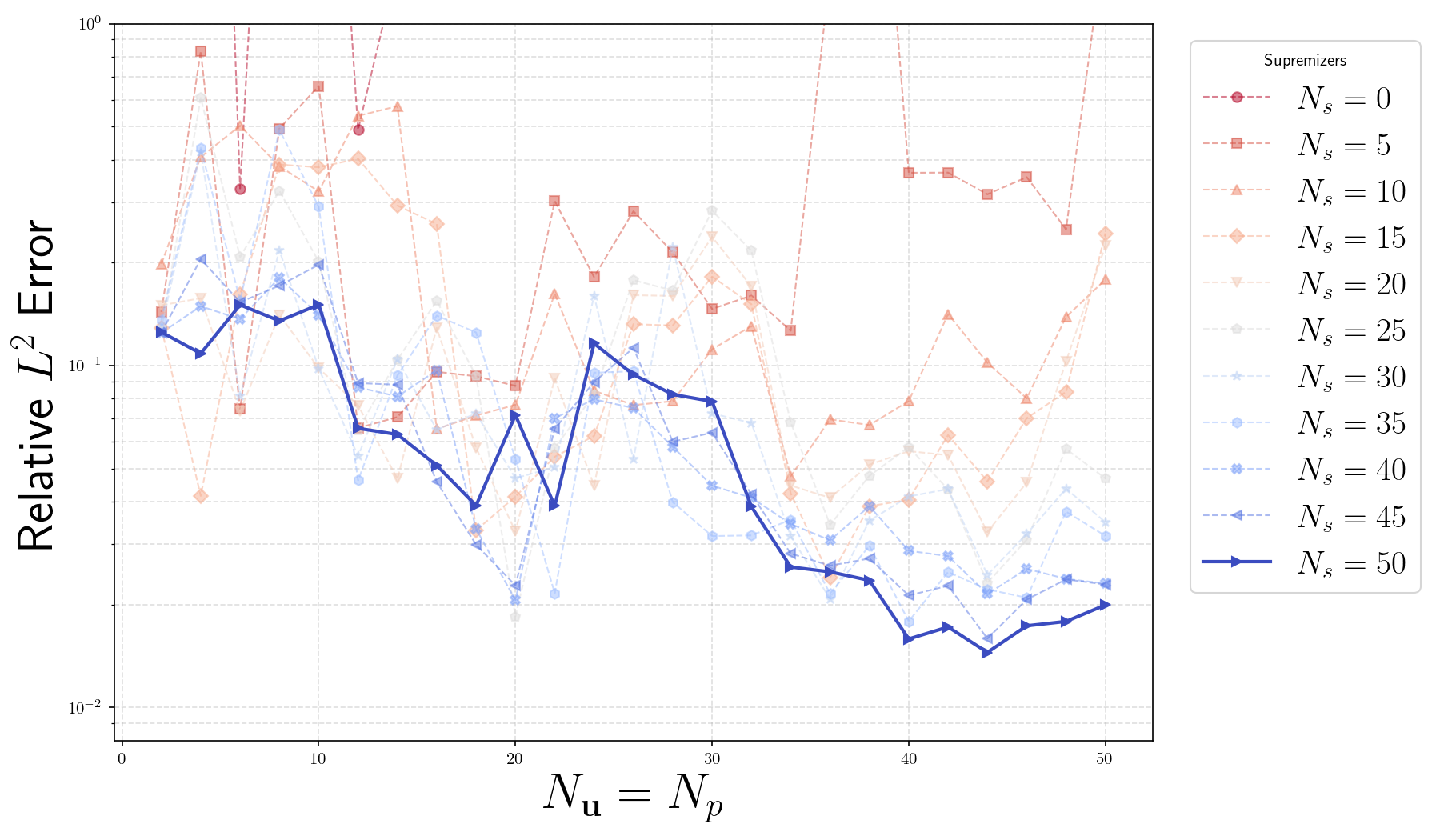}
        \caption{Relative pressure error for $N_{\boldsymbol{u}} = N_p$}
        \label{cylinder - sup_varying_modes - pressure}
    \end{subfigure}
    \caption{Analysis of the relative errors induced by the choice of the number of supremizers in the reduced coupled algorithm. Figures \ref{cylinder - sup_fixed_modes - velocity} and \ref{cylinder - sup_fixed_modes - pressure} present the results obtained by changing the number of velocity modes $N_{\boldsymbol{u}}$, while keeping $N_p$ fixed to $12$. Figures \ref{cylinder - sup_varying_modes - velocity} and \ref{cylinder - sup_varying_modes - pressure} instead stand for the case in which the number of velocity and pressure modes equally vary.}
    \label{cylinder_sup-analysis}
\end{figure}

Regarding the coupled approach, supremizers enrichment is needed due to its saddle-point structure, to avoid spurious pressure solutions. Figure \ref{cylinder_sup-analysis} shows the relative errors for different choices of the number of supremizers $N_s$ depending on the number of modes. Simulations are run in the undeformed geometry and two different frameworks are considered: in Figures \ref{cylinder - sup_fixed_modes - velocity}, \ref{cylinder - sup_fixed_modes - pressure}, the number of pressure modes $N_p$ is kept fixed to 12, while in Figures \ref{cylinder - sup_varying_modes - velocity} and \ref{cylinder - sup_varying_modes - pressure} both $N_{\boldsymbol{u}}$ and $N_p$ vary with $N_{\boldsymbol{u}}=N_p$.\\
As expected, spurious pressure modes are obtained for the lowest $N_{\boldsymbol{u}}$ choices, increasing the relative error. Effects of the higher complexity of the phenomenon can be noticed in this analysis as well. For $N_p = 12$, enlarging the reduced velocity space alone is not anymore sufficient to guarantee the stabilization of the saddle-point problem. A supremizer enrichment has to be performed and the best choice proves to be $N_s = 9$, with both low and not oscillating relative errors. This is not the case for $N_p = N_{\boldsymbol{u}}$, where the increase of the reduced pressure space requires, for larger $N_{\boldsymbol{u}}=N_p$ values, the use of $N_s = 50$, the largest possible value in our analysis.

\begin{figure}[!htb]
    \centering
    \begin{minipage}[c]{0.32\textwidth}
        \centering
        \begin{subfigure}{\linewidth}
            \centering
            \includegraphics[width=\linewidth]{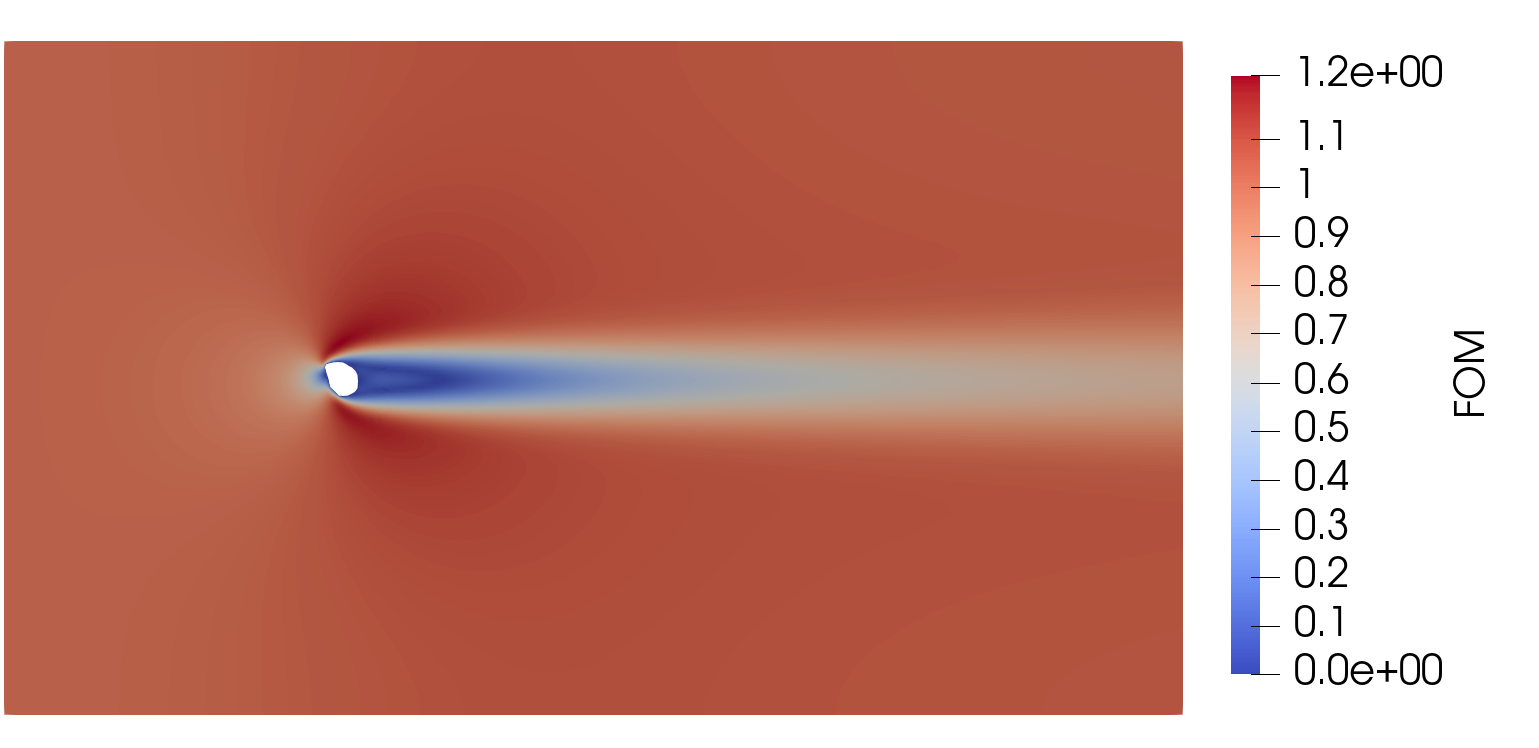}
        \end{subfigure}
    \end{minipage}
    \hfill
    \begin{minipage}[c]{0.65\textwidth}
        \begin{subfigure}{0.48\linewidth}
            \centering
            \includegraphics[width=\linewidth]{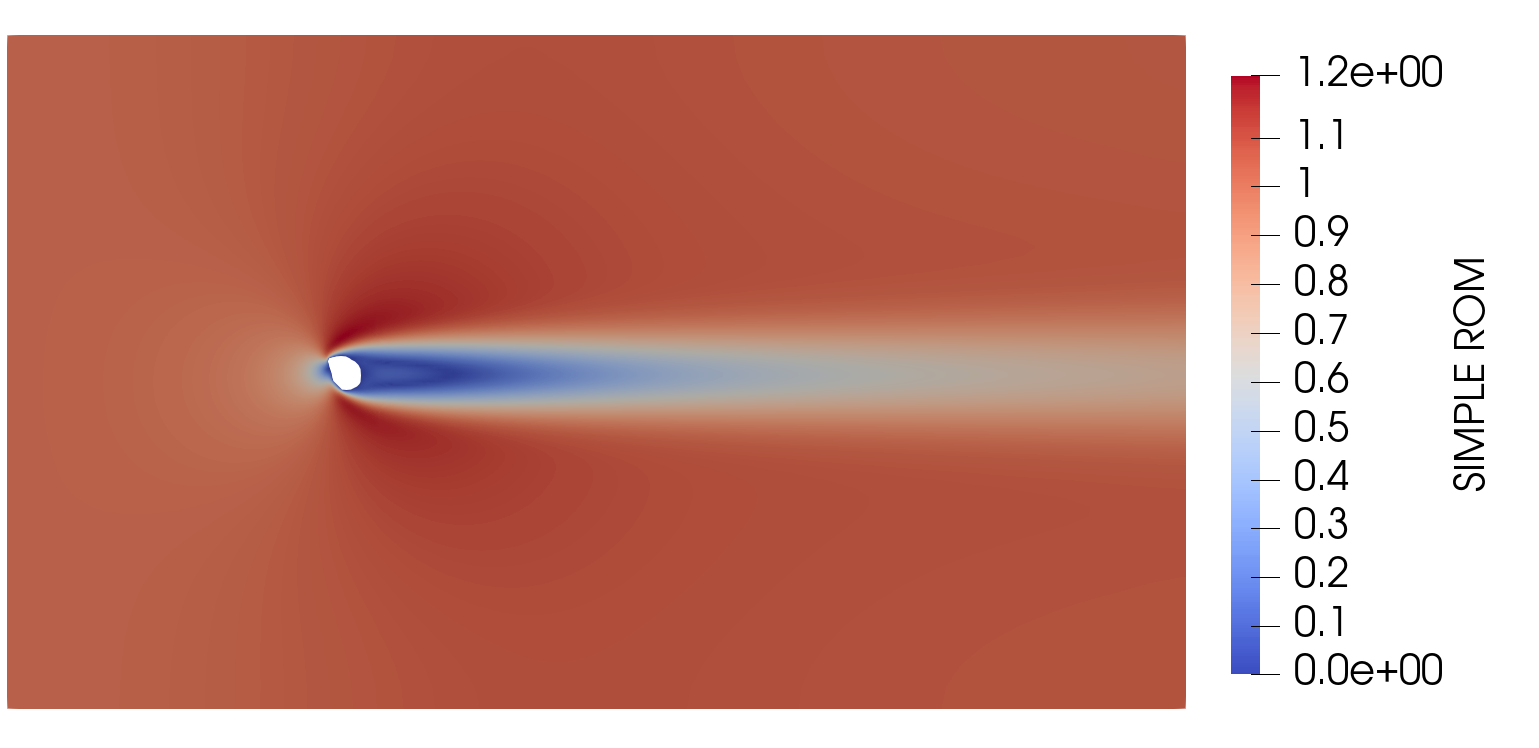}
        \end{subfigure}\hfill
        \begin{subfigure}{0.48\linewidth}
            \centering
            \includegraphics[width=\linewidth]{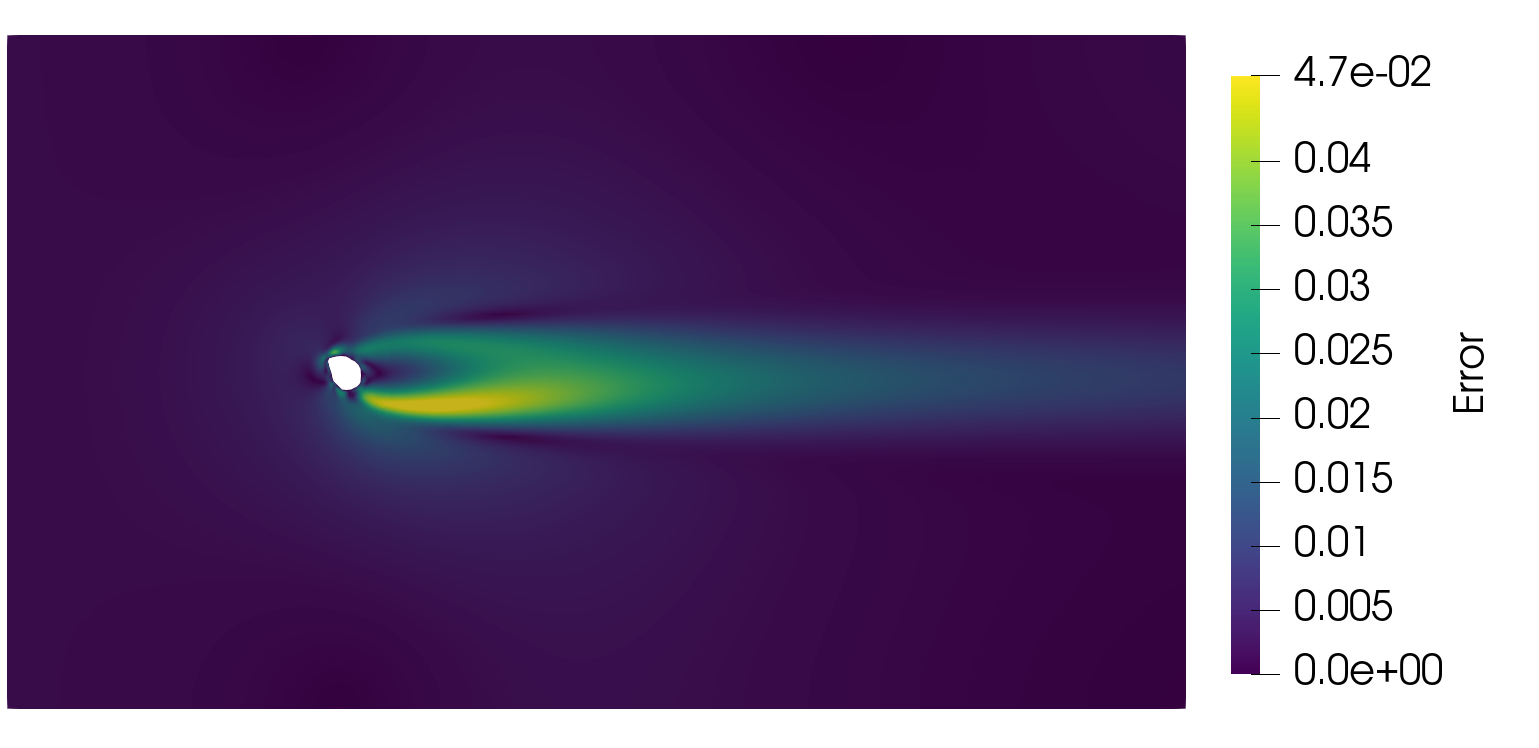}
        \end{subfigure}
        \\[2ex]
        \begin{subfigure}{0.48\linewidth}
            \centering
            \includegraphics[width=\linewidth]{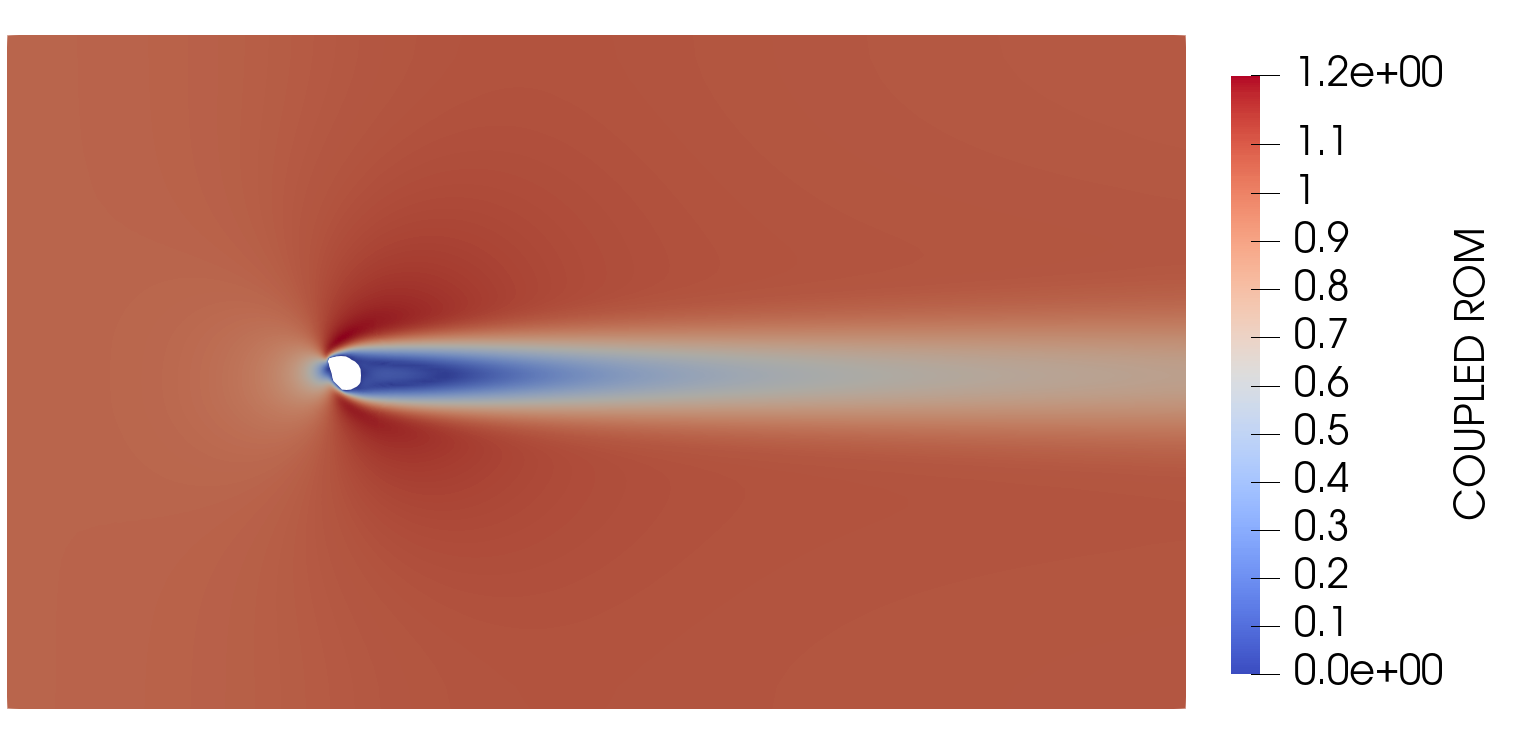}
        \end{subfigure}\hfill
        \begin{subfigure}{0.48\linewidth}
            \centering
            \includegraphics[width=\linewidth]{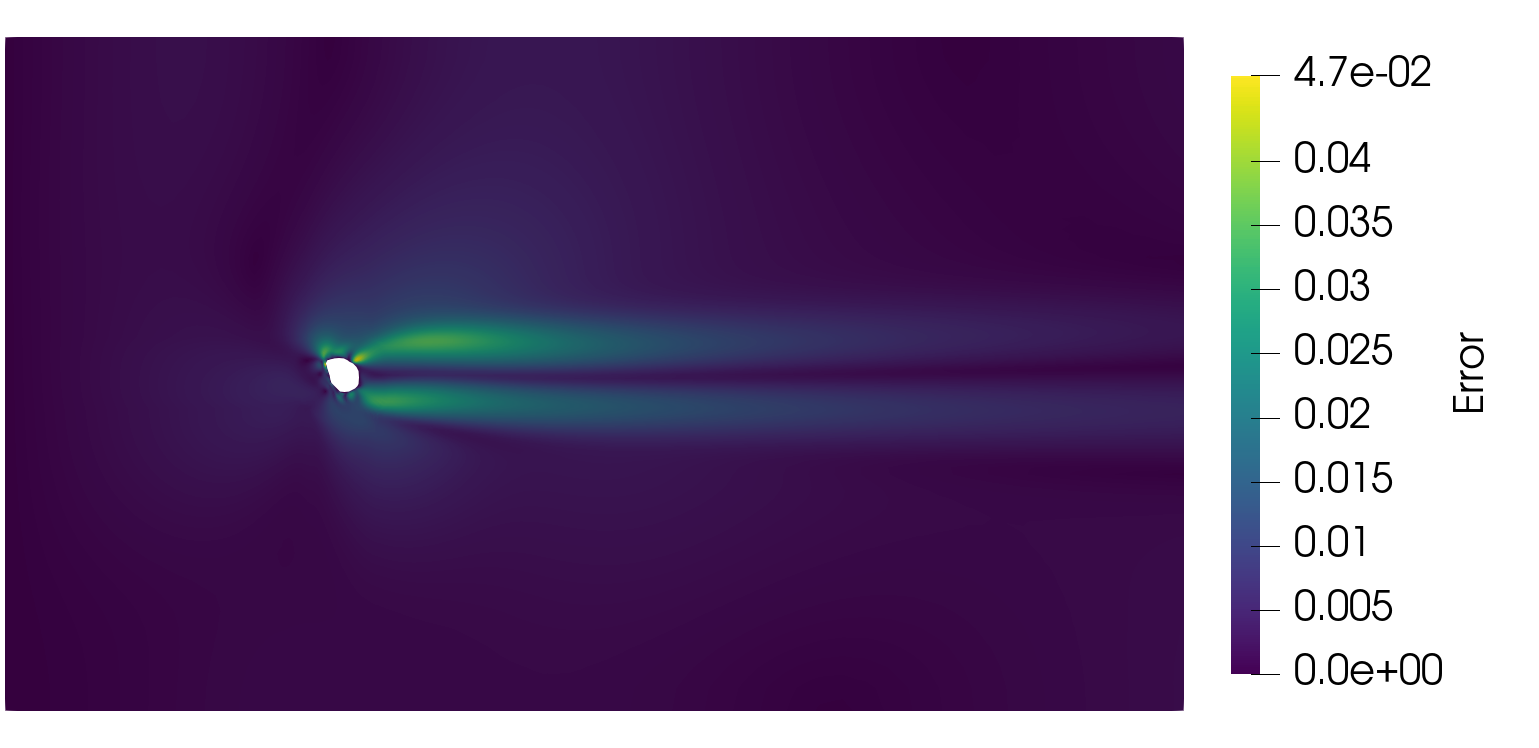}
        \end{subfigure}
    \end{minipage}

    \medskip

    \begin{minipage}[c]{0.32\textwidth}
        \centering
        \begin{subfigure}{\linewidth}
            \centering
            \includegraphics[width=\linewidth]{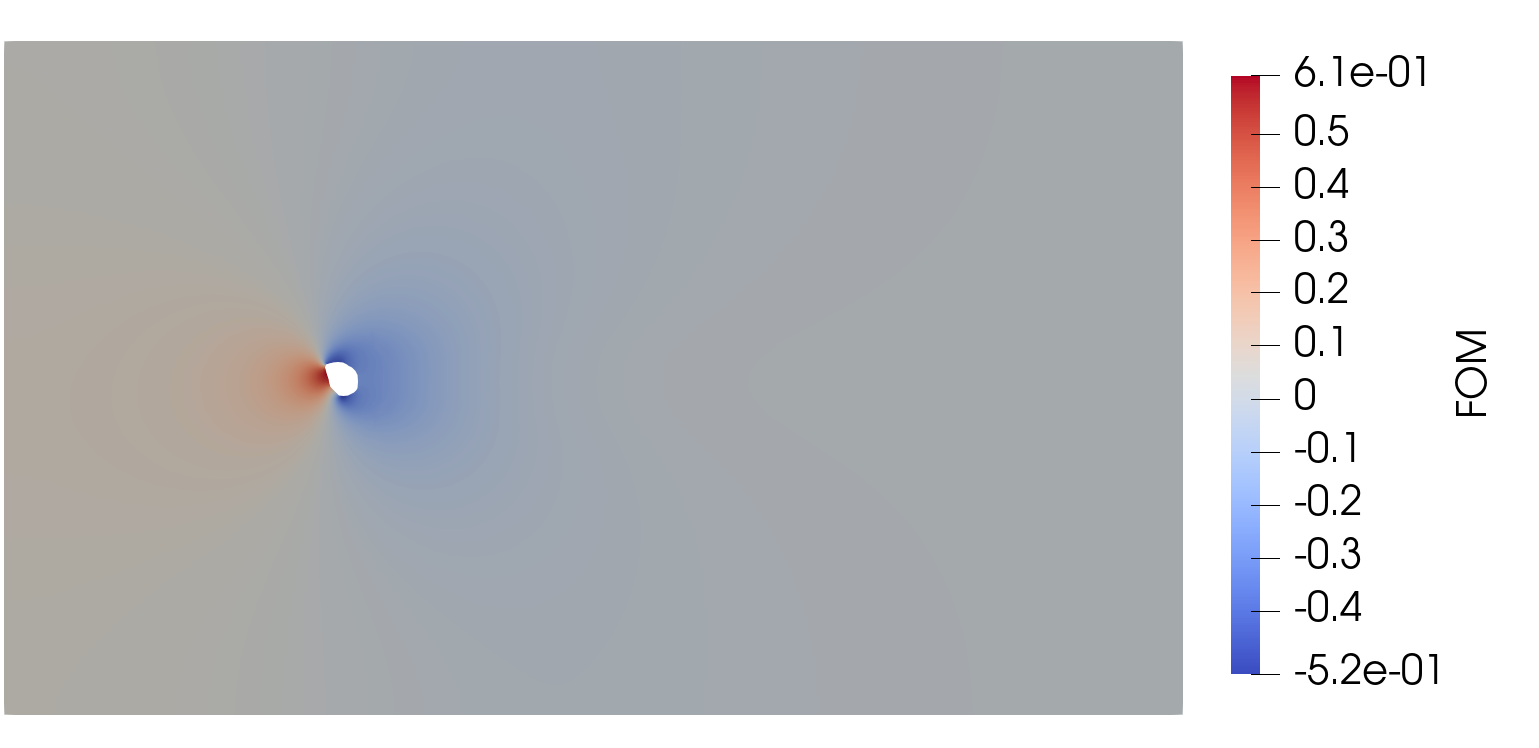}
        \end{subfigure}
    \end{minipage}
    \hfill
    \begin{minipage}[c]{0.65\textwidth}
        \begin{subfigure}{0.48\linewidth}
            \centering
            \includegraphics[width=\linewidth]{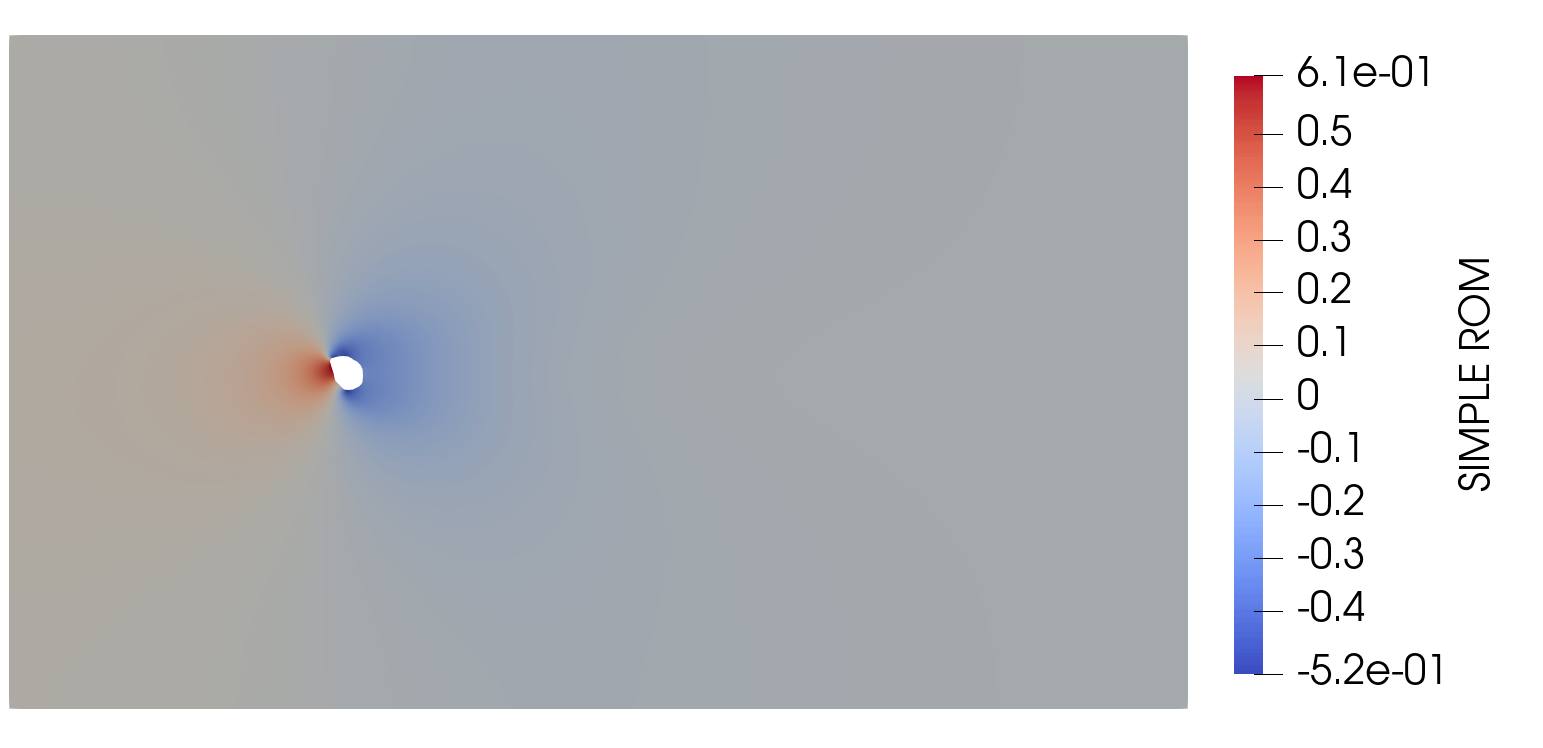}
        \end{subfigure}\hfill
        \begin{subfigure}{0.48\linewidth}
            \centering
            \includegraphics[width=\linewidth]{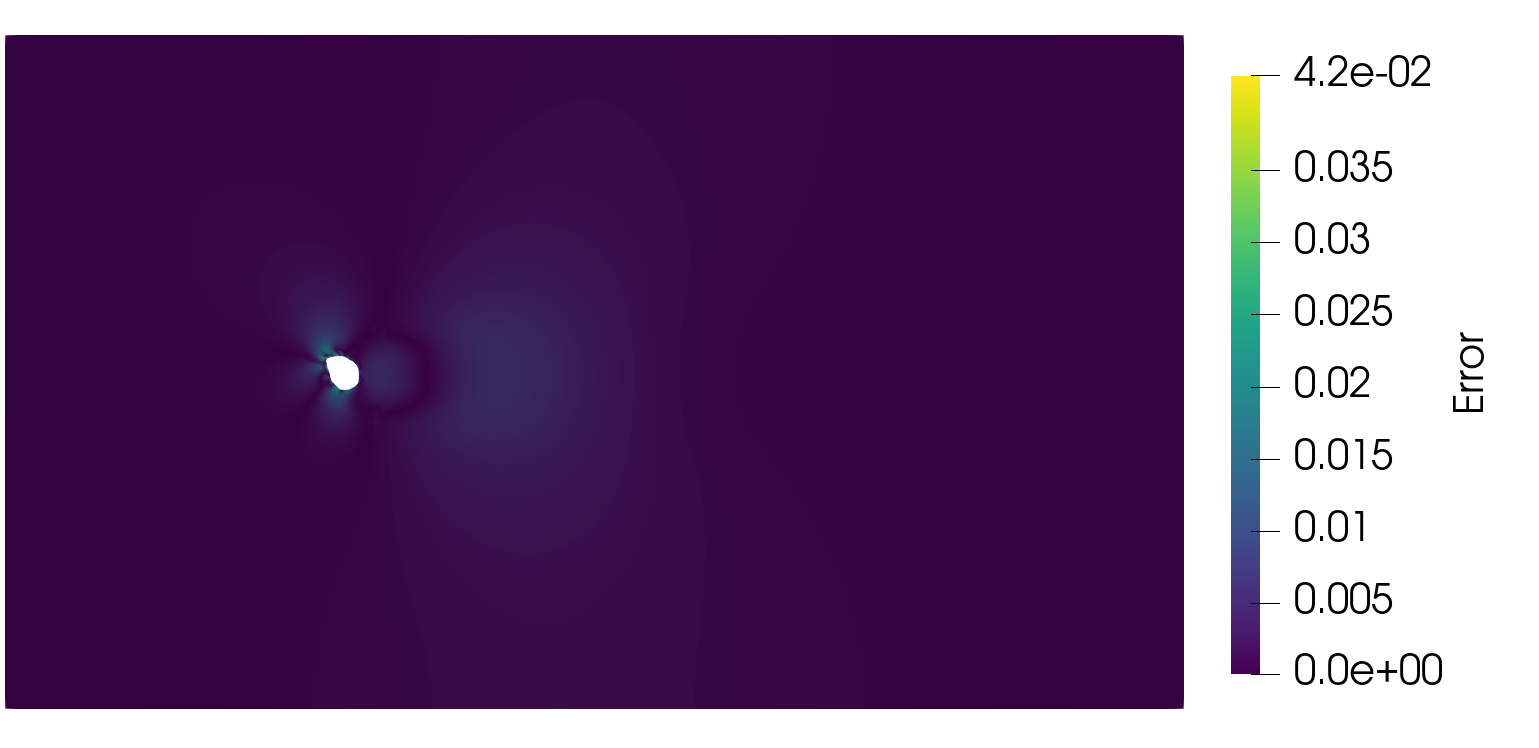}
        \end{subfigure}
        \\[2ex]
        \begin{subfigure}{0.48\linewidth}
            \centering
            \includegraphics[width=\linewidth]{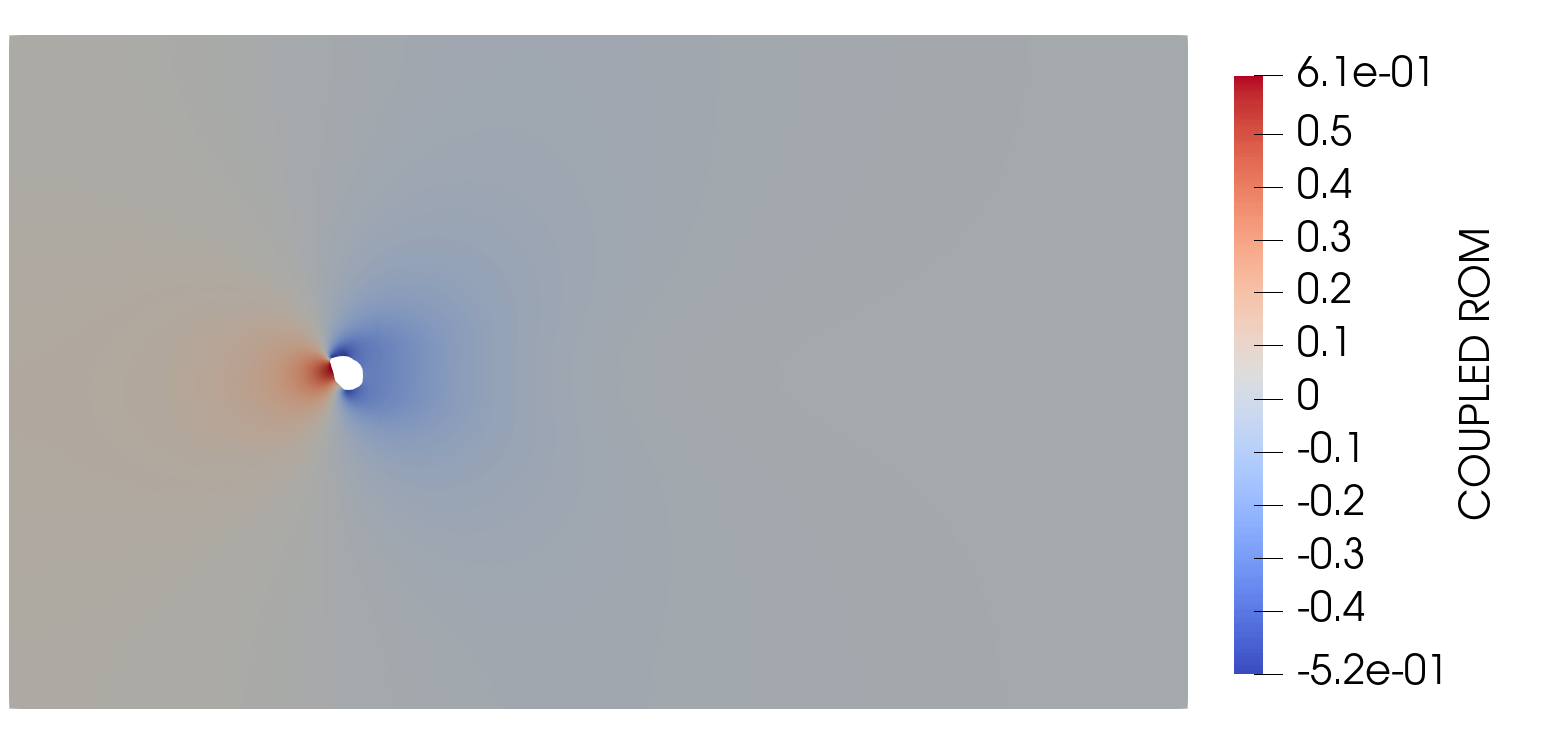}
        \end{subfigure}\hfill
        \begin{subfigure}{0.48\linewidth}
            \centering
            \includegraphics[width=\linewidth]{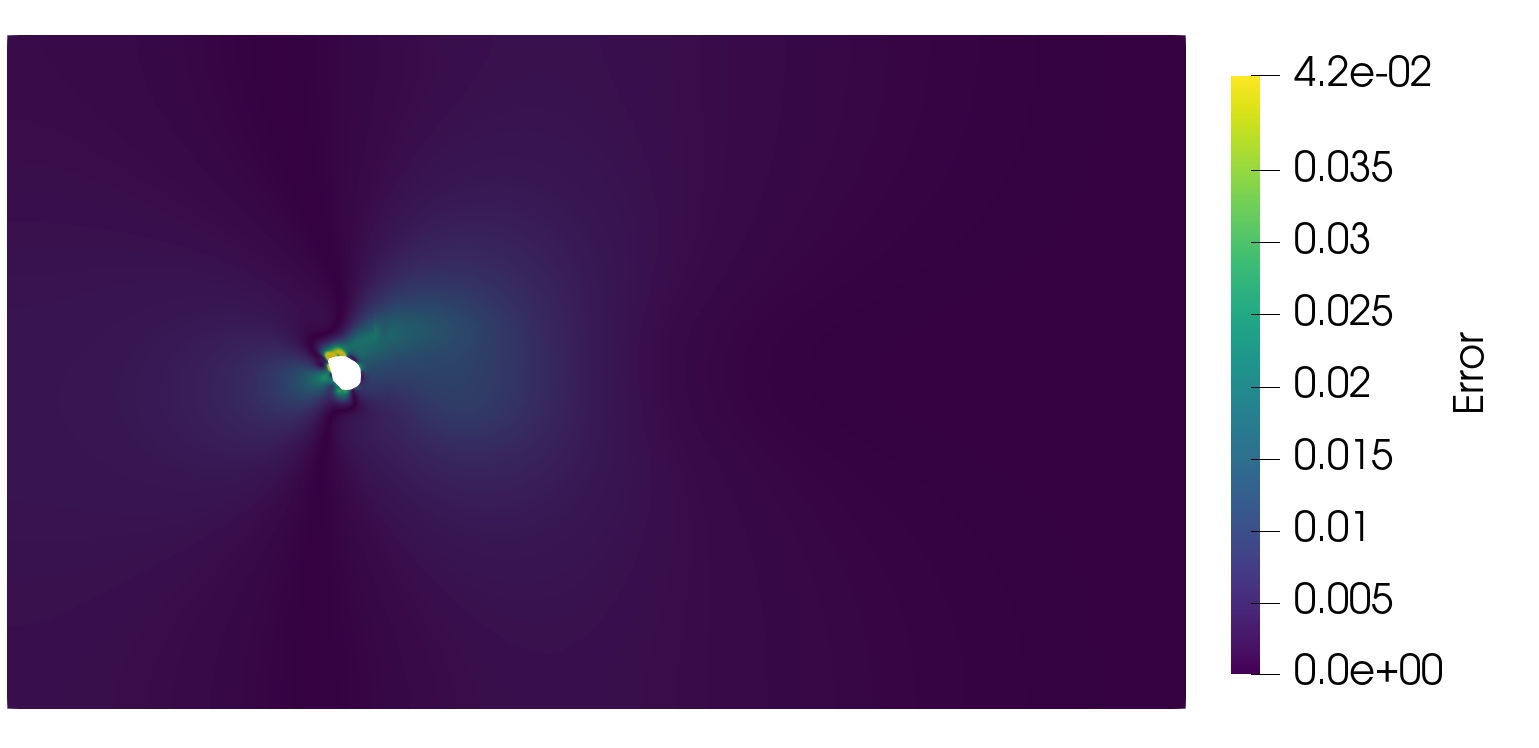}
        \end{subfigure}
    \end{minipage}
    \caption{Qualitative analysis of the absolute errors induced by the model order reduction: we compare the results obtained through the two reduced algorithms with those obtained at the full order level, both in terms of velocity and pressure (first and second set of figures respectively). Simulations have been carried out with $N_{\boldsymbol{u}} = 35$, $N_p = 12$ and $N_s = 9$.}
        \label{Cylinder_absolute_error}
\end{figure}

To investigate the accuracy of both methods, we test them on a new deformed geometry. Figure \ref{Cylinder_absolute_error} shows the predicted fields and absolute errors with respect to the full-order counterpart for one deformed case. Reduced solutions refer to $N_{\boldsymbol{u}} = 35$, $N_p = 12$ and $N_s = 9$. Both ROMs accurately approximate the full-order results, with comparable order of magnitude for the absolute error both in velocity and pressure. Nonetheless, the SIMPLE algorithm proves to be more accurate in the considered deformation. Specifically, relative errors are equal to 6.43\% (pressure) and 0.50\% (velocity) for the coupled algorithm, 3.18\% (pressure) and 0.71\% (velocity) for the SIMPLE one.

\begin{figure}[!htb]
    \begin{subfigure}[t]{0.16\textwidth}
        \vspace{0pt}
        \centering
        \includegraphics[width=\linewidth]{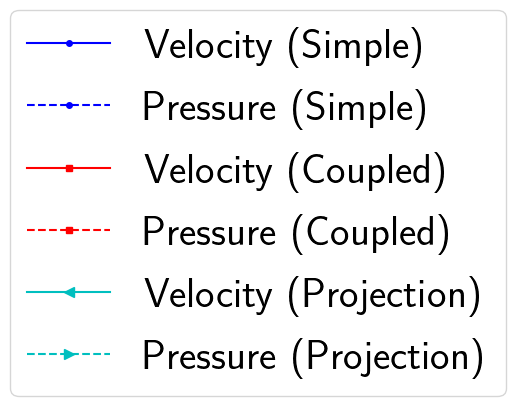}
    \end{subfigure}\hfill
    \begin{subfigure}[t]{0.41\textwidth}
        \vspace{0pt}
        \centering
        \includegraphics[width=\linewidth]{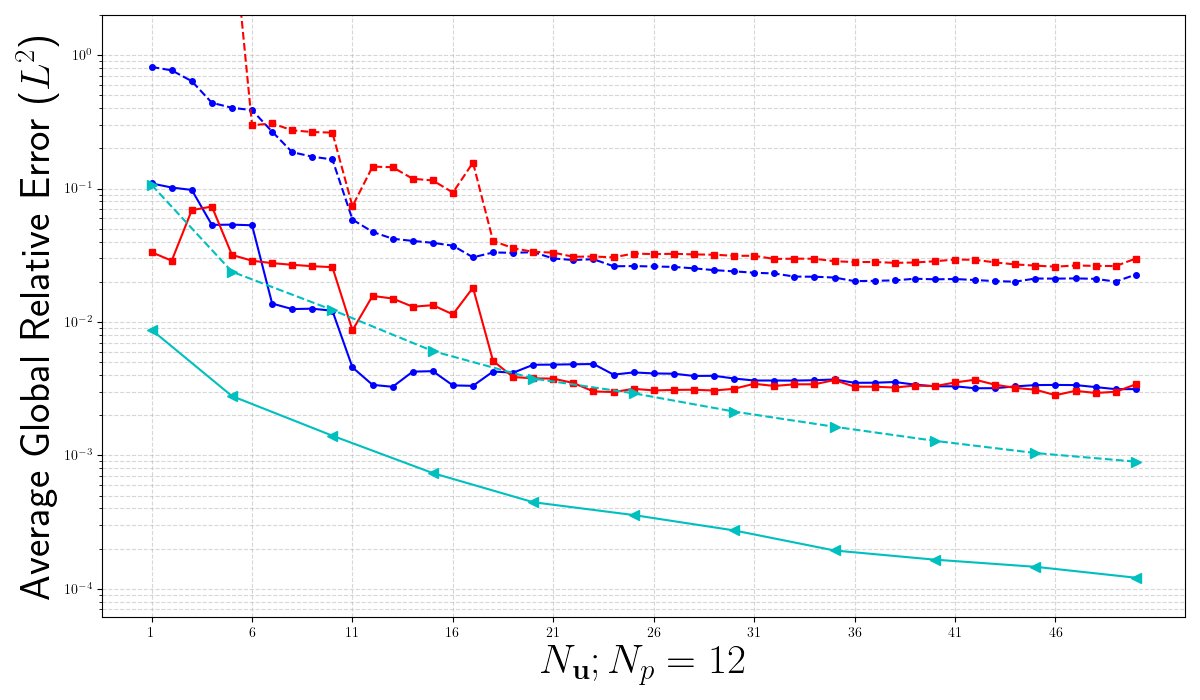}
        \caption{Average relative errors for $N_p = 12$}
        \label{cylinder_pres_fixed}
    \end{subfigure}\hfill
    \begin{subfigure}[t]{0.41\textwidth}
        \vspace{0pt}
        \centering
        \includegraphics[width=\linewidth]{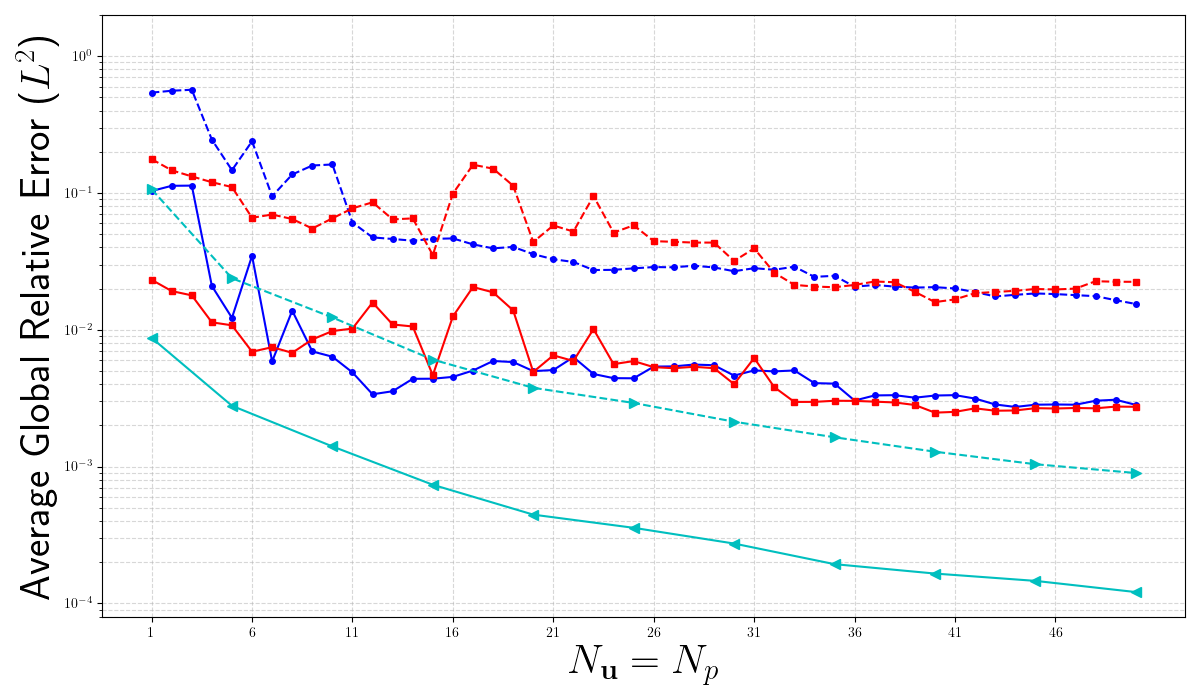}
        \caption{Average relative errors for $N_{\boldsymbol{u}} = N_p$}
        \label{cylinder_pres_varies}
    \end{subfigure}
    \caption{Comparison of the average relative errors induced by the reduced order models against the exact projection error. In Figure \ref{cylinder_pres_fixed}, the number of pressure modes $N_p$ is kept fixed to $12$, whereas it varies accordingly to $N_{\boldsymbol{u}}$ in Figure \ref{cylinder_pres_varies}.}
    \label{relative_errors_cylinder}
\end{figure}

To deepen our analysis, we test both ROMs on 10 new geometries and average the corresponding errors.
Figure \ref{cylinder_pres_fixed} exhibits the average error when fixing $N_p = 12$ and $N_s = 9$. It appears that the SIMPLE algorithm tends to be more stable for $N_{\boldsymbol{u}} \leq 5$, due to the segregated nature of the problem, which circumvents handling a saddle-point problem. Conversely, the coupled approach exhibits in the same region completely out-of-range relative errors, caused by spurious pressure modes. As in the cavity case, the analysis exhibits a middle region, namely for $6 \leq N_{\boldsymbol{u}} \leq 17$, in which the errors due to the coupled algorithm decay to a lower order of magnitude, although they remain moderately worse than those of the SIMPLE scheme.\\
The difference between the two approaches is drastically reduced for $N_{\boldsymbol{u}} \geq 18$. The errors become almost indistinguishable in the velocity case, stabilizing around 0.3\%. For the pressure field, the SIMPLE algorithm still shows slightly better results, with an average error of 2\% compared to the 3\% obtained through the coupled approach.

In Figure \ref{cylinder_pres_varies}, the case for $N_{\boldsymbol{u}} = N_p$ is considered. The coupled algorithm shows a more stable behaviour, enabled by the choice of a large number of supremizer $N_s = 50$, which stabilizes the problem even for lower dimensional spaces, namely for $N_{\boldsymbol{u}} = N_p \leq 10$. The SIMPLE algorithm leads to errors within a comparable order of magnitude, yet with higher instability. The latter is mitigated by including a sufficient amount of pressure information (recall that $N_p = 12$ corresponds to 99.99\% of the total pressure energy), resulting in a monotonically decaying pressure error and a velocity one oscillating around a plateau value of 0.3\%. Conversely, for $10 \leq N_p = N_{\boldsymbol{u}} \leq 30$, the coupled algorithm exhibits higher instabilities, probably due to a combination of an insufficient amount of velocity information (recall that $N_{\boldsymbol{u}} = 35$ corresponds to 99.99\% of velocity information content) and a too large number of supremizers, with respect to the adopted reduced modes. As a matter of fact, larger dimensional spaces ($N_{\boldsymbol{u}} \geq 32$) mitigate this phenomenon, highlighting a common behaviour for the two algorithms. Both show errors stabilizing around 0.3\% for velocity and 2\% for pressure.

Both methods reach relatively high plateaus with respect to exact projection errors, which are computed from the projection of the full-order solutions onto the reduced spaces. This discrepancy is mainly caused by two different factors: on the one hand, projecting the problem with respect to the undeformed $L^2(\Omega)$ scalar product implies some sort of inaccuracy, when dealing with other geometries; on the other hand, considering too many modes add more noise rather than information content to the model, in the aforementioned \textit{saturation} phenomenon. In addition, the linear reduced representation hypothesis introduces an intrinsic error when dealing with non-linear problems \cite{khamlich2025advanced}. A potential remedy is discussed in \cite{stabile2020efficient}, though investigating this methodology is outside the scope of this work.

Finally, from a computational point of view, both algorithms successfully reduce the complexity of the problem, decreasing the degrees of freedom from $4 \, N_h = 36800$ to $N_{SIMPLE} = N_{\boldsymbol{u}} + N_p = 47$ and $N_{coupled} = N_{\boldsymbol{u}} + N_p + N_s = 56$. Conversely, not implementing (D)EIM reduction \cite{stabile2020efficient}, we do not expect speed-ups in terms of computational time, thus we do not compare the algorithms under this aspect. Nevertheless, they present different behaviours when looking at the number of required non-linear iterations. The full-order SIMPLE algorithm reaches convergence in $\mathcal{O}(100)$ iterations, while $\mathcal{O}(1000)$ are needed at the reduced segregated level. This slowdown is mainly caused by the poor, yet unavoidable, choice of the under-relaxation coefficients $\alpha_{\boldsymbol{u}}$ and $\alpha_p$. In contrast, the coupled algorithm takes $\mathcal{O}(10)$ non-linear iterations to converge.

In addition, the coupled and SIMPLE approaches are compared in terms of aerodynamic coefficients prediction \cite{Fossati}. The reduced lift and drag coefficients are computed using ITHACA-FV and the results are compared with those obtained through the SIMPLE algorithm implemented in OpenFOAM. Specifically, the reduced coefficients are evaluated by integrating the reconstructed reduced pressure and velocity fields over the deformed obstacle boundary.\\
The relative errors used in the algorithms' validation are defined as
\begin{equation}
    \varepsilon_d \coloneqq \dfrac{\left|c_{d_{FOM}} - c_{d_{ROM}}\right|}{\left|c_{d_{FOM}}\right|}
    \qquad
    \varepsilon_l \coloneqq \dfrac{\left|c_{l_{FOM}} - c_{l_{ROM}}\right|}{\left|c_{l_{FOM}}\right|},
\end{equation}
where $c_d$ and $c_l$ are the drag and lift coefficients respectively, whereas the subscripts $FOM$ and $ROM$ denote the order of the considered method. The results reported in Table \ref{tab:aerodynamic_coeff_errors} reveal two different scenarios. The prediction of the drag coefficient proves to be quite accurate, with average relative errors of 0.8\% and 1.3\% for the SIMPLE and coupled approaches, respectively. Contrarily, the errors exhibit a significant increase for the lift coefficient, rising, on average, to 25\% for the SIMPLE approach and 44\% for the coupled one.\\
This discrepancy stems from the different physical nature of the two quantities, which is directly reflected in their numerical evaluation \cite{Stabile2017CAIM, stabile2018finite}. As a matter of fact, the lift coefficient is primarily determined by the pressure field, which is subject to higher errors with respect to the velocity one.
Conversely, the drag coefficient is related to the drag force, which is mainly influenced by the viscous effects associated with the velocity field.
\begin{table}[!htb]
    \centering
    \renewcommand{\arraystretch}{1.5}
    \begin{tabular}{|c||c|c|c||c|c|c|}
        \hline
        Case & $c_d$ FOM & $\varepsilon_d$ SIMPLE & $\varepsilon_d$ coupled & $c_l$ FOM & $\varepsilon_l$ SIMPLE & $\varepsilon_l$ coupled\\
        \hline
        \hline
        1 & 1.66231 & $\mathbf{0.00294030}$ & 0.04015954 & 0.423620  & $\mathbf{0.08543021}$ & 0.30777048\\
        \hline
        2 & 1.52408 & $\mathbf{0.00758569}$ & 0.01608261 & -0.248686 & $\mathbf{0.02170471}$ & 0.06180353\\
        \hline
        3 & 1.56932 & $\mathbf{0.00469272}$ & 0.01551904 & -0.058273 & $\mathbf{0.35876334}$ & 0.54246363\\
        \hline
        4 & 1.65254 & 0.01465694 & $\mathbf{0.00158011}$ & 0.393177 & $\mathbf{0.04302356}$ & 0.07831369 \\
        \hline
        5 & 1.73667 & $\mathbf{0.00422336}$ & 0.01407022 & -0.355787 & 0.08350877 & $\mathbf{0.05314789}$\\
        \hline
        6 & 1.66956 & $\mathbf{0.01173725}$ & 0.01319872 & -0.128637 & 0.16934565 & $\textbf{0.15726514}$ \\
        \hline
        7 & 1.76290 & 0.00645612 & $\mathbf{0.00427619}$ & 0.014810 & \textbf{1.56041224} & 2.76155711\\
        \hline
        8 & 1.60486 & 0.01000180 & \textbf{0.00577713} & -0.039248 & \textbf{0.21887420} & 0.35519699\\
        \hline
        9 & 1.76547 & \textbf{0.01229263} & 0.01905569 & -0.124042 & \textbf{0.03299372} & 0.03911873\\
        \hline
        10 & 1.6684 & 0.00551533 & \textbf{0.00360936} & -0.295638 & \textbf{0.01645975} & 0.04337441 \\
        \hline \hline
        Mean & 1.66162 & $\mathbf{0.00801021}$ & 0.01333286 & -0.041870 & $\mathbf{0.25905162}$ & 0.44000116 \\
        \hline
    \end{tabular}
    \caption{Drag and lift prediction relative errors. The reduced aerodynamic coefficients are computed using $N_{\boldsymbol{u}} = 35$ and $N_p=12$ velocity and pressure modes respectively. Bold values correspond to the lowest error among the two ROMs.}
    \label{tab:aerodynamic_coeff_errors}
\end{table}
\section{Conclusions and future work}\label{conclusions}
The main objective of the present work is to compare two different reduced-order approaches within the framework of FV discretization for geometrically parametrized domains in CFD. To the best of the authors' knowledge, the existing literature lacks a direct and explicit comparison between the reduced \textit{monolithic} and \textit{segregated} approaches. The first deals with a so-called \textit{saddle-point problem}, coming from a velocity-pressure coupling. The second decouple the two unknown fields, mimicking the SIMPLE algorithm employed at the full-order level.

In this study, two benchmark cases are tested: a lid-driven cavity flow and a flow past a cylinder. In the former, the cavity geometry is randomly perturbed, whereas in the latter, the cross-section of the cylindrical obstacle is parametrized. The algorithms are compared in terms of field reconstruction accuracy and aerodynamic coefficient prediction (for the cylinder case). 

On the one hand, the coupled approach exhibits higher instabilities, especially for lower-dimensional reduced spaces. Nonetheless, a proper supremizer enrichment proves effective in smoothing out the errors, in particular within larger-dimensional frameworks. Therefore, the optimal enrichment is one that sufficiently enlarges the reduced spaces without compromising the underlying physical information.\\
On the other hand, the segregated approach yields more accurate results, even for lower-dimensional reduced spaces. Nevertheless, this comes at the cost of a slower convergence rate, induced by numerical sensitivities to the under-relaxation factors characterizing this algorithm. In particular, the best choice for the under-relaxation coefficients proves not to be necessarily the same adopted at the full-order level.\\
As the dimension of the reduced spaces increases, the performances of the two approaches become similar, stabilizing at approximately the same error level.

Future extensions of the present work may include:
\begin{itemize}
    \item an in-depth analysis of the under-relaxation strategy in the SIMPLE-ROM to mitigate computational requirements and accelerate convergence;
    \item a rigorous \textit{a priori} investigation of the optimal number of supremizer modes required to guarantee scheme stability;
    \item the application of the considered methods to higher-Reynolds-number flows, incorporating time dependence and/or turbulence modeling (e.g., see \cite{stabile2018finite}, which focuses solely on physical parameterizations);
    \item the implementation of hyper-reduction techniques, such as the Discrete Empirical Interpolation Method (DEIM), to bypass the assembly of full-order matrices during the online stage (see, e.g., \cite{stabile2020efficient});
    \item the extension of the present analysis to three-dimensional flows.
\end{itemize}

\section*{Acknowledgements}
DO and GR acknowledge the support provided bt the INdAM-GNCS (CUP E53C25002010001. Additionally, DO and GR acknowledge the support provided by the European Union-NextGenerationEU, in the framework of the iNEST-Interconnected Nord-Est Innovation Ecosystem (iNEST ECS00000043– CUP G93C22000610007) consortium. Moreover, this study was carried out within the ``20227K44ME - Full and Reduced order modelling of coupled systems: focus on non-matching methods and automatic learning (FaReX)" project – funded by European Union – Next Generation EU  within the PRIN 2022 program (D.D. 104 - 02/02/2022 Ministero dell’Università e della Ricerca). 

\bibliographystyle{plain}
\bibliography{biblio}

\appendix
\section{Cylinder deformations using Bézier curves}\label{Bezier}
Bézier curves are parametric curves, widely used in computer graphics and related fields (see e.g. \cite{farin2014curves}). They are mathematically based on Bernstein polynomials, first introduced in 1912 \cite{bernstein1912demonstration}, and have found extensive application, specifically in automotive design, through the works of Paul de Casteljau \cite{de1959outillages} and Pierre Bézier \cite{bezier2014mathematical}.

Following the procedure described by Viquerat and Hachem \cite{Viquerat_2020}, Bézier curves have been used in this context to smoothly, up to the desired order, deform the original cylindrical obstacle. The algorithm is grounded on the following steps:
\begin{itemize}
    \item [i)]Take 8 control points $p_i$, $i = 1$, \dots $8$ as the intersection of the cylinder's section along the $xy$-plane with the axis $x$ and $y$ and with the quadrants' bisectors $y = x$ and $y = -x$.
    \item [ii)]Modify the distance from the origin (in the $xy$ plane) of each of these points by random factors $\rho_i$, with $\rho_i \in [0.75, 1.25]$, $i = 1$, \dots $8$ (Figure \ref{Moving guide vertices}).
    \item [iii)]Consider the angles $\theta_{i, i+1}$ defined as the minimum between the angles formed by the segment joining two consecutive points $p_i$ and $p_{i+1}$ and the $y$ and $x$ directions respectively. This is done to avoid too concave (and therefore hard to treat) deformations.\\
    For each point $p_i$, compute the mean $\gamma_i$ between the two angles involving it, namely $\theta_{i-1, i}$ and $\theta_{i, i+1}$.
    \item [iv)]For any control point $p_i$, move along the line connecting $p_{i-1}$ and $p_i$ both ``forward" (towards the point $p_{i+1}$) and ``backward" (towards the point $p_{i-1}$) by a fixed factor. In this case, it corresponds to $0.3$ times the length of the segment connecting the considered consecutive points (e.g. $p_ip_{i+1}$ if moving ``forward"). There are now two more points $q_{i,1}$, $q_{i,2}$ for each of the original points $p_i$ (Figure \ref{Preparing control points}).
    \item [v)]Rotate every segment $q_{i,1}q_{i,2}$ by the previously computed angle $\gamma_i$, obtaining the new points $r_{i,1}$, $r_{i,2}$. For each couple of consecutive points $p_i$, $p_{i+1}$, there are now two intermediate points $r_{i,1}$, $r_{i+1, 2}$ (see Figure \ref{Control points}). 8 different Bézier cubic curves $\mathcal{B}_i$ are defined, with ``smoothness" given by the fact that the same tangent direction is imposed to consecutive cubics $\mathcal{B}_{i-1}$, $\mathcal{B}_{i}$, in the common point $p_i$.
    \begin{figure}[!htb]
        \centering
        \begin{minipage}[t]{0.495\linewidth}
            \includegraphics[width=\linewidth]{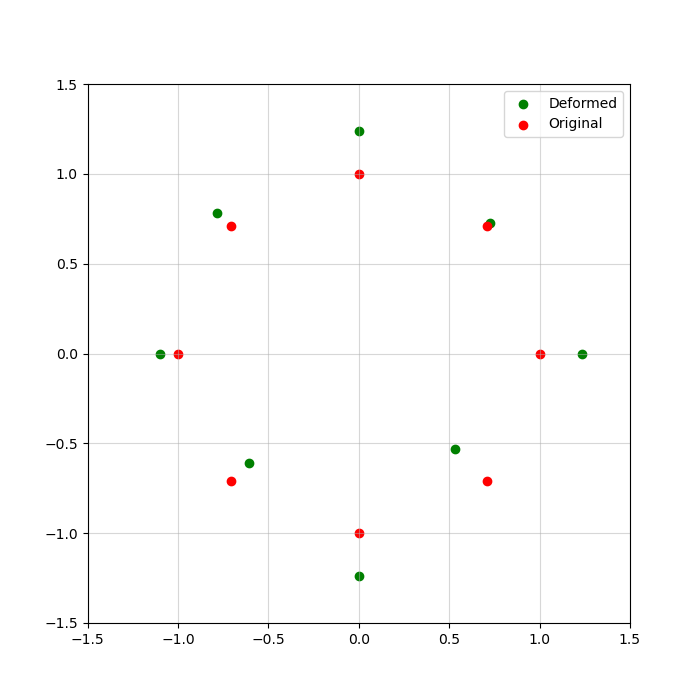}
            \vspace{-1cm}
            \caption{Random deformation of the guide vertices that are moved by 8 independent random factors $\rho_i \in $ $[0.75, 1.25]$.}
            \label{Moving guide vertices}
        \end{minipage}
        \hfill
        \begin{minipage}[t]{0.495\linewidth}
            \includegraphics[width=\linewidth]{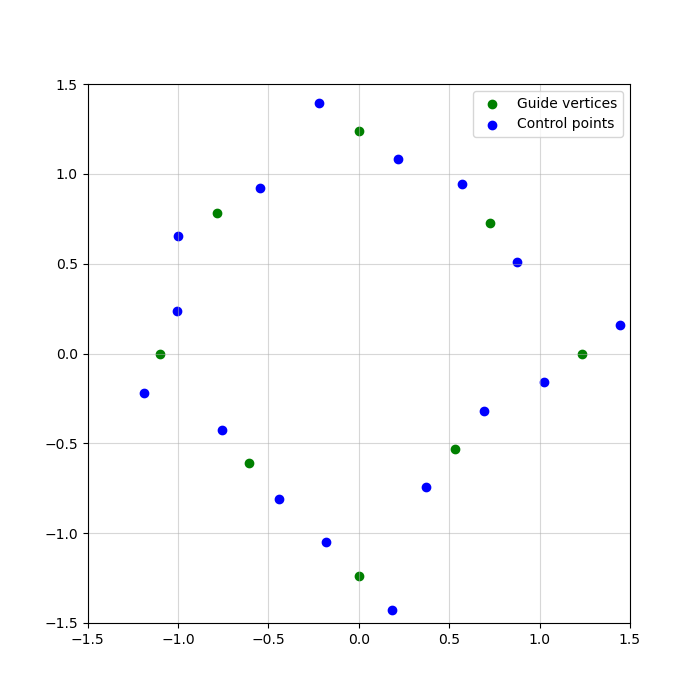}
            \vspace{-1cm}
            \caption{The control points are obtained from the directions connecting couples of consecutive guide vertices.}
            \label{Preparing control points}
        \end{minipage}
    \end{figure}
    \begin{figure}[!htb]
        \centering
        \begin{minipage}[t]{0.495\textwidth}
            \centering
            \includegraphics[width=\linewidth]{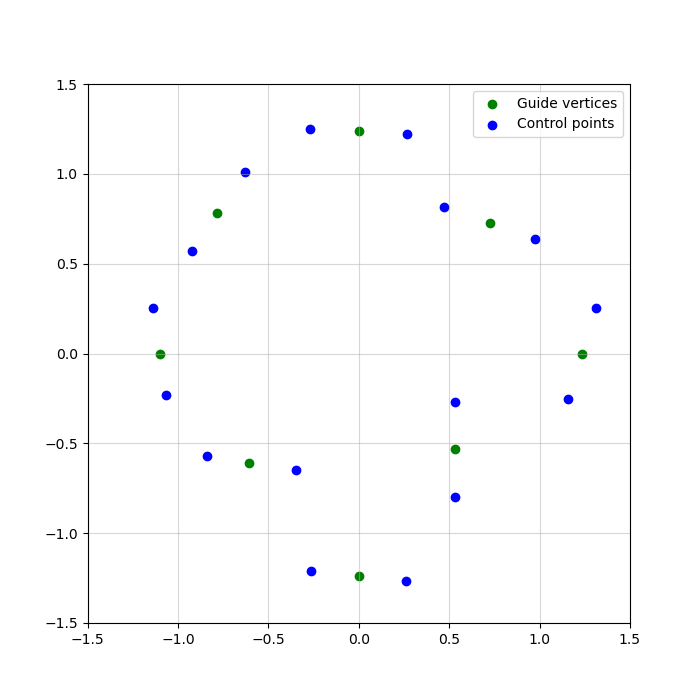}
            \vspace{-1cm}
            \caption{Each couple of control points is rotated with respect to the reference guide vertex by the angles $\gamma_i$.}
            \label{Control points}
        \end{minipage}
        \hfill
        \begin{minipage}[t]{0.495\textwidth}
            \centering
            \includegraphics[width=\linewidth]{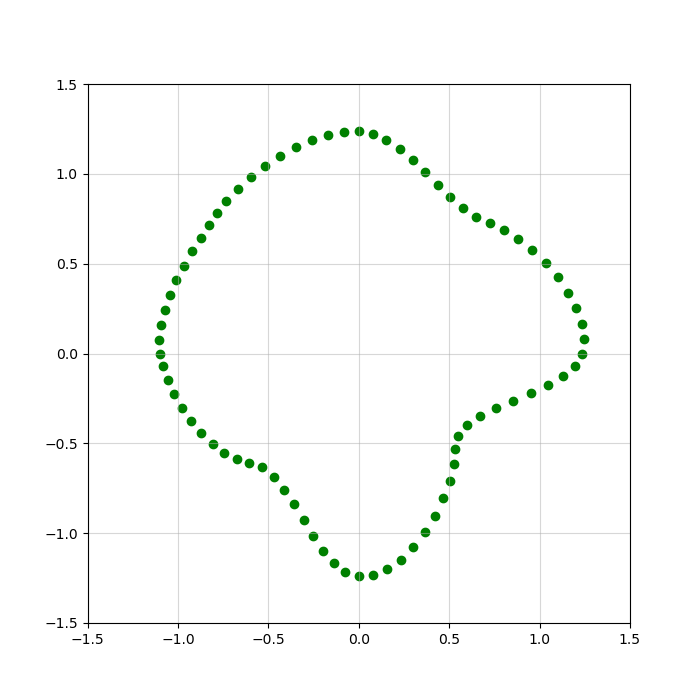}
            \vspace{-1cm}
            \caption{Sampled points on the boundary of the deformed cylinder.}
            \label{Sampling points}
        \end{minipage}
    \end{figure}   
    \item [vi)]Sample each cubic $\mathcal{B}_i$ by 9 intermediate points (apart from the cubic's ends $p_i$ and $p_{i+1}$), as depicted in Figure \ref{Sampling points}. This specific choice of intermediate points lead to a one-to-one correspondence between the deformed and the undeformed meshes at the obstacle's boundary.
\end{itemize}

\section{Radial Basis Functions for mesh deformations}\label{RBF}
As discussed in Section \ref{fvm discretization FOM algorithm}, FV methods rely on the spatial discretization of a physical domain via a computational mesh. When the geometry deforms, this framework requires solving a multi-dimensional \textit{scattered data interpolation problem} to determine the displacement of internal nodes.
\begin{definition}\label{interpolation's def}
    Given a set of data $(\boldsymbol{x}_i, y_i)$, $i = 1$, \dots, $N$, with $\boldsymbol{x}_i \in \mathbb{R}^s$, $y_i \in \mathbb{R}$, an interpolating function of these data is a (continuous) function $\mathcal{P}: \mathbb{R}^s \longmapsto \mathbb{R}$ such that $\mathcal{P}(\boldsymbol{x}_i) = y_i$, for $i = 1$, \dots, $N$.
\end{definition}
A common approach \cite{fasshauer2007meshfree} consists of assuming that the interpolating function $\mathcal{P}$ is a linear combination of basis functions $B_k$
\begin{equation}\label{interpolation function w.r.t. basis B_k}
    \mathcal{P}(\boldsymbol{x}) = \sum_{k=1}^Nc_k B_k(\boldsymbol{x}).
\end{equation}
Under this assumption, solving the interpolation problem described in definition \ref{interpolation's def} consists of solving a linear system of the form
\begin{equation}\label{linear system of data interpolation}
    A \boldsymbol{c} = \boldsymbol{y},
\end{equation}
where $A_{ij} = B_j(\boldsymbol{x}_i)$, $i,j = 1$, .., $N$, $\boldsymbol{c} = [c_1, \dots, c_N]^T$ and $\boldsymbol{y} = [y_1, \dots, y_N]^T$.\\
A particular and widely adopted choice of mesh-dependent functions are Radial Functions \cite{DEBOER2007784}.
\begin{definition}\label{radial function}
A function $\phi: \mathbb{R}^s \longmapsto \mathbb{R}$ is called \textit{radial} if there exists a function $\varphi : [0, + \infty) \longmapsto \mathbb{R}$ such that
\begin{equation*}
    \phi(\boldsymbol{x}) = \varphi(\rho) \: , \qquad \mathrm{where} \quad \rho = \norm{x}.
\end{equation*}
\end{definition}
Radial Basis Function (RBF) interpolation \cite{fasshauer2007meshfree, DEBOER2007784} leverages these radially symmetric kernels centered at the data locations. Given the scattered dataset $\{(\boldsymbol{x}_i, f(\boldsymbol{x}_i))\}_{i=1}^N$, the continuous interpolant $\mathcal{P}_f : \mathbb{R}^s \longrightarrow \mathbb{R}$ is formulated as:
\begin{equation}\label{rbf interpolant}
    \mathcal{P}_f(\boldsymbol{x}) = \sum_{k = 1}^N c_k \varphi\left(\norm{\boldsymbol{x} - \boldsymbol{x}_k}\right),
\end{equation}
where the coefficients $c_k$ are found by imposing the interpolation conditions
\begin{equation*}
    \mathcal{P}_f(\boldsymbol{x}_i) = f(\boldsymbol{x}_i) \:\:\:\:\:, \: i = 1, \, \dots, \, N,
\end{equation*}
i.e. solving the following linear system
\begin{equation}\label{interpolation condition}
    \begin{pmatrix}
        \varphi(\norm{\boldsymbol{x}_1 - \boldsymbol{x}_1}) & \varphi(\norm{\boldsymbol{x}_1 - \boldsymbol{x}_2}) & \cdots & \varphi(\norm{\boldsymbol{x}_1 - \boldsymbol{x}_N}) \\
        \varphi(\norm{\boldsymbol{x}_2 - \boldsymbol{x}_1}) & \varphi(\norm{\boldsymbol{x}_2 - \boldsymbol{x}_2}) & \cdots & \varphi(\norm{\boldsymbol{x}_2 - \boldsymbol{x}_N}) \\
        \vdots & \vdots & \ddots & \vdots \\
        \varphi(\norm{\boldsymbol{x}_N - \boldsymbol{x}_1}) & \varphi(\norm{\boldsymbol{x}_N - \boldsymbol{x}_2})  & \cdots &
        \varphi(\norm{\boldsymbol{x}_N - \boldsymbol{x}_N})
    \end{pmatrix}
    \begin{pmatrix}
        c_1 \\
        c_2 \\
        \vdots \\
        c_N
    \end{pmatrix}
    = 
    \begin{pmatrix}
       f(\boldsymbol{x}_1) \\
       f(\boldsymbol{x}_2) \\
       \vdots \\
       f(\boldsymbol{x}_N)
    \end{pmatrix}.
\end{equation}
In the present mesh-motion framework, the set of interpolation centers is partitioned into two distinct subsets, $\{\boldsymbol{x}_1, \dots, \boldsymbol{x}_N\} = \{\boldsymbol{x}_1, \dots, \boldsymbol{x}_{N_1}\} \cup \{\boldsymbol{x}_{N_1+1}, \dots, \boldsymbol{x}_{N_2}\}$. These correspond to nodes lying on the internal obstacle boundary $\partial\Omega$ and the external far-field boundaries, respectively. These coordinates are mapped to the updated spatial configurations via the displacement function $f$:
\begin{equation}\label{continuous mesh motion definition}
    \boldsymbol{y}_i = f(\boldsymbol{x}_i) = 
    \begin{dcases}
        \boldsymbol{z}_i \:, \quad \text{if } i \in \{1, \dots, N_1\}\\
        \boldsymbol{x}_i \:, \quad \text{if } i \in \{N_1+1, \dots, N_2\}
    \end{dcases},
\end{equation}
where $\boldsymbol{z}_i$ are the mesh points lying on the boundary of the deformed obstacles $\partial\Omega(\mub)$. The grid internal points are relocated subsequently, by evaluating the interpolating function $P_f$ defined in \eqref{rbf interpolant}, obtained through the solution of \eqref{interpolation condition}.\\
Different RBFs can be taken into account in the mesh motion strategy. Given a positive shape parameter $r \in \mathbb{R}^+$, standard options include:
\begin{itemize}
    \item Gaussian splines
    \begin{equation*}
        \varphi(\|\boldsymbol{x}\|) = e^{-\frac{\|{\boldsymbol{x}}\|^2}{r^2}}
    \end{equation*}
    \item Multi-quadratic biharmonic splines
    \begin{equation*}
        \varphi(\|\boldsymbol{x}\|) = \sqrt{\norm{\boldsymbol{x}}^2 + r^2}
    \end{equation*}
    \item Inverse multi-quadratic biharmonic splines
    \begin{equation*}
        \varphi(\|\boldsymbol{x}\|) = (\norm{\boldsymbol{x}}^2 + r^2 )^{-\frac{1}{2}}.
    \end{equation*}
\end{itemize}
In Table \ref{tab:rbf_comparison}, the topological effects of these different RBF kernels on the resulting mesh metrics are quantified. Based on this analysis, multiquadric biharmonic splines with a shape parameter $r = 0.1$ are selected for this study. This choice is dictated by the strict requirement to minimize both the non-orthogonality and the skewness of the internal cells while preserving a compact support scale. Excessively large values of $r$ lead to severe grid deterioration, eventually resulting in computational cell inversion and solver divergence.
\begin{table}[!htb]
    \centering
    \renewcommand{\arraystretch}{1.5}
    \begin{tabular}{|c|c|c|c|c|}
        \hline
        RBF & $r$ & Max non-ortho & Avg. non-ortho & Max skew. \\
        \hline
        \hline
        \textbf{Multi. bihar.} & \textbf{0.1} & \textbf{44.17} & \textbf{10.83} & \textbf{0.57} \\
        \hline
        Multi. bihar. & 0.2 & 44.30 & 10.87 & 0.57 \\
        \hline
        Multi. bihar. & 0.3 & 44.47 & 10.95 & 0.57 \\
        \hline
        Multi. bihar. & 0.4 & 44.70 & 11.03 & 0.57 \\
        \hline
        Multi. bihar. & 0.5 & 45.14 & 11.17 & 0.57 \\
        \hline
        Multi. bihar. & 0.6 & 79.87 & 14.82 & 1.03 \\
        \hline
        Multi. bihar. & 0.7 & 179.71 & 66.54 & 625.21 \\
        \hline
        Gaussian & 0.1 & 43.24 & 10.36 & 1.19 \\
        \hline
        Gaussian & 0.2 & 43.41 & 10.40 & 0.74 \\
        \hline
        Gaussian & 0.3 & 179.34 & 24.43 & 65.27 \\
        \hline
        Inv. Multi. bihar. & 0.1 & 43.31 & 10.45 & 0.72 \\
        \hline
        Inv. Multi. bihar. & 0.2 & 43.26 & 10.51 & 0.60 \\
        \hline
        Inv. Multi. bihar. & 0.3 & 43.23 & 10.56 & 0.56 \\
        \hline
        Inv. Multi. bihar. & 0.4 & 43.19 & 10.61 & 0.56 \\
        \hline
        Inv. Multi. bihar. & 0.5 & 43.23 & 10.66 & 0.57 \\
        \hline
        Inv. Multi. bihar. & 0.6 & 43.29 & 10.73 & 0.57 \\
        \hline
        Inv. Multi. bihar. & 0.7 & 70.84 & 12.68 & 1.14 \\
        \hline
        Inv. Multi. bihar. & 0.8 & 179.87 & 36.96 & 235.42 \\
        \hline
    \end{tabular}
    \caption{Comparison of mesh deformation through different RBFs. The effects of the specific RBF and parameter $r$ are analyzed in terms of induced maximum non-orthogonality, average non-orthogonality and maximum skewness. Bold values represent the best option in terms of parameter magnitude, non-orthogonality and skewness.}
    \label{tab:rbf_comparison}
\end{table}

\end{document}